\documentclass[fleqn,usenatbib]{mnras}

\usepackage{newtxtext,newtxmath}
\usepackage{caption}

\usepackage[T1]{fontenc}

\DeclareRobustCommand{\VAN}[3]{#2}
\let\VANthebibliography\thebibliography
\def\thebibliography{\DeclareRobustCommand{\VAN}[3]{##3}\VANthebibliography}

\usepackage{graphicx}	% Including figure files
\usepackage{amsmath}	% Advanced maths commands
\DeclareFontEncoding{LS1}{}{}
\DeclareFontSubstitution{LS1}{stix}{m}{n}
\DeclareSymbolFont{symbols4}{LS1}{stixbb}{m}{it}
\DeclareMathSymbol{\varhexagonblack}{\mathord}{symbols4}{"DD}
\DeclareMathSymbol{\hexagonblack}   {\mathord}{symbols4}{"DE}

\newcommand{\ewha}{$\text{EW}_{\text{H}\alpha+\text{N}\textsc{ii}}$\,}

\title[JWST's PEARLS: TN J1338--1942 - I]{JWST's PEARLS: TN J1338--1942 - I. Extreme jet triggered star-formation in a $z=4.11$ luminous radio galaxy}
\author[K. J. Duncan et al.]{Kenneth J. Duncan,$^{1}$\thanks{E-mail: kdun@roe.ac.uk (KJD)}
Rogier A. Windhorst,$^{2}$
Anton M. Koekemoer,$^{3}$
Huub J. A. R\"{o}ttgering,$^{4}$\newauthor
Seth H. Cohen,$^{2}$
Rolf A. Jansen,$^{2}$
Jake Summers,$^{2}$
Scott Tompkins,$^{2}$
Taylor A. Hutchison,$^{5}$\newauthor
Christopher J. Conselice,$^{6}$
Simon P. Driver,$^{7}$
Haojing Yan,$^{8}$
Nathan J. Adams,$^{6}$
Cheng Cheng,$^{9}$\newauthor
Dan Coe,$^{3,10,11}$
Jose M. Diego,$^{12}$
Herv\'e Dole,$^{13}$
Brenda Frye,$^{14}$
Hansung B. Gim,$^{15}$
Norman A. Grogin,$^{3}$\newauthor
Benne W. Holwerda,$^{16}$
Jeremy Lim,$^{17}$
Madeline A. Marshall,$^{18,19}$
Mario Nonino,$^{20}$\newauthor
Nor Pirzkal,$^{3}$
Aaron Robotham,$^{7}$
Russell E. Ryan, Jr.,$^{3}$
and Christopher N. A. Willmer$^{114}$
\\
$^{1}$Institute for Astronomy, University of Edinburgh Royal Observatory, Blackford Hill, Edinburgh, EH9 3HJ, UK\\
$^{2}$School of Earth and Space Exploration, Arizona State University, Tempe, AZ 85287-1404, USA\\
$^{3}$Space Telescope Science Institute, 3700 San Martin Drive, Baltimore, MD 21218, USA\\
$^{4}$Leiden Observatory, PO Box 9513, 2300 RA Leiden, The Netherlands\\
$^{5}$Astrophysics Science Division, NASA Goddard Space Flight Center, 8800 Greenbelt Rd, Greenbelt, MD 20771, USA\\
$^{6}$Jodrell Bank Centre for Astrophysics, Alan Turing Building, University of Manchester, Oxford Road, Manchester M13 9PL, UK\\
$^{7}$International center for Radio Astronomy Research (ICRAR) and the International Space center (ISC), The University of Western Australia,\\ M468 35 Stirling Highway, Crawley, WA 6009, Australia\\
$^{8}$Department of Physics and Astronomy, University of Missouri,
Columbia, MO 65211, USA\\
$^{9}$Chinese Academy of Sciences, National Astronomical Observatories, CAS, Beijing 100101, China\\
$^{10}$Association of Universities for Research in Astronomy (AURA) for the European Space Agency (ESA), STScI, Baltimore, MD 21218, USA\\
$^{11}$Center for Astrophysical Sciences, Department of Physics and Astronomy, The Johns Hopkins University, 3400 N Charles St. Baltimore, MD 21218, USA\\
$^{12}$Instituto de F\'isica de Cantabria (CSIC-UC). Avenida. Los Castros s/n. 39005 Santander, Spain\\
$^{13}$Universit\'e Paris-Saclay, CNRS, Institut d'Astrophysique Spatiale, 91405, Orsay, France\\
$^{14}$Steward Observatory, University of Arizona, 933 N Cherry Ave,
Tucson, AZ, 85721-0009, USA\\
$^{15}$Department of Physics, Montana State University, P. O. Box 173840, 
Bozeman, MT 59717, USA\\
$^{16}$Department of Physics and Astronomy, University of Louisville,
Louisville KY 40292, USA\\
$^{17}$Department of Physics, The University of Hong Kong, Pokfulam Road, 
Hong Kong\\
$^{18}$National Research Council of Canada, Herzberg Astronomy \&
Astrophysics Research center, 5071 West Saanich Road, Victoria, BC V9E 2E7, Canada\\
$^{19}$ARC center of Excellence for All Sky Astrophysics in 3 Dimensions
(ASTRO 3D), Australia\\
$^{20}$INAF-Osservatorio Astronomico di Trieste, Via Bazzoni 2, 34124
Trieste, Italy\vspace{-0.5cm}
}

\date{Accepted 2023 April 21. Received 2023 April 20; in original form 2022 December 18}

\pubyear{2023}

\begin{document}
\label{firstpage}
\pagerange{\pageref{firstpage}--\pageref{lastpage}}
\maketitle

\begin{abstract}
We present the first JWST observations of the $z=4.11$ luminous radio galaxy TN J1338--1942, obtained as part of the ``Prime Extragalactic Areas for Reionization and Lensing Science'' (``PEARLS'') project.
Our NIRCam observations, designed to probe the key rest-frame optical continuum and emission line features at this redshift, enable resolved spectral energy distribution modelling that incorporates both a range of stellar population assumptions and radiative shock models.
With an estimated stellar mass of $\log_{10}(M/\text{M}_{\odot}) \sim 10.9$, TN J1338--1942 is confirmed to be one of the most massive galaxies known at this epoch.
Our observations also reveal extremely high equivalent-width nebular emission coincident with the luminous AGN jets that is best fit by radiative shocks surrounded by extensive recent star-formation.
We estimate the total star-formation rate (SFR) could be as high as $\sim1600\,\text{M}_{\odot}\,\text{yr}^{-1}$, with the SFR that we attribute to the jet induced burst conservatively $\gtrsim500\,\text{M}_{\odot}\,\text{yr}^{-1}$.
The mass-weighted age of the star-formation, $t_{\textup{mass}} <4$ Myr, is consistent with the likely age of the jets responsible for the triggered activity and significantly younger than that measured in the core of the host galaxy.
The extreme scale of the potential jet-triggered star-formation activity indicates the potential importance of positive AGN feedback in the earliest stages of massive galaxy formation, with our observations also illustrating the extraordinary prospects for detailed studies of high-redshift galaxies with JWST.
\end{abstract}


%%%%%%%%%%%%%%%%%%%%%%%%%%%%%%%%%%%%%%%%%%%%%%%%%%

%%%%%%%%%%%%%%%%% BODY OF PAPER %%%%%%%%%%%%%%%%%%

\section{Introduction} \label{sec:intro}
The luminous high redshift radio galaxy (HzRG), TN\,J1338--1942 \citep[TNJ1338 hereafter;][$z=4.11$]{DeBreuck1999}, is one of the most powerful radio sources known in the early Universe. 
Initially selected based on its bright radio continuum emission and ultra-steep radio spectral index \citep[][$S_{\nu} \propto\nu^{-1.31}$]{DeBreuck2000}, TNJ1338 was subsequently discovered to reside at the heart of a significant over-density of galaxies extending to Mpc scales \citep{Venemans2002, Miley2004, Zirm2005, Intema2006, Overzier2008, Saito2015}, making it one of the earliest known proto-cluster structures, and a potential progenitor of the most massive clusters (and brightest cluster galaxies) in the local Universe.

As host to extremely luminous extended Ly$\alpha$ emission \citep[$L_{\text{Ly}\alpha} \sim 4\times 10^{44}\,\text{erg\,s}^{-1}$;][]{DeBreuck2000,Venemans2002,Swinbank2015}, potential jet-induced star-formation activity \citep{Miley2004, Zirm2005} and evidence of large active galactic nuclei (AGN) driven outflows \citep{Swinbank2015}, 
TNJ1338 and its wider environment therefore represent a unique laboratory for studying both AGN feedback in action and the earliest stages of proto-cluster formation.
Given these extreme properties and the wealth of multi-wavelength observations available, TNJ1338 therefore represents a prime target for JWST observations, with Cycle 1 programmes in both imaging \citep[GTO 1176, PI: Windhorst;][]{PearlsOverview} and resolved spectroscopy (GO 1964, PIs: Overzier and Saxena) aiming to address a broad range of scientific questions.

One of the key outstanding challenges for radio-loud AGN such as TNJ1338 is understanding how their radio jets interact with their surrounding environments \citep[see e.g.][and references therein]{Fabian2012, King2015}.
For example, how, and to what degree, do the jets trigger or suppress star-formation?
One of the standout observational features of TNJ1338 is the presence of extended optical emission aligned with the most luminous radio lobe \citep{Miley2004, Zirm2005}.
Based on its rest-frame UV emission \citet{Zirm2005} estimated that non-stellar processes (scattered light, synchrotron and inverse-Compton emission) can contribute only a few percent of the rest-frame UV emission aligned with the radio jet; recent star-formation activity is therefore the dominant emission mechanism.

A primary objective of this study is therefore to resolve the nature of the extended emission aligned with the radio jet.
Is star-formation the dominant emission mechanism? If so, how does the age of this star-formation activity compare to the jet activity (and the accretion that powered this)? How significant is the triggered activity in the evolution of the host galaxy?

HzRGs are thought to host some of the most massive super-massive black holes (SMBHs) at a given epoch \citep{Miley2008}.
Crucially, unlike luminous quasars, the AGN in HzRG do not outshine their hosts, allowing detailed studies of the galaxies in which the most massive SMBH reside without being dazzled by the quasar light \citep[e.g.][]{Marshall2021}.
A secondary of this study is to exploit the high spatial resolution and sensitivity at rest-frame optical wavelengths now available from JWST to perform the first resolved study of the physical properties of a HzRG host galaxy, and begin placing high-redshift accretion activity into its wider context.
Specifically, we want to advance beyond estimations of just the total stellar mass, establishing instead \textit{when} and how the host galaxy formed.
%How do the massive HzRG hosts observed at $2 < z < 6$ relate to the first galaxies discovered at $z > 12$ \textbf{[Add JWST $z > 10$ citations]} or the surprisingly evolved galaxies identified $6 < z < 10$ \textbf{[Labbe et al.? And similar]}?

% However, to definitively link the observed emission to the jet activity firstly requires confirming the source of the emission as star-formation, and secondly relating the age of this star-formation activity to that of the radio lobes.

Given these objectives, the filters for our JWST/NIRCam observations were selected to maximise the colour information available for sources at the systemic redshift of TNJ1338 \citep[$z=4.11$, derived from the observed He\textsc{ii} $\lambda1640$ emission line;][]{Swinbank2015}, with medium-band filters constraining key rest-frame optical features.
Fortuitously, at $z\sim4.1$, the NIRCam SW filters at 1.5 to $2.1\mu$m bracket the stellar age-sensitive Balmer/$D_{4000\textup{\AA}}$ break, while the LW medium bands offer a clean measure of the rest-frame optical continuum and the strength of the H$\alpha$+N\textsc{ii} nebular emission line complex.
With resolved constraints on both the older stellar population and recent star-forming activity respectively, NIRCam will enable us to probe the nature of TNJ1338 through detailed spectral energy distribution (SED) modelling in unprecedented detail for a galaxy at this redshift.

The rest of this paper is set out as follows: Section~\ref{sec:data} outlines the NIRCam and accompanying ancillary photometric data used in this study, as well as the methodologies for data reduction and homogenisation.
In Section~\ref{sec:visual_results} we present a qualitative analysis of the rest-frame optical morphology of TNJ1338 revealed by NIRCam.
Next, Section~\ref{sec:resolved_analysis} outlines the methodology for our quantitative analysis of the resolved properties of TNJ1338 through SED modelling.
Section~\ref{sec:results} then presents the results of SED modelling analysis and our discussion.
Finally, in Section~\ref{sec:conclusions}, we summarise the results and our conclusions. 
Throughout this paper, all magnitudes are quoted in the AB system \citep{OkeGunn1983} unless otherwise stated.
We also assume a $\Lambda$ Cold Dark Matter cosmology with $H_{0} = 70$ km\,s$^{-1}$\,Mpc$^{-1}$, $\Omega_{m}=0.3$ and $\Omega_{\Lambda}=0.7$ (corresponding to a scale of $6.87\,\text{kpc\,}$ per \arcsec).\\

\section{Data}\label{sec:data}

\subsection{Optical Imaging}
The Hubble Space Telescope (HST) Advanced Camera for Surveys (ACS) observations of the TNJ1338 field used in our analysis comprise 18 orbits split across four broadband filters: 9,400s in F475W and F625W, 11,700 s in F775W, and 11,800s in F850LP \citep[Proposal 9291;][]{Miley2004, Zirm2005}.
These four filters span the Lyman break and rest-frame UV at $z\sim4.1$, with the F625W-band probing the Ly$\alpha$ emission line.
The reduced ACS observations were retrieved from the Hubble Legacy Archive (HLA)\footnote{\url{https://hla.stsci.edu}}, with processing using the standard HLA reduction pipelines, including appropriate bias and dark current subtraction, flat-fielding and geometric distortion corrections. 
In the HLA processing, the exposures for each band have been \textit{drizzled} onto a consistent 0.04\arcsec\ pixel scale.

% \begin{figure}
%     \includegraphics[width=\columnwidth]{figures/pearls_tnj1338_rgz_acs.pdf}
% 	\caption{Rest-frame UV image of the HzRG TNJ1338. Red, green and blue channels are taken from the HST/ACS \textit{F775W},  \textit{F625W} and \textit{F475W} respectively. Luminous, extended Ly$\alpha$ emission associated with TNJ1338 is clearly visible (green channel). \textbf{[Illustrative: To be updated to rest-frame optical NIRCam image and moved to results section]}}
%     \label{fig:acs_cutouts}
% \end{figure}

In addition to the HST/ACS observations, TNJ1338 has also been observed with ground-based multi-band optical imaging extending over a wider field-of-view (FOV) using the Subaru/Suprime-Cam imager \citep[$B$, $R$, $i'$ and $IA624$ narrow-band,][]{Intema2006, Saito2015}.
% The Suprime-Cam imaging has a FOV of $34\times27$ arcmin and a native pixel scale of 0.202\arcsec, with 5$\sigma$ limiting magnitudes of 26.87, 26.92, 26.56 and 26.94 for the $B$, $R$, $i'$ and $IA624$ bands respectively (within a 1.5\arcsec aperture).
We retrieve the raw Suprime images from the SMOKA archive \citep{Baba2002} and produce a stacked mosaic with a custom pipeline that includes astrometric corrections performed using Gaia EDR3 objects \citep{Gaia2021}, taking into account their proper motion to the epoch of Suprime observations.
The $i'$-band image, reaching a 5$\sigma$ limiting magnitude of 26.56 is used to provide a reference catalogue for astrometric alignment of the JWST/NIRCam imaging as outlined below.
Due to the increased depth and resolution afforded by the HST/ACS imaging, we do not include the Subaru ground-based imaging in our analysis of TNJ1338 directly.
% The ground-based imaging is however used as a 

\subsection{NIRCam Imaging}
The JWST/NIRCam observations of the TNJ1338 field were taken as part of the ``Prime Extragalactic Areas for Reionization and Lensing Science" \citep[PEARLS;][]{PearlsOverview} Guaranteed Time Observations (GTO) programme.
These observations, taken on July 1st 2022, consist of imaging in the $F150W$, $F182M$, $F210M$, $F300M$, $F335M$ and $F360M$ filters with a uniform integration time of 1030.73 seconds in each filter (SHALLOW4 readout pattern and 4-point INTRAMODULEBOX dithers).
For a full description of the data reduction steps, we refer the reader to \citet{PearlsOverview}.
Here we summarise the key steps, including those specific to the processing of the TNJ1338 imaging.

Initial calibration and processing is done using the standard STScI \texttt{CALWEBB}\footnote{\url{https://jwst-docs.stsci.edu/getting-started-with-jwst-data}}.
All analyses and images presented in this paper are based on the latest reductions using Pipeline version 1.7.2 and the context file `jwst\_0995.pmap\_filters' made available in October 2022, which incorporates the on-orbit flat-fielding observations and improved flux calibrations.
Detector level offsets in the SW filters and $1/f$-noise in both SW and LW images are removed in individual frames using the \texttt{ProFound} code \citep{Robotham2017,Robotham2018}.
We also remove stray light ``wisps" in the SW imaging by subtracting a wisp template that best matches the amplitude observed in that specific image.

Before mosaicing, it is essential that individual frames are placed on a consistent and reliable astrometric frame.
However, due to the unexpected early scheduling of the TNJ1338 visit\footnote{The JWST visit for this field was designed to ensure clean imaging of the TNJ1338 radio galaxy, regardless of exact roll-angle, using a positional offset that locates the prime galaxy at the centreof one of the SW detectors. Based on the long-range schedule window of Jan-Mar 2023 (and the associated roll angle range), a positional offset was submitted to maximise the overlap between one NIRCam module and that ACS imaging, whilst also placing the second module over a number of confirmed proto-cluster members \citep{Venemans2002, Saito2015}, the observed roll angle of the NIRCam imaging resulted in reduced overlap with the existing HST/ACS observations described above. The unexpected early scheduling resulted in a field orientation $\sim180\deg$, still centring the TNJ1338 radio galaxy at the centre of a SW detector (by design) but reducing the area with combined HST and JWST coverage.} the observed roll angle of the NIRCam imaging resulted in reduced overlap with the existing HST/ACS observations described above.
Image registration for the NIRCam frames is therefore done using the wide area Subaru imaging and associated catalogues on the Gaia DR3 reference frame \citep{Gaia2021,Gaia2022}, specifically the $i'$-band, which has the greatest sensitivity for redder objects.
Positional offsets (adjusted for proper motion) between the NIRCam images and the reference catalogue are used to adjust each frame's centre and positional angle until the offsets are minimised.
The aligned images are then \emph{drizzled} onto mosaics with a 0.03\arcsec\, pixel scale using the \texttt{AstroDrizzle} package \citep{Koekemoer2013, Avila2015}. 

As measured by \citet{PearlsOverview}, the resulting mosaics reach 5-$\sigma$ limiting magnitudes of 27.7, 27.4 and 27.2 in the SW bands ($F150W$, $F182M$, $F210M$ respectively), and 28.35, 28.25 and 28.16 in the LW bands ($F300M$, $F335M$ and $F360M$ respectively).
In line with other observations, the measured sensitivities represent a small reduction in the SW bands when compared to pre-launch Exposure Time Calculations ($\sim 0.1\text{mag}$).
However, the measured LW bands represent a substantial improvement on expectations, typically reaching $\sim1$ mag deeper.

\begin{figure}
    \includegraphics[width=\columnwidth]{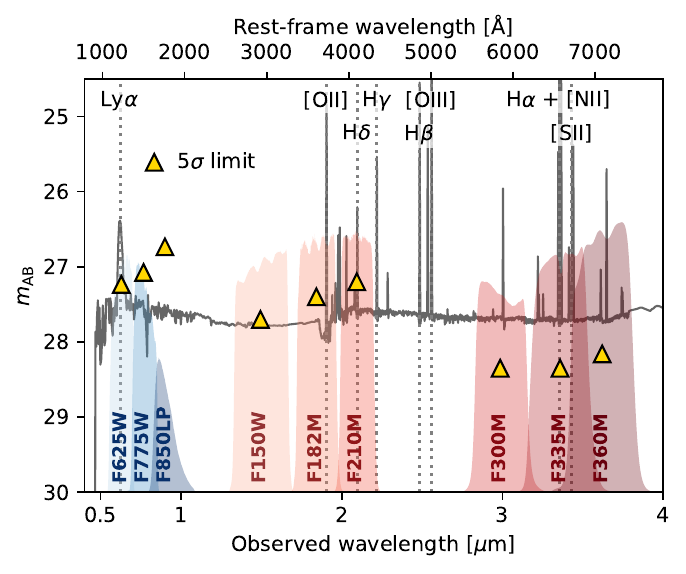}
	\caption{Wavelength range and limiting sensitivies of the photometric observations used in this analysis. Triangles illustrate the 5$\sigma$ limiting magnitudes for the HST/ACS \citep{Zirm2005} and JWST/NIRCam \citep{PearlsOverview} bands, with the corresponding  filter transmission profiles shown as blue and red shaded regions respectively. Also shown for reference is a canonical stellar population as seen at $z=4.11$ (100 Myr old continuous star-formation history, solar metallicity and zero dust attenuation) with a total stellar mass of $10^{8}\,\text{M}_{\odot}$. Vertical dotted lines illustrate key UV and optical emission lines, with individual species (or pairs) labelled.}
    \label{fig:filters}
\end{figure}

Finally, using the $F335M$ NIRCam mosaic as our reference, we use the \texttt{reproject} package to place the ACS observations onto the unified pixel grid defined by the NIRCam mosaics.
Comparing the positional offsets of sources detected in both the ACS and NIRCam imaging, we confirm agreement between the reprojected images and the drizzled mosaics is at the sub-pixel level.
The combined HST/ACS and JWST/NIRCam filter coverage and corresponding limiting sensitivity is illustrated in  Fig.~\ref{fig:filters}, as well as an illustrative galaxy spectral energy distribution (SED) at $z=4.11$ - highlighting both the key spectral features probed by the NIRCam bands and the unprecedented sensitivity at 1-4$\mu$m.

\subsection{Additional ancillary data}\label{sec:ancillary}
Due to its extreme nature, TNJ1338 and its surrounding protocluster environment have also been observed across the full electromagnetic spectrum.
Following its initial detection in the low-frequency radio \citep{DeBreuck2000}, TNJ1338 has also been observed with higher frequency radio continuum observations.
In this study, we primarily make use of the highest available resolution imaging provided by the \emph{Karl G. Jansky} Very Large Array (VLA) X-band observations presented by \citet{Pentericci2000}.
The observations are centreed at a frequency of 8.46 GHz, reaching an RMS sensitivity of 25$\mu\text{Jy\,beam}^{-1}$, and have a spatial resolution of 0.23\arcsec, sufficient to resolve the jet structure on the scale of the observed optical emission \citep{Zirm2005}.

Additional observations spanning the infrared include 24$\mu$m \emph{Spitzer}/MIPS \citep{DeBreuck2010}, 100-500$\mu$m \emph{Herschel}/PACS and SPIRE \citep{Drouart2016}, $450$/850$\mu$m JCMT/SCUBA \citep[][]{DeBreuck2004}, 1.2mm IRAM/MAMBO observations \citep[][]{DeBreuck2004}, and ALMA Band-3 \citep{Falkendal2019}.
However, none of these observations have sufficient spatial resolution to incorporate within any resolved analysis on the properties of TNJ1338.
Nevertheless, together they can still place valuable constraints on the integrated IR emission from TNJ1338.

% \begin{itemize}
%     \item SCUBA-2 m and IRAM MAMBO 1.2mm observations \citep{DeBreuck2004}: not essential, but valuable constraint for integrated SED fitting. 
%     % \item VLT ISAAC K: As above, not essential given the depth NIRCam medium bands - however the broadband $K_{\text{S}}$ will add additional colour information for bright targets (i.e. TN 1338). 
% \end{itemize}

\subsection{Point-spread function homogenisation}
To ensure accurate pixel by pixel colours and enable studies of the resolved SED, we homogenise all of the available ACS and NIRCam imaging to a common point-spread function (PSF).
Empirical PSFs are first generated from stacked stars within the field, with the initial selection of bright stars from Gaia \citep{Gaia2022} visually inspected to exclude saturated sources, stars with close neighbours and stars near the edge of the field with incomplete coverage within the cutouts.
Before generating the final stacked PSF, we use the \texttt{Photutils.ePSF} package \citep{PhotutilsASCL, Photutils1.5.0} to iteratively re-centre each star using a pixel oversampling of four.
Due to the small number of suitable stars available in the field, we found that the full \texttt{Photutils.ePSF} effective PSF modelling approach was not able to accurately model the full structure of the JWST/NIRCam PSF. 
We therefore generate the final PSFs from a simple median stack of the re-centred star images with each star first normalised by the flux contained within a 0.18\arcsec\, diameter aperture.

To derive the convolution kernels required to transform the measured PSFs to a given target PSF we employ the \texttt{pypher} package \citep{Boucaud2016}.
Nominally, the F360M filter exhibits the broadest PSF full width half-maximum (FWHM) within our filter set and would be a suitable target PSF for JWST/NIRCam-only studies.
However, we find that that difference in structure between the wings of the HST/ACS and JWST/NIRCam PSFs can result in excess noise within the convolution kernel.
Our target PSF is therefore defined to be a Moffat profile ($f(r) \propto (1 + (r/\alpha)^{2})^{- \beta}$) with $\beta=2$ and $\alpha=3.63$ px, resulting in a $\textup{FWHM} = 0.14\arcsec$ fractionally larger than that measured for our largest empirical PSF (F360M; $0.13\arcsec$).

\begin{figure}
\centering
    \includegraphics[width=\columnwidth]{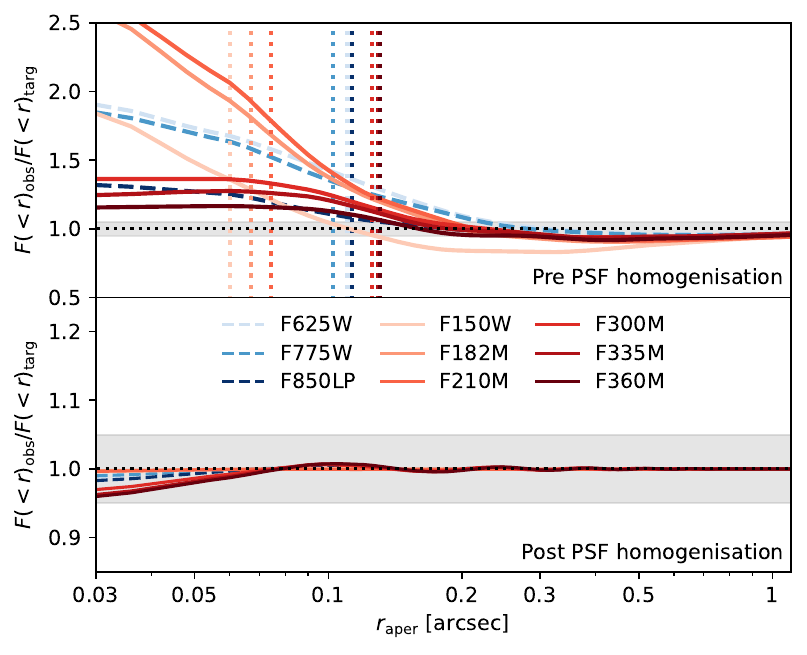}
	\caption{Enclosed flux as a function of aperture radius (i.e. the curve of growth) for the filters used in this analysis ($F(<r)_{\text{obs}}$) relative to that of the target Moffat PSF ($F(<r)_{\rm{targ}}$). Top and bottom panels show the measurements before and after PSF homogenisation. In the top panel, we also illustrate the FWHM of the measured PSFs based on fitting a 2D Moffat Profile to the stacked PSF image. In both panels the shaded region illustrates $\pm5\%$ around the optimal value of unity.}
    \label{fig:psf_homogenisation}
\end{figure}

\begin{figure}%\vspace{-0.5cm}
    \includegraphics[width=\columnwidth]{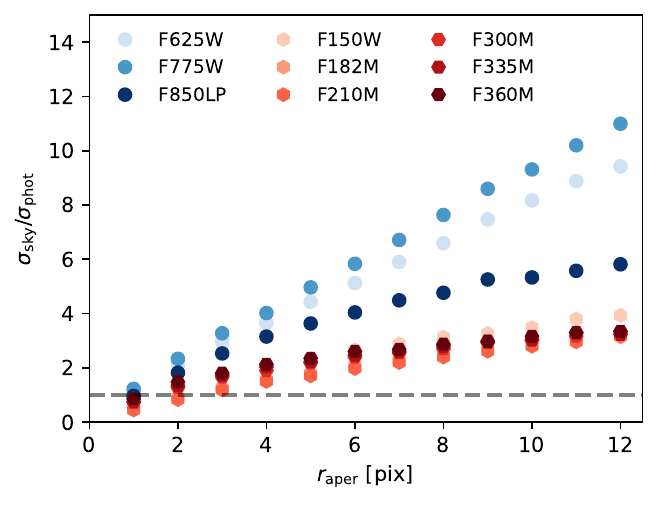}
	\caption{Aperture-dependent correction applied to measured uncertainties to account for correlated noise in the drizzled and PSF-homogenised images. Corrections for HST/ACS broadband filters are shown as circles with a blue colour-scale, while correction for JWST/NIRCam medium- and broadband filters are plotted as hexagons with a red colour-scale.}
    \label{fig:correlated_noise}
\end{figure}
Measuring the curves of growth in circular apertures for all HST/ACS and JWST/NIRCam bands after PSF homogenisation, we find that the PSFs of the convolved images agree to within $<5\%$ at all radii (see Fig.~\ref{fig:psf_homogenisation}).

\subsection{Correcting uncertainties for correlated noise}\label{sec:correlated_noise}
While both the HST and JWST imaging allow calculation of uncertainties within an aperture or region based on the noise maps, it is well understood that uncertainties derived by common photometry tools can be underestimated in the presence of correlated noise \citep{Leauthaud2007}.
Since the convolution of the drizzled images with a PSF kernel will further increase any existed correlation between pixels, uncertainties derived from the HST/ACS or JWST/NIRCam error maps can therefore underestimate the total uncertainty for larger regions (i.e. apertures, isophotes or source segments).

We estimate the corrections required to account for correlated noise in our subsequent photometry steps following an approach similar to other works in the literature \citep[][see also \citeauthor{Trenti2011}~\citeyear{Trenti2011} for an alternative approach]{Bielby2012,Laigle2016,Mehta2018}. 
Firstly, for each band we derive an initial source catalogue from the PSF-homogenised image, requiring a 5$\sigma$ detection threshold with a minimum of 10 connected pixels.
We then define $10^4$ random empty positions within the image, where the segmentation map derived from the detected sources is first dilated to ensure that the selected empty sky regions do not fall within extended sources.
The correlated noise correction for a given aperture size is then estimated by comparing the 3$\sigma$-clipped scatter measured in empty sky apertures with the median uncertainty measured for sources using the same size aperture. 
Since our resolved analysis of TNJ1338 will employ flux measurements over a range of different photometric aperture or segment sizes, we follow the approach of \citet{Mehta2018} and derive the correlated noise uncertainty corrections as a function of aperture radius (area).
The resulting measurements are shown in Fig.~\ref{fig:correlated_noise}.
For an aperture with radius $r=0.14\arcsec$ ($\sim5$ pixels; i.e. the FWHM of the PSF homogenised images), we find that the uncertainties are underestimated by a factor of $2-5$, with a filter dependence in line with naive expectations (i.e. that bands with greater oversampling with respect to their native resolution require greater corrections).

\begin{figure*}
\centering{
    \includegraphics[width=\textwidth,trim={0.2cm 0 0.1cm 0},clip]{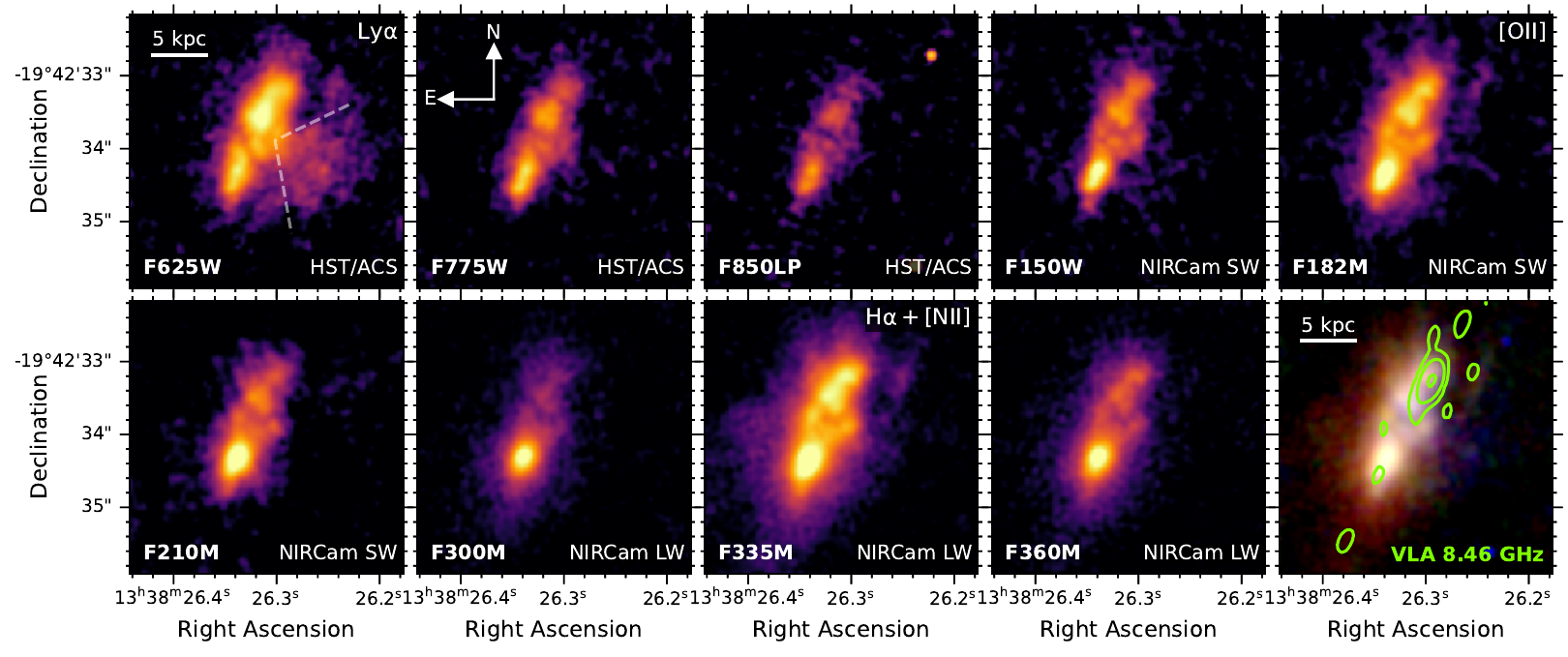}
	\caption{PSF-homogenised $3.75\arcsec\times3.75\arcsec$ cutouts around TNJ1338 for each of the bands used in this analysis, where all images are oriented per the annotation in the 2nd panel. The colour-scale for each image is based on the local noise properties, with an \emph{asinh} scaling between 1 and $50\times$ or $120\times$ the local $\sigma$-clipped noise for the ACS or NIRCam SW filters and NIRCam LW filters respectively. In the first panel ($F625W$), dashed lines indicate the approximate bounds of the extended Ly$\alpha$ emission feature labeled as ``the Wedge" by \citet{Zirm2005}. Filters containing prominent nebular emission lines have the the relevant line noted in the upper right of each panel (see also Fig.~\ref{fig:filters}). The RGB colour image in the lower-right panel combines the ACS imaging for the blue channel, NIRCam/SW imaging for the green channel and NIRCam/LW imaging for the red channel. Green contours show the VLA 8.46 GHz radio continuum emission associated with the galaxy, with contours starting at 4$\sigma$ (increasing by $4^n$).}
    \label{fig:rgb_cutouts}}
\end{figure*}

In our subsequent analysis, we calculate the required uncertainty correction for a given segment, isophotal or Kron flux measurement by interpolating the correlated uncertainty measurements based on the segment area.
We note that due to the extreme rest-frame optical luminosity of TNJ1338 and the sensitivity of the JWST observations, the limiting uncertainty on fluxes and derived colours is the remaining systematic uncertainty in the zeropoints between bands \citep[see Appendix B1 of][]{PearlsOverview}.
Nevertheless, since these zeropoint uncertainties are systematic within a given detector and will apply uniformly to TNJ1338, accounting for the correlated noise during SED fitting analysis is necessary to ensure that resolved studies correctly account for signal to noise variations across the host galaxy.

\section{JWST's view of TN J1338-1942}\label{sec:visual_results}

% \subsection{Stellar continuum and H$\alpha$ morphology of TNJ1338}
% \noindent\textit{-- What is the rest-frame optical morphology of TNJ1338?\\
% -- Is there any evidence of recent/ongoing merger activity?\\
% -- How does morphology of recent SF (H$\alpha$) compare to stellar continuum?\\
% -- How does morphology correlate with that of radio emission? (e.g. evidence of positive feedback)
% }

The combination of the physical extent of TNJ1338 with the high spatial resolution and sensitivity of JWST enables us to conduct the first spatially-resolved study of the rest-frame optical properties of an AGN host galaxy in the early Universe.
Fig.~\ref{fig:rgb_cutouts} shows the combined PSF-homogenised imaging of TNJ1338, from the rest-frame UV probed by HST/ACS to the rest-frame optical probed by the NIRCam/LW filters.

A number of key features are immediately revealed by the NIRCam imaging. 
Firstly, the JWST observations confirm the clear presence of a bright galaxy core located at the Southern side of the source.
Although present in VLT/ISAAC $K_{S}$ broadband imaging \citep{Zirm2005}, the ground-based imaging was not able to fully resolve the morphology of the rest-frame optical emission, or sensitive enough to rule out more extensive obscured emission.

% \begin{itemize}
%     \item 
%     \item Core versus extended emission show clear difference in colour. Jet aligned emission significantly bluer and evidence for emission line dominated SED.
%     \item Most clear through the $F300M - F335M$ colour... Fig.~\ref{fig:ew_map}
% \end{itemize}

\begin{figure}
\centering{
    \includegraphics[width=1.05\columnwidth]{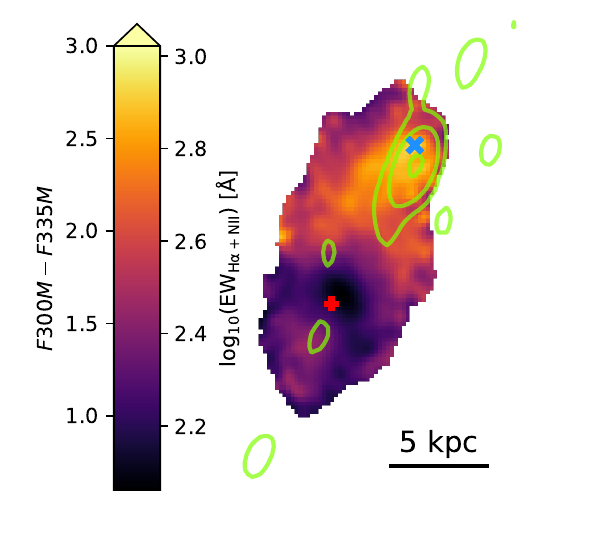}
	\caption{Observed $F300M - F335M$ colour across the extended emission in TNJ1338. The twinned colour scaling illustrates the corresponding H$\alpha$+N\textsc{ii} equivalent width inferred based on the simple assumption of a flat spectrum (see text). Only pixels with flux $>10\sigma$ above the local noise in both band are shown. Symbols illustrate the locations of the peak $F300M$ flux (red plus) and maximum $F300M - F335M$ colour (peak $\text{EW}_{\text{H}\alpha+\text{N}\textsc{ii}}$; blue cross). Green contours show the VLA 8.46 GHz radio continuum emission, with contours starting at 4$\sigma$ (increasing by $4^n$).}
    \label{fig:ew_map}}
\end{figure}

In the filters probing rest-frame wavelengths above the Balmer break ($F210M$ to $F360M$; see Fig.~\ref{fig:filters}) there are significant differences in colour for the galaxy core and jet-aligned emission (where the jet emission is illustrated in the Fig.~\ref{fig:rgb_cutouts}).
The extended rest-frame optical emission is visibly clumpy, with some of the  distinct regions previously identified in Ly$\alpha$ visible more clearly in the $F182M$ and $F335M$, filters that are dominated by the bright [O\textsc{II}] and H$\alpha$ emission lines respectively (see annotations in Fig.~\ref{fig:rgb_cutouts}).
% The difference is seen most starkly in $F300M - F335M$, indicating that the extended emission is dominated by nebular emission.

To illustrate the variation and amplitude of the optical emission line strengths across TNJ1338, in Fig.~\ref{fig:ew_map} we show the observed $F300M - F335M$ colour for the individual pixels with significant detections in both filters ($>10\times$ the local RMS background in both images).
To convert the measured colours into a physically meaningfully quantity independent of assumptions on the associated emission mechanism, we also estimate the $\text{H}\alpha+\text{N}\textsc{ii}$ emission line equivalent widths (EW) implied by the observed colours.
For the simplifying assumption of a flat intrinsic optical continuum, where the emission line-free colour would result in $F300M-F335M=0$, the EW of the combined emission line complex can be estimated as:
\begin{equation}
    \text{EW}_{\text{H}\alpha+\text{N}\textsc{ii}} = \frac{W_{\text{eff}}}{(1+z)} (10^{-0.4\Delta_{\text{mag}}} - 1)\,\text{\AA},
\end{equation}
where $\Delta_{\text{mag}}$ is the change in magnitude in the $F335M$ filter attributed to the $\text{H}\alpha+\text{N}\textsc{ii}$ emission line complex \citep[see e.g.][with $\Delta_{\text{mag}} = F335M - F300M$ for our assumption]{MarmolQueralto2016}, and the effective width, $W_{\text{eff}}$, for the $F335M$ filter is taken to be 3389.42\AA. 
We caution that the resulting \ewha estimates are only approximate, however as is illustrated later in the paper (see Figs~\ref{fig:sed_shocks} and \ref{fig:sed_galaxy}), the assumption of a flat intrinsic continuum is well motivated for much of the extended emission.

There is clear evidence for a strong variation in \ewha along the axis between the centre of the host galaxy (shown by the red plus) and the northern radio lobe, with the peak \ewha (blue cross) situated extremely close to the peak of the 8.46 GHz continuum emission.
Fig.~\ref{fig:ew_map} also reveals that there is also an increase in \ewha in the direction of the knots of high-frequency radio emission in the south east, albeit with substantially lower amplitude than the northern lobe.
The region of higher \ewha emission south-east of the galaxy core is visible as a knot of bluer rest-UV emission in the original HST/ACS imaging and NIRCam $F150W$.
The alignment of this higher \ewha region relative to the southern radio lobe and the centre of the rest-frame optical continuum add additional evidence for jet activity driving the nebular line emission. 
However, given the systematic offset in astrometry for the VLA observations could be as large as 0.3\arcsec \citep{Zirm2005}, the original interpretation of the peak in 8.46 GHz emission as a radio core remains plausible (we note that such an offset would also place the peak of the northern lobe closer to the peak in \ewha).
 
Another noteworthy morphological feature for which the physical interpretation is informed by the confirmed galaxy position is the extended cone of Ly$\alpha$ emission, previously dubbed ``the Wedge" by \citet{Zirm2005}.
This feature, visible in the HST/ACS $F625W$ image in Fig.~\ref{fig:rgb_cutouts} (with dashed white lines illustrating the approximate bounds), was hypothesised to be the result of starburst driven outflows emanating from the galaxy.
However, we see no equivalent feature in the $F335M$ filter that directly probes H$\alpha$, either in the raw imaging or when smoothing to enhance low surface-brightness emission.
Furthermore, the focal point of the wedge (see annotation in Fig.~\ref{fig:rgb_cutouts}) does not coincide with the confirmed location of the host galaxy and so is unlikely to be directly driven by a central starburst.
We therefore interpret this emission to be a feature of the extended Ly$\alpha$ halo surrounding TNJ1338. 
However, given the observed extent of the Ly$\alpha$ halo \citep[$>100$ kpc;][]{Venemans2002}, its complex kinematics \citep{Swinbank2015} and the lack of additional information for this feature provided by the NIRCam imaging, we do not include any further analysis of ``the Wedge" in this study.

\section{Resolved SED modelling}\label{sec:resolved_analysis}
While the variation in observed colours seen in Fig.~\ref{fig:ew_map} indicates the presence of significant nebular contribution to TNJ1338's rest-frame optical emission, more quantitative analysis of the full SED is necessary to place constraints on the nature of the emission and how it relates to the properties of the host galaxy.
We therefore perform detailed SED modelling for individual regions within the galaxy.

\subsection{Segmentation photometry}
Given the irregular, clumpy and strongly wavelength dependent morphology seen in Fig.~\ref{fig:rgb_cutouts}, reliably decomposing the observed emission with parametric morphological models is not optimal.
We therefore use a multi-step process to segment the TNJ1338 into relevant physical regions for SED analysis.
Our photometry methodology is intended to balance maximising the spatial information available with the optimal signal-to-noise ratio (SNR) necessary to derive detailed stellar population properties with multi-band photometry \citep[see e.g.][]{Leja2019}.

We use the F335M medium band as our detection and segmentation image due to the presence of the H$\alpha$ emission line within the band-pass, allowing extremely high SNR identification of multiple distinct clumps within the host galaxy, as well as the luminous galaxy core.
Segmentation of the galaxy into these regions of interest is done in multiple stages using the \texttt{Photutils} package 
\citep[v.1.5;][]{PhotutilsASCL, Photutils1.5.0}.
Firstly, the main galaxy core is identified using a detection threshold of $>120\times$ the local background RMS.
After masking the full core, the peak of the jet-aligned emission and the galaxy outer core are selected with a lower noise threshold of $>43\times$ local RMS (where the threshold was iteratively selected to ensure that the two areas were maximised without becoming connected).
We further split the outer core in two relative to the peak of the radio continuum emission to allow us to compare the properties of the northern and southern edges of the host galaxy and the variation in physical properties relative to the radio jet.
Similarly, to allow a more detailed study of how the SED varies close to the northern radio lobe, we split the secondary peak into four segments by first de-blending into two components and then splitting each of those in two relative to the peak of the radio emission.

Finally, the remaining galaxy emission is split into bins following the optimised Voronoi binning procedure of \citet{Cappellari2003}.
Starting at a high target SNR per bin of $>250$, we iteratively apply the \citet{Cappellari2003} algorithm, lowering the target SNR until the smallest Voronoi bin falls below an area of 50 pixels.
The starting target SNR is set to this high threshold based on two factors: firstly, the Voronoi binning optimisation only makes use of the noise provided by the noise map, which does not account for the correlated uncertainties as outlined in Section~\ref{sec:correlated_noise}, and secondly the fact that the emission in F335M is substantially brighter than the others and by setting a high threshold we aim to ensure that the final SNR accounting for correlated uncertainties remains significant in all bands.
% The iterative binning step procedure yields a final target $\text{SNR} > 196$ for a smallest Voronoi segment of 46 pixels.
\begin{figure}
\centering{
    \includegraphics[width=1.0\columnwidth]{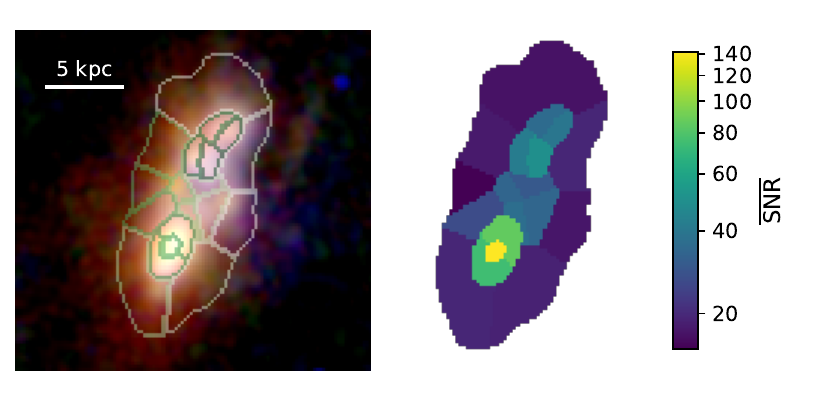}
	\caption{Image regions identified for quantitative analysis following the method outlined in the text. Left: segment boundaries overlaid on the combined RGB colour image presented in the Fig.~\ref{fig:rgb_cutouts}, where the colour-scale is arbitrary. Right: Mean signal to noise ratio ($\overline{\text{SNR}}$) for each source region, including corrections for correlated noise as outlined in Section~\ref{sec:correlated_noise}.}
    \label{fig:segment_overlay}}
\end{figure}
The final multi-band photometry is performed by constructing a combined segmentation map containing all of the regions defined above; segmentation photometry is then performed on each of the PSF-homogenised HST/ACS and JWST/NIRCam images.
Uncertainties measured for each segment are corrected for correlated noise based on the segment area and the corresponding calculations outlined in Section~\ref{sec:correlated_noise}.
In Fig.~\ref{fig:segment_overlay} we illustrate the resulting image segments compared to the galaxy emission.
Also shown in Fig.~\ref{fig:segment_overlay} is the mean signal-to-noise in each segment across all bands used in the SED fitting analysis ($\overline{\text{SNR}}$).
The final photometry catalogue consists of 19 segments, with the mean SNR across the observed filters ranging from $\sim 7.5$ in F775W to $\sim 105$ in F335M. 

\subsection{SED modelling}
\subsubsection{Stellar population models}
Modelling of the resolved TNJ1338 SED is performed using the \texttt{Prospector} Bayesian SED modelling code \citep{Prospector}, which incorporates a full range of stellar population models from the Flexible Stellar Population Synthesis \citep[FSPS;][]{Conroy2009, Conroy2010} package \citep[and the \texttt{python-FSPS} bindings;][]{pythonFSPS}.
Nebular line and continuum emission associated with the stellar population are implemented in \texttt{Prospector} following the method outlined in \citet{Byler2017}, which uses the photo-ionization models of \texttt{Cloudy} \citep{Cloudy} to self-consistently model the total nebular emission for the ionising spectra of the FSPS stellar populations.

One of the key advantages of \texttt{Prospector} is its implementation of non-parametric star-formation histories (SFH), which have been demonstrated to provide more accurate constraints on the SFH derived from both photometric and spectroscopic measurements of galaxy SEDs, producing less biased measurements and more accurate estimates of the corresponding uncertainties \citep{Leja2019}.
However, as highlighted by \citet{Prospector} and \citet{Leja2019}, the assumed priors can have a strong impact on the inferred SFH and hence the corresponding stellar mass estimate.
The impact of this prior is maximised in the presence of very young stellar populations ($< 30$ Myr) due to the outshining of older stars, and also at higher redshifts where constraints on the rest-frame near-infrared stellar emission are not available \citep{Tacchella2022b, Topping2022, Whitler2022a, Whitler2022b}.

From Fig.~\ref{fig:ew_map} and additional inspection of the measured SEDs in a number of regions within TNJ1338, there is clear evidence for extremely high \ewha emission, potentially indicative of a very young stellar population.
For our analysis we therefore follow the approach of recent studies \citep[see e.g.][]{Endsley2022} and assume two sets of SFH parameters designed to bracket the potential extremes in stellar mass.
Firstly, we assume a non-parametric SFH  using the `continuity' prior, with 5 age bins: 
\begin{equation}
\begin{split}
    0 < t < 10\, \text{Myr},\\
    10 < t < 100\, \text{Myr},\\
    100 < t < 331\, \text{Myr},\\
    331\,\text{Myr} < t < 1.09\,\text{Gyr},\\
    1.09 < t < 1.29\,\text{Gyr}.
\end{split}
\end{equation}

Secondly, we assume a constant SFH (CSFH) where the time since the onset of SF, $t_{\text{age}}$, is allowed to vary between 1 Myr and 1.29 Gyr (i.e. a formation redshift of $z=20$) with a uniform prior in $\log_{10}(t_{\text{age}}$/Gyr).

For both types of SFH, we use a \citet{Kroupa2001} initial mass function (IMF).
Stellar metallicity is allowed to vary in the range $-2 < \log_{10}(Z/Z_{\odot}) < 0.19$ and the gas-phase metallicity (for nebular emission) is fixed to that of the stellar metallicity.
Additionally, for the nebular emission component of the FSPS models \citep{Byler2017}, we include both line and continuum emission, allowing the ionisation parameter, $U$, to vary with a log-uniform prior in the range $-4 < \log_{10}(U) < -1$.
We choose to fit $\log_{10}(U)$ as a free parameter after initial tests demonstrated that lower $\log_{10}(U)$ values than the assumed default ($=-2.5$) were necessary to reproduce the $F182M$ observations in many regions.
Dust attenuation is assumed to be simple foreground attenuation with varying optical depth, assuming either \citet{Calzetti2000} or Small Magellanic Cloud \citep[SMC;][]{Gordon2003} dust attenuation laws.
We allow the optical depth at 5500\AA\, to vary in the range $0 < \tau_{5500\text{\AA}} < 3$ with a uniform prior.
Limitations and caveats of our SED fitting model assumptions will be discussed further in Section~\ref{sec:results} within the context of the inferred physical properties of TNJ1338.

% Finally, we note that \texttt{Prospector} optimisation is performed using the \texttt{Dynesty} nested sampling code \citep{Dynesty} with default stopping criteria ($10^{4}$ effective samples).
When fitting the observed photometry, we include an additional 5\% fractional uncertainty (added in quadrature to the measured uncertainties) to account for the estimated remaining systematic uncertainty in the NIRCam photometric zeropoints \citep{Boyer2022}.
Because of the significant potential for scattering of Ly$\alpha$ photons into and out of the large extended Ly$\alpha$ emission around TNJ1338 \citep{Venemans2002, Swinbank2015}, we do not include the ACS $F625W$ filter in the SED fitting analysis (however we show the observed and predicted magnitudes in subsequent figures for reference).

%\textsc{Prospector}-$\alpha$ \citet{Leja2017}
    
\subsubsection{Radiative shock models}
In a number of powerful HzRGs, the jet-aligned emission has been attributed to radiative shocks \citep[][in particular, in AGN with jets $< 150$ kpc such as TNJ1338]{Best2000, Inskip2002, Moy2002}.
To explore the potential for this mechanism driving the high-EW emission in TNJ1338, we also compare the observed photometry to predictions from the \texttt{MAPPINGS III} Library of Fast Radiative Shock Models \citep{MappingsIII}. 
\texttt{MAPPINGS III} contains model predictions for radiative shocks and associated photo-ionised precursor for five different abundance patterns: SMC, LMC, \citet{Dopita2005}, Solar and $2\times$ Solar, with shock velocities ranging from 100 to 1000 km\,s$^{-1}$. 

To fit the \texttt{MAPPINGS III} library to the observed photometry, we convolve the shock and shock+precursor SEDs with the HST/ACS and JWST/NIRCam filter profiles at the fixed redshift of $z=4.11$ to obtain the predicted colours.
The optimal scaling for each model to match the observed photometry from $F775W$ to $F360M$ for a given segment of TNJ1338 is then calculated using the standard analytic optimisation \citep[see e.g. Eq.\,4 of ][]{Duncan2019}, with the best-fitting model identified through $\chi^{2}$ minimisation of the full model set. 
For consistency with stellar population modelling above, we include an additional 5\% fractional uncertainty added in quadrature to the observed photometric uncertainties.
We note that following \citet{MappingsIII}, the radiative shock models are assumed to be effectively dust free and no additional attenuation is applied to the shock model predictions.

    % \item Target properties:\\
    % -- Total stellar mass (and masses of any individual sub-components), accounting for emission line contributions\\
    % -- SED derived star-formation rate(s), accounting for rest-UV, H$\alpha$ and far-IR/sub-mm (total SFR only)\\
    % -- Parametric/Non-parametric SFH constraints (c.f. cosmic star-formation history and evolution of LBGs)\\

% \begin{figure}
% \centering
%     \includegraphics[width=\columnwidth]{figures/tnj1338_mass_sfr_map.pdf}
% 	\caption{.}
%     \label{fig:mass_sfr}
% \end{figure}

% \begin{figure}
% \centering
%     \includegraphics[width=\columnwidth]{figures/tnj1338_sed_posterior_mass_sfr.pdf}
% 	\caption{.}
%     \label{fig:posteriors_mass_sfr}
% \end{figure}

% \begin{figure*}
% \centering
%     \includegraphics[width=0.9\textwidth]{figures/tnj1338_sed_posterior_core_clump.pdf}
% 	\caption{.}
%     \label{fig:posteriors}
% \end{figure*}

\section{Resolved physical properties of TNJ1338}\label{sec:results}
\begin{figure}
\centering
    \includegraphics[width=0.9\columnwidth]{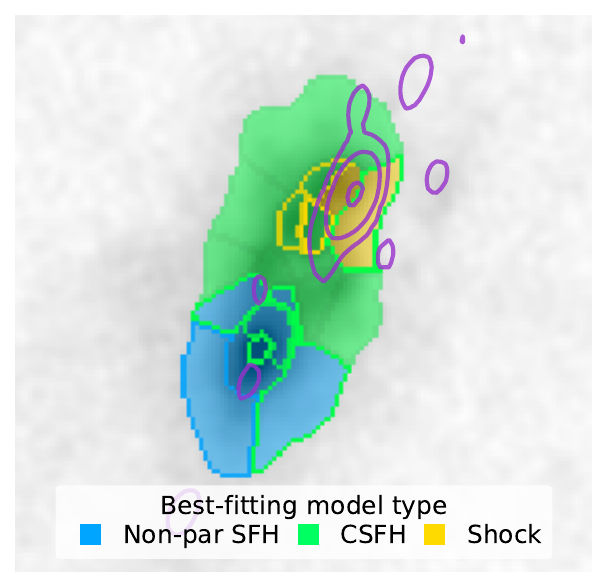}
	\caption{TNJ1338 source regions coloured by the best-fitting model from SED fitting, with additional boundary colours indicating alternative models statistically consistent with the observed photometry (see text). White contours show the VLA 8.46 GHz radio continuum emission associated with the galaxy, with contours starting at 4$\sigma$ (increasing by $4^n$). Regions closest to the galaxy core are well modeled by stellar populations (blue/green) while those closest to the peak of the radio emission are more consistent with self-ionising shock models (yellow shaded or edged regions).}
    \label{fig:sed_bestmodel}
\end{figure}

In total, our resolved SED modelling incorporates 5 different sets of assumptions (two sets of star-formation histories each with two types of dust attenuation, plus one set of radiative shock models).
To summarise the results of the SED fitting analysis, Fig.~\ref{fig:sed_bestmodel} illustrates the best-fitting model type for each region within TNJ1338.
Colour shading shows which SFH assumption or shock model produces the best-fit for that region when assuming an SMC dust law, based on the $\chi^{2}$ of the maximum a posteriori (MAP) value from \texttt{Prospector} or the minimum-$\chi^{2}$ for the radiative shock models.
Since the number of free parameters varies significantly between the assumed models, we also calculate the corresponding Bayesian Information Criterion (BIC) value for each model fit.
Different colours around the edges of regions illustrate where the best model either changes for a Calzetti dust law \emph{or} there is no significant evidence for the best model over the second best, i.e. where the difference in BIC between the two models is $\Delta\text{BIC} < 4$, with the edge colour corresponding to the alternative possible model.

Supporting the qualitative picture presented in Section~\ref{sec:visual_results}, Fig.~\ref{fig:sed_bestmodel} demonstrates that there are distinct regions within TNJ1338.
The source regions closest to the host galaxy in the southern half of TNJ1338 show evidence for a more extended and complex SFH, marginally favouring the non-parametric SFH assumption. 
The SEDs of the region closest to the jet are instead more consistent with the radiative shock models or an extremely recent burst of SF activity.
We note however that only one region (in which the peak of the 8.46 GHz radio continuum emission is located) strongly favours the radiative shock model regardless of the SFH or dust models assumed for the stellar population model (see Section~\ref{sec:host_properties} below).
Regardless of the source of ionising photons, the SEDs in this region are clearly dominated by strong nebular emission lines.

\begin{figure}
\centering
    \includegraphics[width=0.49\textwidth]{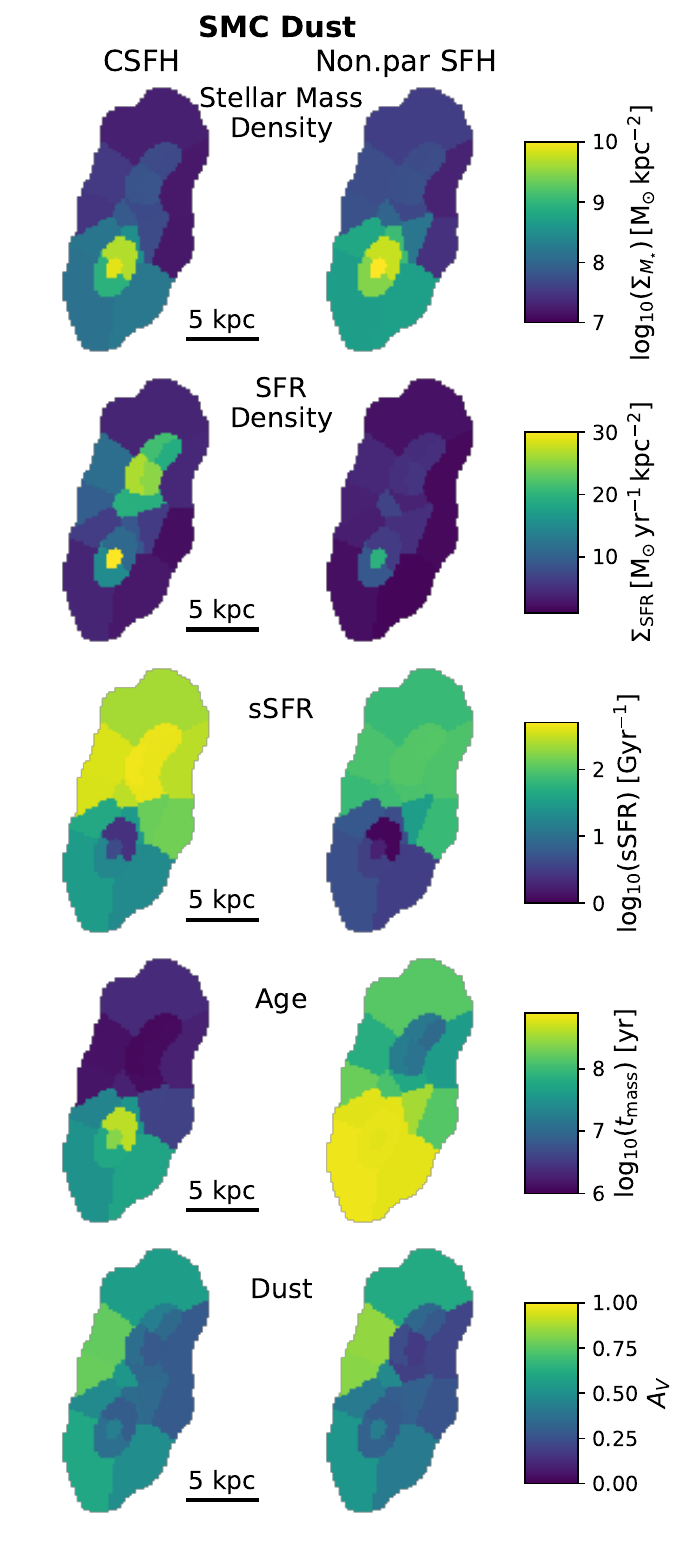}
	\caption{Maps of the resolved physical properties inferred from \texttt{Prospector} SED fitting assuming SMC dust, where colours correspond to the median of the respective posterior distributions. The left-hand columns show the results for the continuous SFH, with the right-hand column showing the results assuming non-parametric SFH. Each column of five plots shows the stellar mass surface density ($\Sigma_{\textup{M}_{\star}}$), SFR surface density ($\Sigma_{\textup{SFR}}$), sSFR, mass-weighted age ($t_\textup{mass}$) and dust extinction ($A_{\textrm{V}}$) for each segment. SFR values used to derive $\Sigma_{\textup{SFR}}$ and sSFR are averaged over 10 Myr, or the maximum age of the CSFH model in that region if less than 10 Myr.}
    \label{fig:resolved_properties}
\end{figure}

Fig.~\ref{fig:resolved_properties} presents two sets of resolved maps of key physical properties inferred from the stellar population modelling.
The maps further highlight the distinction between the northern and southern halves of TNJ1338, with the stellar mass surface density, $\Sigma_{\textup{M}_{\star}}$, for the southern half of TNJ1338 substantially greater (by up to two dex) than the jet region, with a clear concentration around the galaxy core.
A similar picture is seen in the inferred mass-weighted ages, $t_\textup{mass}$, where there is a clear dichotomy across the galaxy. 
Crucially, we observe the same spatial variation across the galaxy regardless of the assumed SFH or dust models.
However, the exact normalisation of some properties (e.g. $\Sigma_{\textup{SFR}}$ or $t_\textup{mass}$) do systematically vary with the assumed dust attenuation law.
In the following subsections, we explore the details of Fig.~\ref{fig:sed_bestmodel} and \ref{fig:resolved_properties} and the corresponding properties of the jet-aligned emission and host galaxy in greater detail.

\subsection{Nature of the jet-aligned emission}
In their previous study of TNJ1338, \citet{Zirm2005} attribute the extended optical emission (rest-frame UV) to SF activity triggered by the radio jet.
As illustrated in Fig.~\ref{fig:sed_bestmodel}, the NIRCam observations reveal a more complex picture with multiple potential emission mechanisms.
Resolved spectroscopic observations are required to fully disentangle the emission-line contributions from star-formation, shocks and AGN photo-ionisation.
Nevertheless, the resolved photometric SEDs are consistent with TNJ1338 being a massive galaxy with radiatively driven shocks and extensive recent star-formation spatially correlated with the luminous radio jets.
% Here, we further explore in detail the nature of the jet-aligned optical emission and the potential role of the jet activity in driving it.

\begin{figure}
\centering
    \includegraphics[width=\columnwidth]{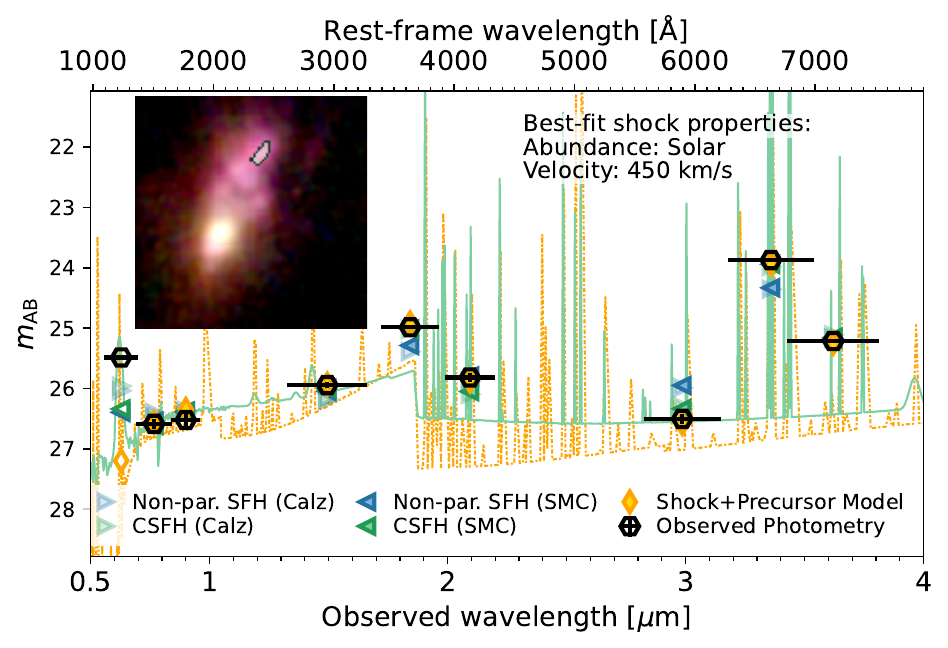}
\includegraphics[width=\columnwidth]{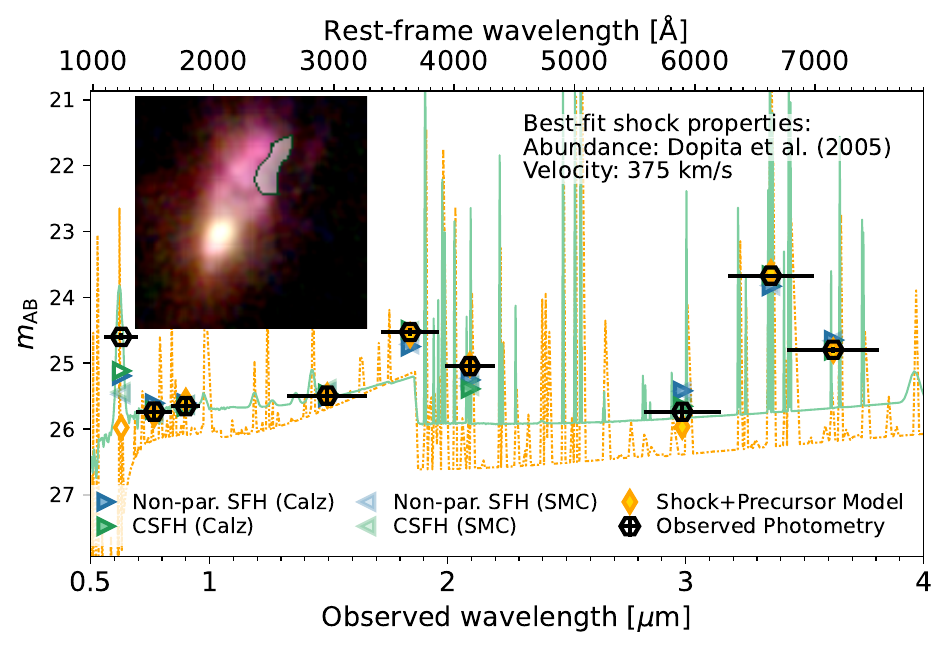}
\includegraphics[width=\columnwidth]{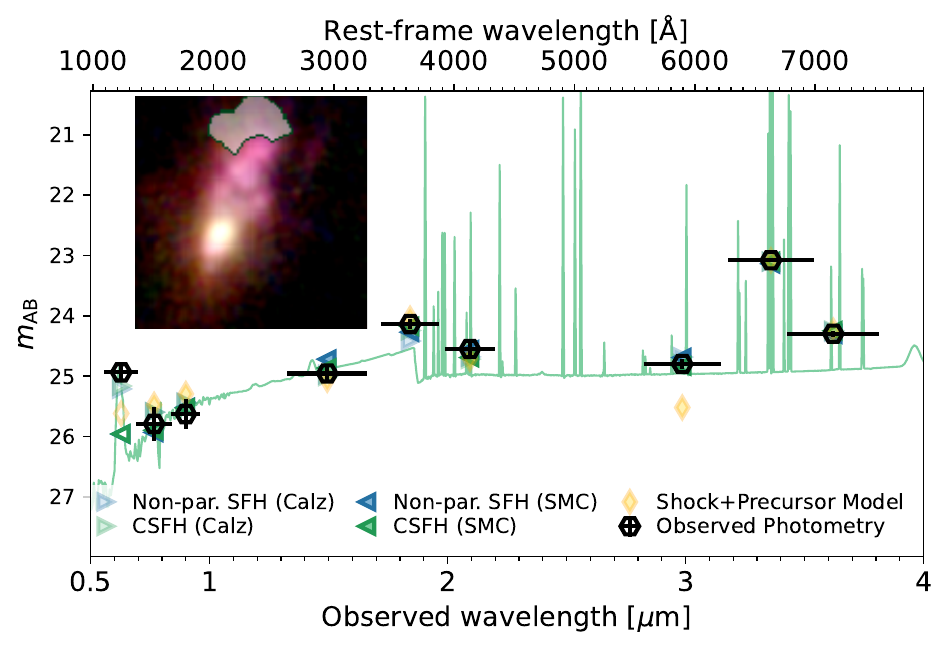}
	\caption{Observed SED (black hexagons) and best-fitting model photometry for the non-parametric (blue) or continuous (green) SFH assumptions, assuming \citet[][right-pointing triangles]{Calzetti2000} or SMC \citep[][left-pointing triangles]{Gordon2003} dust attenuation laws, for the two regions of the jet-aligned emission most consistent with radiative shock emission and a neighboring region where a simple young stellar population is preferred. The model spectrum corresponding to the best-fitting SFH/dust law assumption is shown as a solid line. Yellow diamonds show the best-fitting model photometry for the radiative shock models, with the corresponding spectrum plotted as the yellow dot-dashed line. The $F625W$ filter is not included in any model fit but is shown for reference. The inset RGB image illustrates the galaxy region corresponding to the observed photometry and associated fits.}
    \label{fig:sed_shocks}
\end{figure}

In Fig.~\ref{fig:sed_shocks}, we present the observed SED of the two regions closest to the peak of radio continuum emission and the adjacent region beyond the peak.
For each region we show the corresponding best-fit stellar population (green lines and triangles), with the best-fitting radiative shock models (yellow dash-dotted line and diamonds) shown if consistent with the observations.
% In both regions a shock model that includes the emission from the shock precursor is preferred.

\begin{figure}
\centering
    \includegraphics[width=\columnwidth]{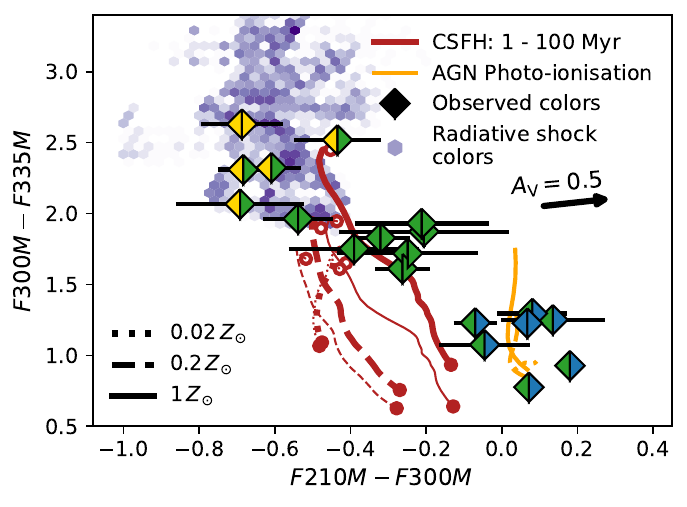}
	\caption{Observed $F210M-F300M$ and $F300M-F335M$ colours with corresponding uncertainties including additional 5\% flux uncertainty (filled symbols and associated error bars) compared to the parameter space probed by different models explored in this study. Background hexagonal bins show the distribution of colours present within the radiative shock (+pre-cursor) models, where the colour shading reflects the number of models with a given colour for illustrative purposes. Red lines show the colours of dust-free CSFH models with range of metallicities, maximum ages ranging from 1 Myr (open circles) to 100 Myr (filled circles) and ionisation parameters $\log_{10}(U) = -2.5$ (thin lines) and $\log_{10}(U) = -3.5$ (thick lines). Orange lines show predictions for AGN photo-ionisation models assuming a harder ionizing spectrum \citet{Mathews1987}, with ionisation parameters varying from $\log_{10}(U) = -2.5$ to $\log_{10}(U) = -1.5$. For both CSFH and AGN photo-ionisation models, dotted, dashed and solid lines correspond to 0.02, 0.2 and 1$\times Z_{\odot}$ metallicities respectively. The median colour vector corresponding to $A_{V} = 0.5$ extinction \citep{Calzetti2000} is illustrated by the black arrow.}
    \label{fig:shock_colours}
\end{figure}

Because of the wide dynamic range present within the observed SEDs, visually identifying the colours that are driving the statistical preference for radiative shocks within these regions is non-trivial.
Fig.~\ref{fig:shock_colours} therefore illustrates the observed distribution of $F210M-F300M$ and $F300M-F335M$ colours across TNJ1338 compared to illustrations of the parameter space covered by the radiative shocks and stellar population models.
Symbols for each region are coloured following the same convention as in Fig.~\ref{fig:sed_bestmodel}.
We can see that $F210M-F300M$ is the colour driving the preference for shocks over SF models in the most extreme regions.
Although a young stellar population can reproduce the most extreme observed $F300M-F335M$ colour, the corresponding SEDs cannot produce sufficiently blue $F210M-F300M$ colours even in the absence of dust.
Without detailed measurements of the emission line ratios, we cannot make robust constraints on the properties of the shock itself. 
However, we note that the best-fitting shock velocities for the regions shown in Fig.~\ref{fig:sed_shocks}, 450 and 375\,km\,s$^{-1}$, are consistent with range of shock velocities measured in the best local analogues of TNJ1338 \citep{Cresci2015, Capetti2022}.

% Given that the regions most consistent with radiative shocks are those closest to the peak of the radio emission, 

% To quantify whether the jet activity provides sufficient power to drive the observed shock emission, we can estimate the jet power associated with the Northern radio lobe following \cite{Hardcastle2018}.
% The jet power, $Q$, is estimated to scale with 150 MHz luminosity ($L_{150\text{MHz}}$) as
% \begin{equation}
%    Q = \left ( \frac{L_{150\text{MHz}}}{3\times10^{27} \text{W\,Hz}^{-1}} \right)\times 10^{38} \text{W}. 
% \end{equation}
% For a flux density of $S_{\nu, 4.6\text{GHz}} = 20.6$mJy \citep{Pentericci2000, Falkendal2019} for the Northern lobe and an assumed spectral slope of $\alpha=-1.31$ \citep{DeBreuck2000}, the resulting radio luminosity of $L_{150\text{MHz}} = 1.8\text{W\,Hz}^{-1}$ yields an approximate jet power of $Q = 6\times10^{39} \text{W}$ ($ 6\times10^{46} \text{erg\,s}^{-1}$).
% Combining the integrated best-fit radiative shock SED of all five potential shock regions (yellow shading or edges in Fig.~\ref{fig:sed_bestmodel}), the maximum total energy being emitted by the radiative shocks equals $1.5\times10^{45} \text{erg\,s}^{-1}$.

Beyond the shock-dominated area, the galaxy regions outside of the galaxy core have SEDs consistent with extremely young stellar populations, with CSFH mass-weighted ages $t_\textup{mass} <4$ Myr ($\log_{10}(t_{\text{mass}}/\text{yr}) < 6.6$; Fig.~\ref{fig:resolved_properties}).
Although we cannot definitively demonstrate a direct causal link between the radio jet and the recent starburst activity, the measurements presented in this study do provide additional support for the scenario of jet-triggered activity.
Firstly, our photometric observations of TNJ1338 provide a qualitatively similar picture to that observed in local known examples of jet-triggered SF.
In the luminous radio galaxy Coma A, resolved spectroscopic observations \citep{Capetti2022} show a region of highly ionised gas extending along the jet axis with emission lines indicative of radiative shocks surrounded by young stellar populations.
Similarly, in simulations of the jet-triggered SF observed in Centaurus A \citep{Crockett2012}, \citet{Gardner2017} demonstrate that the induced star-formation activity occurs in the transverse bow-shock of the radio jet. 
Such a scenario could explain the picture seen in Fig.~\ref{fig:sed_bestmodel}, where the region beyond the shock has an SED consistent only with a young recent burst \citep[see also][]{Bicknell2000}.

Secondly, the tight constraints on the mass-weighted age of the recent starburst spatially correlated with the jet ($t_\textup{mass} <4$ Myr) remain fully consistent with the expected timescale of that jet activity \citep[and the local analogues such as Centaurus A;][]{Crockett2012}.
% Setting a comparable timescale for the age of the jet that could have induced this SF activity is hampered by the poor correlation between jet age and projected size \citep{Hardcastle2018}.
Comprehensively modelling the jet lifetime given the observed radio spectral information and losses from inverse-Compton scattering is beyond the scope of this work.
However, for radio jets with projected sizes similar to that seen in TNJ1338 ($\sim7$ kpc from galaxy core to the peak of the radio emission) and comparable luminosities, inferred jet ages range from $\sim1$ to $>10$ Myr \citep[see e.g. Fig.~14 of][]{Hardcastle2018}, consistent with the observed burst of SF.

\begin{figure}
\centering
    \includegraphics[width=\columnwidth]{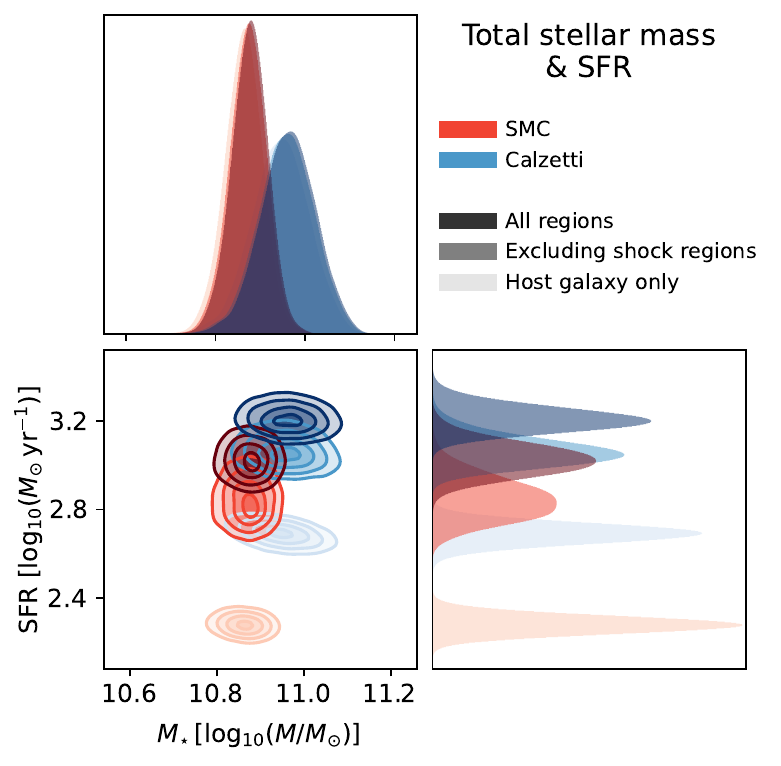}
	\caption{Posterior distributions for the total stellar mass, $M_{\star}$, and ongoing SFR in TNJ1338 inferred from resolved SED fitting, assuming SMC \citep[][red hued distributions]{Gordon2003} or \citet[][blue hued distributions]{Calzetti2000} dust attenuation laws. Upper and right-hand panels show the marginalised distributions for $M_{\star}$ and SFR respectively. Lighter shaded distributions show the combined mass/SFR estimates excluding any regions possibly associated with shock emission (yellow shaded or edged regions in Fig.~\ref{fig:sed_bestmodel}), while the lightest distributions include only those regions attributed to the older host galaxy (blue shaded regions in Fig.~\ref{fig:sed_bestmodel}).}
    \label{fig:total_posteriors}
\end{figure}

Where TNJ1338 differs compared to examples of jet-induced SF activity at lower redshift \citep{Best1997, Croft2006, Crockett2012} is the scale of the star-formation potentially triggered by the jet.
In Fig.~\ref{fig:total_posteriors} we present the combined SED fitting posteriors on the total stellar mass and star-formation measured in TNJ1338 (see also Table~\ref{tab:mass_sfr}).
To produce the combined posteriors we make 3000 random draws from the \texttt{Prospector} nested sampling chain for each region, summing the total stellar mass and SFR across the subset of regions of interest.
For regions associated with the host galaxy (blue shading in Fig.~\ref{fig:sed_bestmodel}) we use the non-parametric SFH assumption, while for regions in the jet-aligned emission (green or yellow shaded regions) we use the fits from the CSFH assumption.

If we assume all regions are dominated by stellar emission (dark blue and red posteriors and histograms), the total estimated SFR amounts to $\gtrsim1000\,\text{M}_{\odot}\,\text{yr}^{-1}$, specifically $\log_{10}(\text{SFR}/\text{M}_{\odot}\,\text{yr}^{-1}) = 3.21^{+0.06}_{-0.04}$ for Calzetti dust, $\log_{10}(\text{SFR}/\text{M}_{\odot}\,\text{yr}^{-1}) = 3.02^{+0.07}_{-0.06}$ for SMC (see Table~\ref{tab:mass_sfr})\footnote{Comparing the total flux contained within the resolved galaxy segments used in this analysis (Fig.~\ref{fig:segment_overlay}) to the total Petrosian flux measured for TNJ1338, we find that the resolved segments contain 88 and 89\% of the total $F300M$ and $F335M$ flux respectively. Total SFR estimates may therefore be underestimated by up to $\sim 0.06\,\text{dex}$, with stellar masses significantly less than this (due to the concentration of the stellar mass within the galaxy core).}.
Even excluding any region that could be dominated by radiative shock emission instead of SF (yellow shaded or bordered areas in Fig.~\ref{fig:sed_bestmodel}) the total estimated SFR is reduced only by 0.18 dex, resulting in rates of $\sim700-1000\,\text{M}_{\odot}\,\text{yr}^{-1}$.
The SFR estimated from the host galaxy regions alone amounts to only 30\% of the total SFR (42\% excluding shock regions) for the \citet{Calzetti2000} dust assumption (18 and 27\% for SMC), meaning that the on-going jet-triggered star-formation could be at a rate of at least $\sim500\,\text{M}_{\odot}\,\text{yr}^{-1}$, independent of the assumed dust attenuation law.

\begin{figure}
\centering
    \includegraphics[width=\columnwidth]{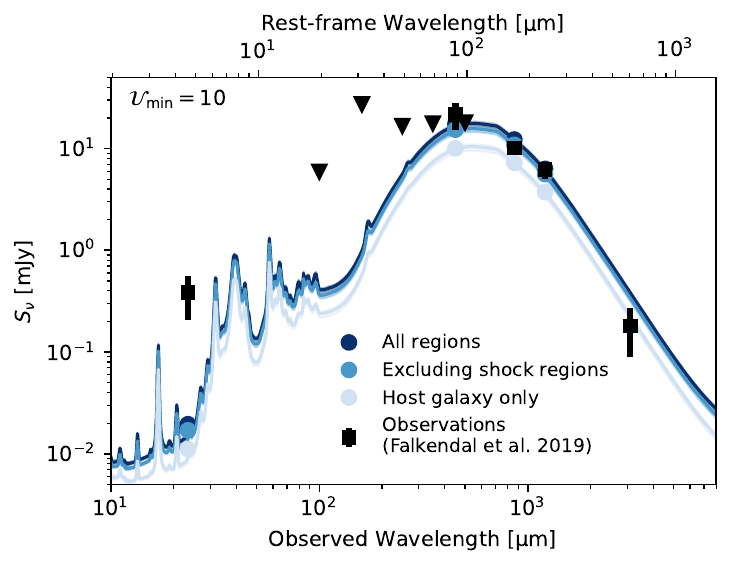}
	\caption{Observed far-IR to sub-mm photometry emission from TNJ1338 (black data points), with $3\sigma$ upper limits plotted as downward triangles. Also shown are the posterior predictions combining all regions of TNJ1338 (dark blue circles and line), only regions not consistent with shock dominated SEDs (lighter blue) and the host galaxy regions only (lightest blue). Observations are taken from the compilation presented in \citet[][see Section~\ref{sec:ancillary} for full details]{Falkendal2019}. The significant excess at 24$\mu$m is indicative of hot-dust emission associated with obscured AGN activity \citep[see the joint AGN+SF fits of][]{Falkendal2019}.}
    \label{fig:fir_predictions}
\end{figure}

As noted in Section~\ref{sec:ancillary}, a number of FIR to sub-mm observations have been taken of TNJ1338 that provide additional constraints on the total ongoing SFR.
To explore whether our inferred physical properties are consistent with the existing long wavelength observations, in Fig.~\ref{fig:fir_predictions} we show the mid-IR to sub-mm SED predicted by the model assumptions that yield the most extreme estimated SFR.
For the \citet{Draine2007} IR model implemented in \texttt{Prospector}, we can broadly reproduce the observed IR SED when assuming a minimum interstellar radiation field, $\mathcal{U}_{\textup{min}} = 10$ (in units of the Milky Way radiation field).\footnote{We use $\mathcal{U}_{\textup{min}}$ here to ensure distinction from the ionisation parameter, $U$, whilst remaining consistent with the nomenclature defined in \citet{Draine2007}.}
The photometric observation that cannot be reproduced by the ongoing SF activity is the 24$\mu$m measurement, with the observed flux for TNJ1338 over 1 magnitude greater than predicted by the SF emission.
Fitting only the integrated IR SED, \citet{Falkendal2019} attribute the 24$\mu$m emission entirely to hot dust associated with AGN activity (not included in our predicted SED).
As our NIRCam observations do not probe the rest-frame near to mid-IR wavelengths where emission from an AGN IR power-law can dominate the observed galaxy SED \citep{Donley2012}, additional resolved mid-IR observations (i.e. JWST/MIRI) are required before placing further constraints on the obscured accretion activity.
 
Given the large systematic uncertainties in robustly constraining dust properties at high redshift \citep[see e.g.][for a recent review]{Hodge2020}, we can not claim that the FIR observations provide additional evidence \emph{for} the extreme SFRs estimated from our resolved SED modelling.
Fig.~\ref{fig:fir_predictions} does however illustrate that a jet-triggered starburst with $\text{SFR}\gtrsim1000\,\text{M}_{\odot}\,\text{yr}^{-1}$ can be fully consistent with the existing FIR emission.

Although extreme relative to low redshift examples, our estimates for the induced SF in TNJ1338 are comparable to those measured in the other known high-redshift examples of jet-triggered star-formation, 4C 41.17 \citep[$z=3.8$;][]{Dey1997,Steinbring2014}, which has a total SFR of $\sim650\,\text{M}_{\odot}\,\text{yr}^{-1}$ based on resolved mm observations \citep{Nesvadba2020}.
Furthermore, as highlighted in Fig.~\ref{fig:total_posteriors} (and implicit in the stellar mass estimates shown in Table~\ref{tab:mass_sfr}), despite the extremely high ongoing star-formation observed, the very short lifetime of the recent burst ($t_\textup{mass} <4$ Myr) means that the total stellar mass added to the galaxy by the jet-triggered activity is negligible compared to the mass in-situ.
It therefore remains unclear whether the potential positive feedback from luminous radio jets such as those in TNJ1338 play a significant role in shaping the evolution of their host galaxies.
% In future, resolved mid-IR (e.g. JWST/MIRI) and millimeter (ALMA) observations could enable

% \begin{itemize}

% \item As shown in Fig.~\ref{fig:resolved_properties}, although the recent burst of SF is extreme in amplitude, its short lifetime means that the stellar mass surface density within the triggered SF total contribution to the overall stellar mass is negligible.

% \item Jet induced SF in other radio AGN (notes for write-up/discussion):
% \begin{itemize}
%     \item Centaurus A - \citet{Crockett2012}
%     \item Coma A - \citet{Capetti2022}: Ionization cone with line ratios consistent with nuclear photoionization. Shock velocities in the range of 300 - 500 km/s
%     \item 3C 34 - Object A - $z=0.69$ \citep{Best1997}
%     \item Minkowski's Object - \citet{Croft2006}
%     \item 4C 41.17: $z=3.8$ - \citet{Dey1997, Steinbring2014}
% \end{itemize}

% \end{itemize}

\begin{table}
	\centering
	\caption{Marginalised total stellar mass, $M$, and ongoing SFR in TNJ1338 inferred from resolved SED fitting, assuming SMC \citep[][]{Gordon2003} or \citet[][]{Calzetti2000} dust attenuation laws.}
	\label{tab:mass_sfr}
	\begin{tabular}{lcc} % four columns, alignment for each
		\hline
          & Stellar Mass & SFR\\
          & $\log_{10}(M/\text{M}_{\odot})$ & $\log_{10}(\text{SFR}/\text{M}_{\odot}\,\text{yr}^{-1})$\\
         \hline
        & \multicolumn{2}{c}{SMC}\\
        \hline
        All regions & $10.88^{+0.04}_{-0.04}$ & $3.02^{+0.07}_{-0.06}$\\
        Excluding shock regions & $10.87^{+0.04}_{-0.04}$ & $2.84^{+0.1}_{-0.08}$\\
        Host galaxy only & $10.86^{+0.04}_{-0.04}$ & $2.28^{+0.03}_{-0.03}$\\
		\hline
  		 & \multicolumn{2}{c}{Calzetti}\\
        \hline
		All regions & $10.97\pm0.06$ & $3.21\pm0.05$ \\
		Excluding shock regions & $10.96\pm0.06$ & $3.06\pm0.06$ \\
		Host galaxy only & $10.95^{+0.06}_{-0.07}$ & $2.69^{+0.04}_{-0.03}$\\
        \hline
	\end{tabular}
\end{table}

\subsubsection{Limitations and systematic uncertainties on stellar population properties}
Although the choice of models and associated priors used in our SED fitting analysis are well motivated for the available photometric information and source redshift, they nevertheless represent an implicit prior on the resulting inferred physical properties. 
A number of known systematic uncertainties could potentially alter our interpretation of the mechanisms powering the jet-aligned emission.

Firstly, the assumed \citet{Kroupa2001} IMF implemented in \texttt{FSPS} is limited to $<120\, \textup{M}_{\odot}$ mass stars, while recent evidence suggests that stars as massive as $300\,\textup{M}_{\odot}$ exist in local star-forming complexes \citep{Crowther2010}.
Secondly, the stellar isochrones and spectral libraries we assume (MIST; \citeauthor{MIST}~\citeyear{MIST}, and MILES; \citeauthor{MILES}~\citeyear{MILES}, respectively) do not include the impact of interacting binaries on the resulting SED.
Including both higher mass stars and the effects of binary interaction leads to both bluer intrinsic stellar continuum SEDs and an increase in the production of ionising photons at a given population age \citep{Stanway2016}.

The use of alternative stellar population models such as the Binary Population and Spectral Synthesis code \citep[BPASS;][]{Eldridge2017}, which incorporates both of these changes, could therefore lead to older inferred mass-weighted ages, higher permitted \ewha and allow for bluer rest-frame optical colours (potentially leading to SF models compatible with the SED currently only fit by radiative shock models). 
The amount of recent SF required to produce the observed \ewha could also be significantly lower than our current estimates (Fig.~\ref{fig:total_posteriors}).
Nevertheless, given the uncertainties on jet lifetimes, such changes would not rule out that the SF activity is induced by the jet.

\subsubsection{Alternative emission mechanisms}
A key scenario not explored in the preceding analysis is the possibility that the jet-aligned emission is powered by AGN photo-ionisation \citep[e.g.][]{Tsvetanov1996, Keel2015}.
Using the He\textsc{ii} $\lambda1640$ emission line observed in long-slit spectroscopy as a proxy for H$\beta$, \citet{Zirm2005} previously estimated that the total nebular continuum emission from photo-ionised gas could only produce a total emission of $m_{F850LP} = 27.4$ over the whole galaxy, significantly fainter than most individual components (see Figs.~\ref{fig:sed_shocks}, \ref{fig:sed_core} and \ref{fig:sed_galaxy}).
Additionally, in Fig.~\ref{fig:shock_colours}, we illustrate the NIRCam colours predicted for a pure AGN photo-ionisation model representative of emission from a potential AGN ionisation cone.
The AGN photo-ionisation tracks show \citet{Cloudy} predictions for the nebular continuum and emission line SED produced by a harder ionising spectrum \citep{Mathews1987} over a range of ionisation parameters ($-2.5 < \log_{10}(U) < -1.5$) and for the same range in metallicities as shown for the CSFH predictions.
The AGN photo-ionisation emission models are unable to reproduce the colours observed for the regions with the highest \ewha, exhibiting redder $F210M - F300M$ colours than the young stellar populations and radiative shock models.

Alternatively, the H$\alpha$ nebulae associated with cooling flows in massive clusters at $z < 1$ have been observed to correlate spatially with radio jets within the cluster \citep{McDonald2010, Tremblay2015}.
The exact nature of these H$\alpha$ nebulae is however still an on-going debate, with evidence for both shock and star-formation driven sources of nebular emission within the known population of cooling flow nebulae \citep{McDonald2012}, as well as additional ionisation from X-rays \citep{McDonald2010}.
This scenario offers a potential mechanism for the observed high \ewha emission in TNJ1338 to be spatially associated with, but not necessarily directly triggered by, the jet activity.
While it is unlikely that a cooling flow such as those seen in virialised clusters could be in place as early as $z=4.11$ \citep{Santos2008, McDonald2013}, the same phenomenon may also arise in the cold accretion flows thought to fuel massive galaxy growth at high redshift \citep{Keres2005,Dekel2009}. 

Nevertheless, without spectroscopic observations to measure detailed line ratios and disentangle different kinematic components (e.g. narrow lines, out-flowing gas and shock components), we cannot exclude an additional contribution from the AGN within either the host galaxy emission or the jet aligned emission.
Further JWST observations with JWST/NIRSpec IFU (GO Programme 1964) should enable conclusive and detailed measurements on the relative contributions of stellar, shock and AGN driven emission across TNJ1338, as well as resolved kinematic information \citep[see, for example, the complex scenario revealed by NIRSpec around a luminous $z\sim3$ red quasar;][]{Vayner2023}.
Additional detailed analysis from our NIRCam photometry alone is therefore not warranted at this time.

\subsection{TNJ1338 host galaxy properties}\label{sec:host_properties}
As shown in Fig~\ref{fig:total_posteriors}, the total stellar mass we infer for TNJ1338 (including all regions) is $\log_{10}(M/\text{M}_{\odot}) = 10.97\pm0.06$ assuming a \citet{Calzetti2000} dust attenuation law \citep[or $\log_{10}(M/\text{M}_{\odot}) = 10.88\pm0.04$ assuming SMC;][]{Gordon2003}.
Excluding any region with emission that could be dominated by a radiative shock only reduces the total mass by 0.02 dex.
Despite the significant differences in input data and modelling assumptions, these values are in agreement with the previous mass estimates for TNJ1338 of $\log_{10}(M/\text{M}_{\odot}) \sim 11$ presented by \citet{Overzier2009}.
Our NIRCam observations therefore confirm that TNJ1338 is amongst the most massive galaxies at this epoch, with a mass significantly above the knee of the galaxy stellar mass function \citep[$\log_{10}(M^{\star}/\text{M}_{\odot})\approx10.5$;][]{Duncan2014,Song2016}.
For a typical field galaxy, this stellar mass would place TNJ1338 in a dark matter halo with mass $\log_{10}(M_{h}/\text{M}_{\odot}) \sim 12.5$ \citep{Legrand2019}.
However, with TNJ1338 also residing with a substantial galaxy over-density \citep{Venemans2002, Overzier2008, Saito2015}, this may represent a conservative estimate.
The full properties of the TNJ1338 proto-cluster revealed by NIRCam, the associated dark matter halo(s), and corresponding cosmological context will be explored in subsequent studies.

\begin{figure}
\centering
    \includegraphics[width=\columnwidth]{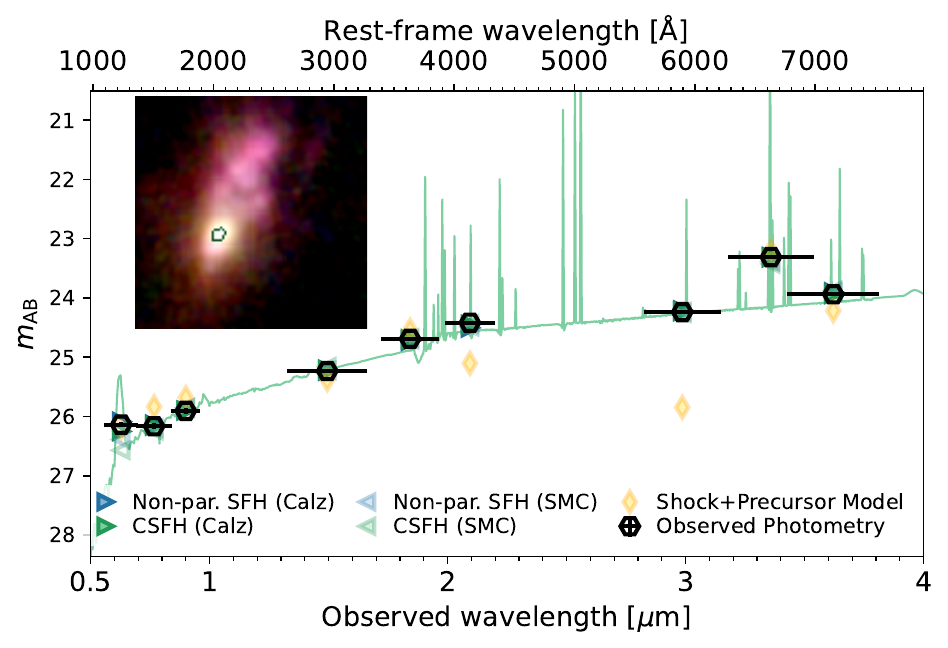}
\includegraphics[width=\columnwidth]{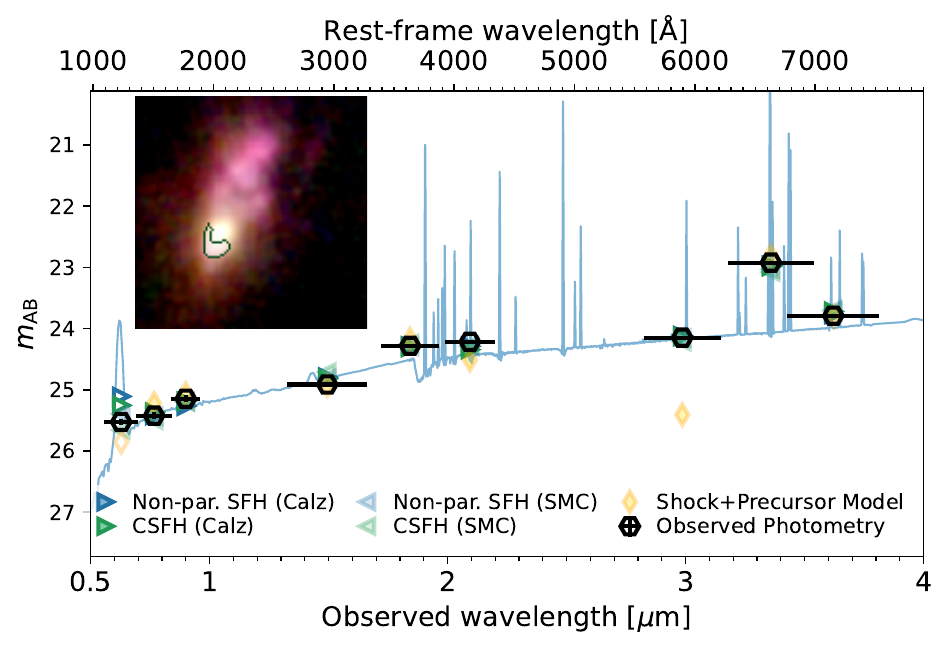}
\includegraphics[width=\columnwidth]{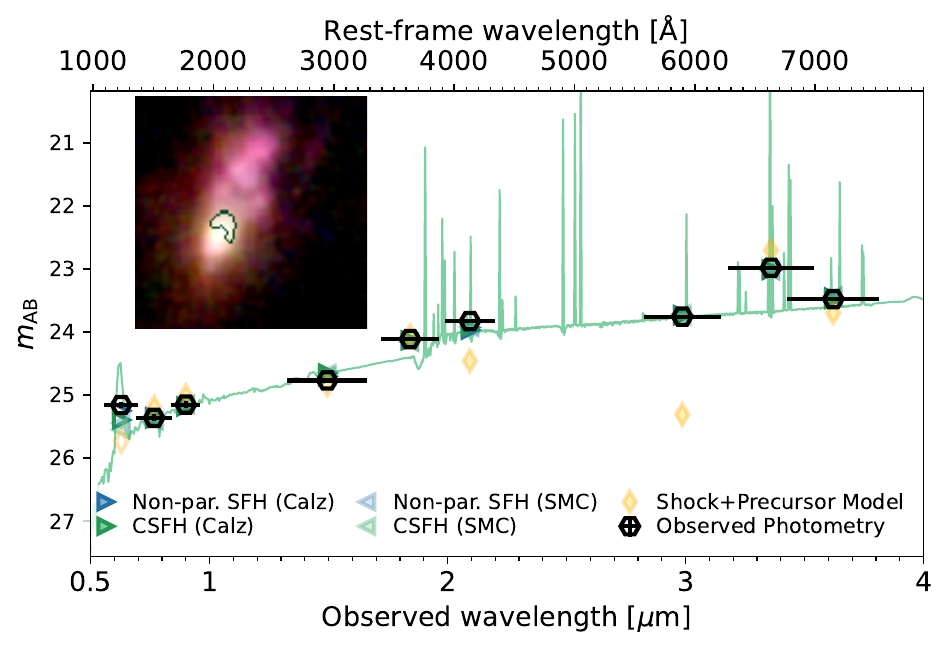}
	\caption{Observed SED for the three main regions associated with the TNJ1338 host galaxy. Plot symbols and colours are defined as in Fig.~\ref{fig:sed_shocks}.}
    \label{fig:sed_core}
\end{figure}

%\textbf{[Formation history - inside out etc ]}

\begin{figure}
\centering
    \includegraphics[width=\columnwidth]{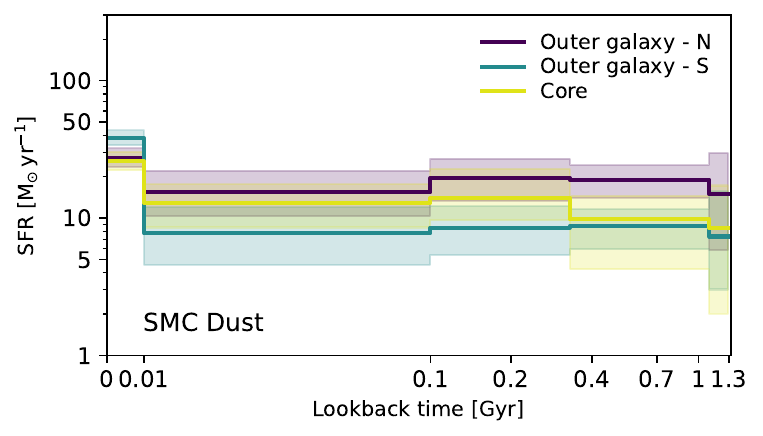}
    \includegraphics[width=\columnwidth]{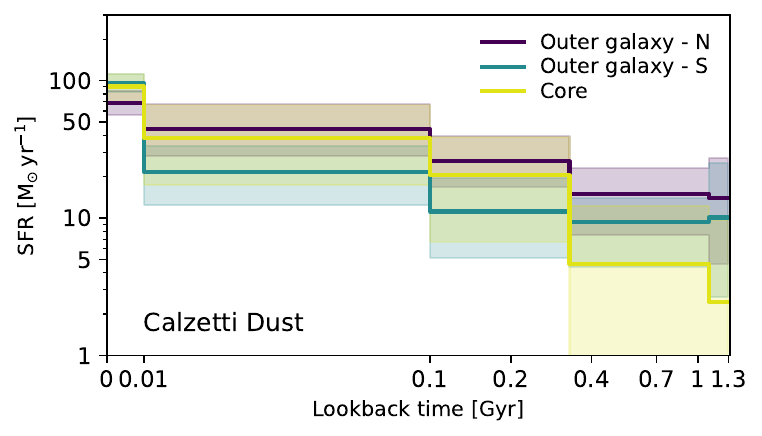}
	\caption{Inferred non-parametric SFH for the core and outer regions of the host galaxy (see Fig.~\ref{fig:sed_core}) for the two dust attenuation law assumptions; SMC \citep[][top]{Gordon2003} and \citet[][bottom]{Calzetti2000}. Solid lines and shaded regions illustrate the median and 16-84th percentiles of the \texttt{Prospector} SED fitting posteriors.}
    \label{fig:combined_sfh}
\end{figure}

We do not observe a large variation in the overall SED across the three regions that contain the bulk of the host galaxy stellar mass (see Fig.~\ref{fig:sed_core}). 
This consistency can also be seen in the non-parametric SFHs inferred for the three galaxy regions (Fig.~\ref{fig:combined_sfh}, which all support a relatively flat or smoothly rising SFH until a recent burst of SF within the most recent age bin (consistent with the potentially jet-induced SF activity).
However the tight constraints on \ewha provided by the NIRCam medium bands are able to reveal subtle differences in the young stellar populations present in these regions.
The reddening in $F300M-F335M$ along the southern radio jet axis illustrated in Fig.~\ref{fig:ew_map} can be seen in the integrated SEDs (Fig.~\ref{fig:sed_core}) with a significant enhancement in \ewha compared to the core or northern galaxy.
This stronger emission line contribution results in a much greater upturn in recent SF activity for the $0 < t < 10$ Myr bin, with the SFR in the south galaxy region increasing by a factor $\sim5$ compared to $\sim2$ in the core.
We also note that the southern side of the galaxy core is the only region of the host galaxy that has strong evidence favouring the complex SFH regardless of dust assumption (Fig.~\ref{fig:sed_bestmodel}).

Given the co-location of the bright radio continuum emission with the enhancement of recent SF in the host galaxy, we interpret these observations as further evidence supporting jet-induced SF as the dominant mechanism driving the extended high \ewha emission.
As above, we defer a more complete analysis of the formation history of TNJ1338 in the context of both its surrounding protocluster and the wider galaxy population at $z\sim4$ to subsequent studies (Duncan et al., in prep).

% \begin{figure}
% \centering
%     \includegraphics[width=\columnwidth]{figures/tnj1338-1942_20221014_galaxy_total_calz.pdf}
%     \includegraphics[width=\columnwidth]{figures/tnj1338-1942_20221014_galaxy_total_smc.pdf}
%     %\includegraphics[width=\columnwidth]{figures/sfh_plot_s8.pdf}
% 	\caption{Inferred non-parametric SFH for the core and outer regions of the host galaxy (see Fig.~\ref{fig:sed_sfh}). Solid lines and shaded regions illustrate the median and 16-84th percentiles of the \texttt{Prospector} SED fitting posteriors.}
%     \label{fig:total_sfh}
% \end{figure}

% \subsection{Integrated stellar population properties of TNJ1338}

% \noindent\textit{-- What is the star-formation history of TNJ1338?\\
% -- How does its evolution compare to the most massive/lumious galaxies observed at $z > 6$?\\
% -- What is the current star-formation rate of TNJ1338? And what does this mean for its fuelling/feedback within the context of the protocluster environment?
% }

\section{Summary and future prospects}\label{sec:conclusions}
We have used JWST/NIRCam to study the resolved rest-frame optical properties of the $z=4.11$ high-redshift radio galaxy TNJ1338.
Our results are summarized as follows:
\begin{itemize}
    \item Deep NIRCam imaging confirms the clear presence of a bright galaxy core in the South of the emission region.
    We observe significant variation in the $F300M - F335M$ colour (a proxy for \ewha) across the associated jet-aligned emission, with enhanced emission line contribution associated with the Northern luminous radio lobe and the fainter Southern jet.
    
    \item JWST's unique combination of resolution and sensitivity allow us to study the detailed rest-frame optical properties of the jet-aligned emission and host galaxy for the first time. Resolved SED modelling incorporating both a range of stellar population assumptions and shock models reveal that the spectral energy distribution of the emission closest to the jet is best-fit by radiative shock models, surrounded by regions more consistent with an extremely young burst of star-formation.
    
    \item Thanks to the tight constraints on the rest-frame \ewha provided by our NIRCam medium-bands, we are able to constrain the mass-weighted ages of the star-formation associated with the jet activity to $t_{\textup{mass}} <4$ Myr, consistent with the expected lifetime of radio jets with the projected size observed in TNJ1338. 
    Together, these new observations support the previously hypothesised physical picture of TNJ1338 being a massive galaxy undergoing significant obscured AGN activity that is powering the luminous radio jets, which are themselves driving radiatively driven shocks and extensive recent star-formation in its wake.
    However, additional detailed spectroscopic observations are required to conclusively confirm this scenario.
    
    \item Under the assumption that all of the jet aligned emission is dominated by stellar populations, the total current SFR estimated ranges from $\log_{10}(\text{SFR}/\text{M}_{\odot}\,\text{yr}^{-1}) = 3.02^{+0.07}_{-0.06}$ to $\log_{10}(\text{SFR}/\text{M}_{\odot}\,\text{yr}^{-1}) = 3.21^{+0.06}_{-0.04}$ depending on the assumed dust attenuation law.
    Even conservatively excluding any region of the jet aligned optical emission that may be dominated by radiative shocks, the SFR attributed to the jet-triggered starburst amounts to at least $\sim500\,\text{M}_{\odot}\,\text{yr}^{-1}$ ($\sim5\times$ than previously estimated from rest-frame UV emission alone).
    
    \item Our NIRCam observations confirm previous estimates placing TNJ1338 as one of the most massive galaxies at this epoch, with total stellar mass estimates of $\log_{10}(M/M_{\odot}) = 10.88\pm0.04$ to $\log_{10}(M/M_{\odot}) = 10.97\pm0.06$ assuming SMC \citep{Gordon2003} or \citet{Calzetti2000} dust attenuation respectively. 
\end{itemize}

The scale of the jet-triggered SF activity revealed by our NIRCam observations highlights the role that positive AGN feedback could play in the formation of massive galaxies.
The observations presented here, and the complementary studies \citep{PearlsOverview, Cheng2022}, represent only a fraction of the full potential contained in the NIRCam observations.
In the second paper in this series (Duncan et al., in prep.) we will exploit the extraordinary sensitivity of NIRCam to H$\alpha$ emission at $z\sim4.1$ to produce a complete census of the proto-cluster environment surrounding TNJ1338.
Additionally, Cycle 1 NIRSpec IFU observations (GO Programme 1964) will add a wealth of new information, including kinematics and resolved maps of emission line diagnostics, that will allow conclusive information on the precise physical mechanisms within the jet-aligned emission.
Full spectro-photometric SED modelling \citep{Prospector} combining the NIRCam and NIRSpec observations should enable more detailed constraints on the formation history of the TNJ1338 host galaxy, alleviating existing degeneracies between the dust, stellar ages and emission line contributions. 

Finally, we highlight that these observations demonstrate the extraordinary potential for further studies of HzRGs extending to higher redshifts and into the epoch of reionization. 
Forthcoming massively-multiplexed spectroscopic surveys such as WEAVE-LOFAR \citep{WEAVELOFAR} will soon provide large homogeneous samples of luminous HzRG out to $z > 6$, that combined with the power of JWST will enable systematic and statistical studies of SMBH host galaxies and the role of AGN feedback in shaping the most massive galaxies in the Universe.

% \begin{itemize}
%     \item Growth of structure ($z=4.11$ protocluster): 
%     \begin{itemize}
%         \item H$\alpha$ + dropout selection of protocluster galaxies
%         \item Additional HST/ACS (+WFC3) mid-cycle observations to extend detailed stellar population studies to protocluster
%     \end{itemize}
%     \item Host galaxy: 
%     % \begin{itemize}
%     %     \item Depending on the results of this paper and NIRSpec observations, additional observations may also answer different questions\\
%     % - MIRI Cycle 2? (Additional NIRCam?)\\
%     % - ALMA
%     % \end{itemize}
%         % \item Future HzRGs: These observations will be an excellent pathfinder for further studies of HzRGs extending into the EoR. With WEAVE-LOFAR commencing large-scale spectroscopic follow-up of low-frequency selected radio sources, we expect large samples of radio AGN out to $z > 6$ that will enable systematic and statistical studies across a representative range of luminosity/redshift.  \\
% \end{itemize}


\section*{Acknowledgements}
KJD thanks Philip Best, Dan Smith, Roderik Overzier, Aayush Saxena, Rebecca Larson and Fergus Cullen for valuable discussions and input in the course of this analysis.
KJD acknowledges funding from the European Union's Horizon 2020 research and innovation programme under the Marie Sk\l{}odowska-Curie grant agreement No. 892117 (HIZRAD) and support from the STFC through an Ernest Rutherford Fellowship (grant number ST/W003120/1).
This work is based on observations made with the NASA/ESA/CSA James Webb Space
Telescope. The data were obtained from the Mikulski Archive for Space
Telescopes at the Space Telescope Science Institute, which is operated by the
Association of Universities for Research in Astronomy, Inc., under NASA
contract NAS 5-03127 for JWST. These observations are associated with JWST
programmes 1176 and 2738.
RAW, SHC, and RAJ acknowledge support from NASA JWST Interdisciplinary
Scientist grants NAG5-12460, NNX14AN10G and 80NSSC18K0200 from GSFC. Work by
CJC acknowledges support from the European Research Council (ERC) Advanced
Investigator Grant EPOCHS (788113). BLF thanks the Berkeley Center for
Theoretical Physics for their hospitality during the writing of this paper.
MAM acknowledges the support of a National Research Council of Canada Plaskett
Fellowship, and the Australian Research Council center of Excellence for All
Sky Astrophysics in 3 Dimensions (ASTRO 3D), through project number CE17010001.
CNAW acknowledges funding from the JWST/NIRCam contract NASS-0215 to the
University of Arizona.
TAH is supported by an appointment to the NASA Postdoctoral Program (NPP) at NASA Goddard Space Flight Center, administered by Oak Ridge Associated Universities under contract with NASA.
We also acknowledge the indigenous peoples of Arizona, including the Akimel
O'odham (Pima) and Pee Posh (Maricopa) Indian Communities, whose care and
keeping of the land has enabled us to be at ASU's Tempe campus in the Salt
River Valley, where much of our work was conducted.
For the purpose of open access, the author has applied a Creative Commons Attribution (CC BY) licence to any Author Accepted Manuscript version arising from this submission.

%For the purpose of open access, the author has applied a Creative Commons Attribution (CC BY) licence to any Author Accepted Manuscript version arising from this submission.

%%%%%%%%%%%%%%%%%%%%%%%%%%%%%%%%%%%%%%%%%%%%%%%%%%
\section*{Data Availability}
The raw data underlying this article will be freely available on the Mikulski Archive for Space Telescopes (\url{https://mast.stsci.edu}) following the 12 month exclusive access period.
In the interim, raw and processed imaging data as well as resolved photometry catalogues will be shared on request to the principal investigator of the PEARLS programme (R. A. Windhorst) and the corresponding author (KJD).

\bibliography{tnj1338_hzrg}{}

\begin{thebibliography}{}
\expandafter\ifx\csname natexlab\endcsname\relax\def\natexlab#1{#1}\fi
\providecommand{\url}[1]{\href{#1}{#1}}
\providecommand{\dodoi}[1]{doi:~\href{http://doi.org/#1}{\nolinkurl{#1}}}
\providecommand{\doeprint}[1]{\href{http://ascl.net/#1}{\nolinkurl{http://ascl.net/#1}}}
\providecommand{\doarXiv}[1]{\href{https://arxiv.org/abs/#1}{\nolinkurl{https://arxiv.org/abs/#1}}}

\bibitem[{{Allen} {et~al.}(2008){Allen}, {Groves}, {Dopita}, {Sutherland}, \&
  {Kewley}}]{MappingsIII}
{Allen}, M.~G., {Groves}, B.~A., {Dopita}, M.~A., {Sutherland}, R.~S., \&
  {Kewley}, L.~J. 2008, \apjs, 178, 20, \dodoi{10.1086/589652}

\bibitem[{{Avila} {et~al.}(2015){Avila}, {Hack}, {Cara}, {Borncamp}, {Mack},
  {Smith}, \& {Ubeda}}]{Avila2015}
{Avila}, R.~J., {Hack}, W., {Cara}, M., {et~al.} 2015, in Astronomical Society
  of the Pacific Conference Series, Vol. 495, Astronomical Data Analysis
  Software an Systems XXIV (ADASS XXIV), ed. A.~R. {Taylor} \& E.~{Rosolowsky},
  281.
\newblock \doarXiv{1411.5605}

\bibitem[{{Baba} {et~al.}(2002){Baba}, {Yasuda}, {Ichikawa}, {Yagi}, {Iwamoto},
  {Takata}, {Horaguchi}, {Taga}, {Watanabe}, {Ozawa}, \& {Hamabe}}]{Baba2002}
{Baba}, H., {Yasuda}, N., {Ichikawa}, S.-I., {et~al.} 2002, in Astronomical
  Society of the Pacific Conference Series, Vol. 281, Astronomical Data
  Analysis Software and Systems XI, ed. D.~A. {Bohlender}, D.~{Durand}, \&
  T.~H. {Handley}, 298

\bibitem[{{Best} {et~al.}(1997){Best}, {Longair}, \& {Rottgering}}]{Best1997}
{Best}, P.~N., {Longair}, M.~S., \& {Rottgering}, H.~J.~A. 1997, \mnras, 286,
  785, \dodoi{10.1093/mnras/286.4.785}

\bibitem[{{Best} {et~al.}(2000){Best}, {R{\"o}ttgering}, \&
  {Longair}}]{Best2000}
{Best}, P.~N., {R{\"o}ttgering}, H.~J.~A., \& {Longair}, M.~S. 2000, \mnras,
  311, 23, \dodoi{10.1046/j.1365-8711.2000.03028.x}

\bibitem[{{Bicknell} {et~al.}(2000){Bicknell}, {Sutherland}, {van Breugel},
  {Dopita}, {Dey}, \& {Miley}}]{Bicknell2000}
{Bicknell}, G.~V., {Sutherland}, R.~S., {van Breugel}, W. J.~M., {et~al.} 2000,
  \apj, 540, 678, \dodoi{10.1086/309343}

\bibitem[{{Bielby} {et~al.}(2012){Bielby}, {Hudelot}, {McCracken}, {Ilbert},
  {Daddi}, {Le F{\`e}vre}, {Gonzalez-Perez}, {Kneib}, {Marmo}, {Mellier},
  {Salvato}, {Sanders}, \& {Willott}}]{Bielby2012}
{Bielby}, R., {Hudelot}, P., {McCracken}, H.~J., {et~al.} 2012, \aap, 545, A23,
  \dodoi{10.1051/0004-6361/201118547}

\bibitem[{{Boucaud} {et~al.}(2016){Boucaud}, {Bocchio}, {Abergel}, {Orieux},
  {Dole}, \& {Hadj-Youcef}}]{Boucaud2016}
{Boucaud}, A., {Bocchio}, M., {Abergel}, A., {et~al.} 2016, \aap, 596, A63,
  \dodoi{10.1051/0004-6361/201629080}

\bibitem[{{Boyer} {et~al.}(2022){Boyer}, {Anderson}, {Gennaro}, {Geha},
  {Wingfield McQuinn}, {Tollerud}, {Correnti}, {Brenner Newman}, {Cohen},
  {Kallivayalil}, {Beaton}, {Cole}, {Dolphin}, {Kalirai}, {Sandstrom},
  {Savino}, {Skillman}, {Weisz}, \& {Williams}}]{Boyer2022}
{Boyer}, M.~L., {Anderson}, J., {Gennaro}, M., {et~al.} 2022, Research Notes of
  the American Astronomical Society, 6, 191, \dodoi{10.3847/2515-5172/ac923a}

\bibitem[{{Bradley} {et~al.}(2016){Bradley}, {Sipocz}, {Robitaille},
  {Tollerud}, {Deil}, {Vin{\'\i}cius}, {Barbary}, {G{\"u}nther}, {Bostroem},
  {Droettboom}, {Bray}, {Bratholm}, {Pickering}, {Craig}, {Pascual}, {Greco},
  {Donath}, {Kerzendorf}, {Littlefair}, {Barentsen}, {D'Eugenio}, \&
  {Weaver}}]{PhotutilsASCL}
{Bradley}, L., {Sipocz}, B., {Robitaille}, T., {et~al.} 2016, {Photutils:
  Photometry tools}, Astrophysics Source Code Library, record ascl:1609.011.
\newblock \doeprint{1609.011}

\bibitem[{{Bradley} {et~al.}(2022){Bradley}, {Sip{\H{o}}cz}, {Robitaille},
  {Tollerud}, {Vin{\'\i}cius}, {Deil}, {Barbary}, {Wilson}, {Busko}, {Donath},
  {G{\"u}nther}, {Cara}, {Lim}, {Me{\ss}linger}, {Conseil}, {Bostroem},
  {Droettboom}, {Bray}, {Andersen Bratholm}, {Barentsen}, {Craig}, {Rathi},
  {Pascual}, {Perren}, {Georgiev}, {De Val-Borro}, {Kerzendorf}, {Bach},
  {Quint}, \& {Souchereau}}]{Photutils1.5.0}
{Bradley}, L., {Sip{\H{o}}cz}, B., {Robitaille}, T., {et~al.} 2022,
  {astropy/photutils: 1.5.0}, 1.5.0, Zenodo,  Zenodo,
  \dodoi{10.5281/zenodo.6825092}

\bibitem[{{Byler} {et~al.}(2017){Byler}, {Dalcanton}, {Conroy}, \&
  {Johnson}}]{Byler2017}
{Byler}, N., {Dalcanton}, J.~J., {Conroy}, C., \& {Johnson}, B.~D. 2017, \apj,
  840, 44, \dodoi{10.3847/1538-4357/aa6c66}

\bibitem[{{Calzetti} {et~al.}(2000){Calzetti}, {Armus}, {Bohlin}, {Kinney},
  {Koornneef}, \& {Storchi-Bergmann}}]{Calzetti2000}
{Calzetti}, D., {Armus}, L., {Bohlin}, R.~C., {et~al.} 2000, \apj, 533, 682,
  \dodoi{10.1086/308692}

\bibitem[{{Capetti} {et~al.}(2022){Capetti}, {Balmaverde}, {Tadhunter},
  {Marconi}, {Venturi}, {Chiaberge}, {Baldi}, {Baum}, {Gilli}, {Grandi},
  {Meyer}, {Miley}, {O'Dea}, {Sparks}, {Torresi}, \& {Tremblay}}]{Capetti2022}
{Capetti}, A., {Balmaverde}, B., {Tadhunter}, C., {et~al.} 2022, \aap, 657,
  A114, \dodoi{10.1051/0004-6361/202141965}

\bibitem[{{Cappellari} \& {Copin}(2003)}]{Cappellari2003}
{Cappellari}, M., \& {Copin}, Y. 2003, \mnras, 342, 345,
  \dodoi{10.1046/j.1365-8711.2003.06541.x}

\bibitem[{{Cheng} {et~al.}(2022){Cheng}, {Huang}, {Smail}, {Yan}, {Cohen},
  {Jansen}, {Windhorst}, {Ma}, {Koekemoer}, {Willmer}, {Willner}, {Diego},
  {Frye}, {Conselice}, {Ferreira}, {Petric}, {Yun}, {Gim}, {Polletta},
  {Duncan}, {Honor}, {Holwerda}, {R{\"o}ttgering}, {Hathi}, {Kamieneski},
  {Adams}, {Coe}, {Broadhurst}, {Summers}, {Tompkins}, {Driver}, {Grogin},
  {Marshall}, {Pirzkal}, {Robotham}, \& {Ryan}}]{Cheng2022}
{Cheng}, C., {Huang}, J.-S., {Smail}, I., {et~al.} 2022, arXiv e-prints,
  arXiv:2210.08163.
\newblock \doarXiv{2210.08163}

\bibitem[{{Choi} {et~al.}(2016){Choi}, {Dotter}, {Conroy}, {Cantiello},
  {Paxton}, \& {Johnson}}]{MIST}
{Choi}, J., {Dotter}, A., {Conroy}, C., {et~al.} 2016, \apj, 823, 102,
  \dodoi{10.3847/0004-637X/823/2/102}

\bibitem[{{Conroy} \& {Gunn}(2010)}]{Conroy2010}
{Conroy}, C., \& {Gunn}, J.~E. 2010, \apj, 712, 833,
  \dodoi{10.1088/0004-637X/712/2/833}

\bibitem[{{Conroy} {et~al.}(2009){Conroy}, {Gunn}, \& {White}}]{Conroy2009}
{Conroy}, C., {Gunn}, J.~E., \& {White}, M. 2009, \apj, 699, 486,
  \dodoi{10.1088/0004-637X/699/1/486}

\bibitem[{{Cresci} {et~al.}(2015){Cresci}, {Marconi}, {Zibetti}, {Risaliti},
  {Carniani}, {Mannucci}, {Gallazzi}, {Maiolino}, {Balmaverde}, {Brusa},
  {Capetti}, {Cicone}, {Feruglio}, {Bland-Hawthorn}, {Nagao}, {Oliva},
  {Salvato}, {Sani}, {Tozzi}, {Urrutia}, \& {Venturi}}]{Cresci2015}
{Cresci}, G., {Marconi}, A., {Zibetti}, S., {et~al.} 2015, \aap, 582, A63,
  \dodoi{10.1051/0004-6361/201526581}

\bibitem[{{Crockett} {et~al.}(2012){Crockett}, {Shabala}, {Kaviraj},
  {Antonuccio-Delogu}, {Silk}, {Mutchler}, {O'Connell}, {Rejkuba}, {Whitmore},
  \& {Windhorst}}]{Crockett2012}
{Crockett}, R.~M., {Shabala}, S.~S., {Kaviraj}, S., {et~al.} 2012, \mnras, 421,
  1603, \dodoi{10.1111/j.1365-2966.2012.20418.x}

\bibitem[{{Croft} {et~al.}(2006){Croft}, {van Breugel}, {de Vries}, {Dopita},
  {Martin}, {Morganti}, {Neff}, {Oosterloo}, {Schiminovich}, {Stanford}, \&
  {van Gorkom}}]{Croft2006}
{Croft}, S., {van Breugel}, W., {de Vries}, W., {et~al.} 2006, \apj, 647, 1040,
  \dodoi{10.1086/505526}

\bibitem[{{Crowther} {et~al.}(2010){Crowther}, {Schnurr}, {Hirschi}, {Yusof},
  {Parker}, {Goodwin}, \& {Kassim}}]{Crowther2010}
{Crowther}, P.~A., {Schnurr}, O., {Hirschi}, R., {et~al.} 2010, \mnras, 408,
  731, \dodoi{10.1111/j.1365-2966.2010.17167.x}

\bibitem[{{De Breuck} {et~al.}(1999){De Breuck}, {van Breugel}, {Minniti},
  {Miley}, {R{\"o}ttgering}, {Stanford}, \& {Carilli}}]{DeBreuck1999}
{De Breuck}, C., {van Breugel}, W., {Minniti}, D., {et~al.} 1999, \aap, 352,
  L51.
\newblock \doarXiv{astro-ph/9909178}

\bibitem[{{De Breuck} {et~al.}(2000){De Breuck}, {van Breugel},
  {R{\"o}ttgering}, \& {Miley}}]{DeBreuck2000}
{De Breuck}, C., {van Breugel}, W., {R{\"o}ttgering}, H.~J.~A., \& {Miley}, G.
  2000, \aaps, 143, 303, \dodoi{10.1051/aas:2000181}

\bibitem[{{De Breuck} {et~al.}(2004){De Breuck}, {Bertoldi}, {Carilli},
  {Omont}, {Venemans}, {R{\"o}ttgering}, {Overzier}, {Reuland}, {Miley},
  {Ivison}, \& {van Breugel}}]{DeBreuck2004}
{De Breuck}, C., {Bertoldi}, F., {Carilli}, C., {et~al.} 2004, \aap, 424, 1,
  \dodoi{10.1051/0004-6361:20035885}

\bibitem[{{De Breuck} {et~al.}(2010){De Breuck}, {Seymour}, {Stern}, {Willner},
  {Eisenhardt}, {Fazio}, {Galametz}, {Lacy}, {Rettura}, {Rocca-Volmerange}, \&
  {Vernet}}]{DeBreuck2010}
{De Breuck}, C., {Seymour}, N., {Stern}, D., {et~al.} 2010, \apj, 725, 36,
  \dodoi{10.1088/0004-637X/725/1/36}

\bibitem[{{Dekel} {et~al.}(2009){Dekel}, {Birnboim}, {Engel}, {Freundlich},
  {Goerdt}, {Mumcuoglu}, {Neistein}, {Pichon}, {Teyssier}, \&
  {Zinger}}]{Dekel2009}
{Dekel}, A., {Birnboim}, Y., {Engel}, G., {et~al.} 2009, \nat, 457, 451,
  \dodoi{10.1038/nature07648}

\bibitem[{{Dey} {et~al.}(1997){Dey}, {van Breugel}, {Vacca}, \&
  {Antonucci}}]{Dey1997}
{Dey}, A., {van Breugel}, W., {Vacca}, W.~D., \& {Antonucci}, R. 1997, \apj,
  490, 698, \dodoi{10.1086/304911}

\bibitem[{{Donley} {et~al.}(2012){Donley}, {Koekemoer}, {Brusa}, {Capak},
  {Cardamone}, {Civano}, {Ilbert}, {Impey}, {Kartaltepe}, {Miyaji}, {Salvato},
  {Sanders}, {Trump}, \& {Zamorani}}]{Donley2012}
{Donley}, J.~L., {Koekemoer}, A.~M., {Brusa}, M., {et~al.} 2012, \apj, 748,
  142, \dodoi{10.1088/0004-637X/748/2/142}

\bibitem[{{Dopita} {et~al.}(2005){Dopita}, {Groves}, {Fischera}, {Sutherland},
  {Tuffs}, {Popescu}, {Kewley}, {Reuland}, \& {Leitherer}}]{Dopita2005}
{Dopita}, M.~A., {Groves}, B.~A., {Fischera}, J., {et~al.} 2005, \apj, 619,
  755, \dodoi{10.1086/423948}

\bibitem[{{Draine} \& {Li}(2007)}]{Draine2007}
{Draine}, B.~T., \& {Li}, A. 2007, \apj, 657, 810, \dodoi{10.1086/511055}

\bibitem[{{Drouart} {et~al.}(2016){Drouart}, {Rocca-Volmerange}, {De Breuck},
  {Fioc}, {Lehnert}, {Seymour}, {Stern}, \& {Vernet}}]{Drouart2016}
{Drouart}, G., {Rocca-Volmerange}, B., {De Breuck}, C., {et~al.} 2016, \aap,
  593, A109, \dodoi{10.1051/0004-6361/201526880}

\bibitem[{{Duncan} {et~al.}(2014){Duncan}, {Conselice}, {Mortlock}, {Hartley},
  {Guo}, {Ferguson}, {Dav{\'e}}, {Lu}, {Ownsworth}, {Ashby}, {Dekel},
  {Dickinson}, {Faber}, {Giavalisco}, {Grogin}, {Kocevski}, {Koekemoer},
  {Somerville}, \& {White}}]{Duncan2014}
{Duncan}, K., {Conselice}, C.~J., {Mortlock}, A., {et~al.} 2014, \mnras, 444,
  2960, \dodoi{10.1093/mnras/stu1622}

\bibitem[{{Duncan} {et~al.}(2019){Duncan}, {Conselice}, {Mundy}, {Bell},
  {Donley}, {Galametz}, {Guo}, {Grogin}, {Hathi}, {Kartaltepe}, {Kocevski},
  {Koekemoer}, {P{\'e}rez-Gonz{\'a}lez}, {Mantha}, {Snyder}, \&
  {Stefanon}}]{Duncan2019}
{Duncan}, K., {Conselice}, C.~J., {Mundy}, C., {et~al.} 2019, \apj, 876, 110,
  \dodoi{10.3847/1538-4357/ab148a}

\bibitem[{{Eldridge} {et~al.}(2017){Eldridge}, {Stanway}, {Xiao}, {McClelland},
  {Taylor}, {Ng}, {Greis}, \& {Bray}}]{Eldridge2017}
{Eldridge}, J.~J., {Stanway}, E.~R., {Xiao}, L., {et~al.} 2017, \pasa, 34,
  e058, \dodoi{10.1017/pasa.2017.51}

\bibitem[{{Endsley} {et~al.}(2022){Endsley}, {Stark}, {Whitler}, {Topping},
  {Chen}, {Plat}, {Chisholm}, \& {Charlot}}]{Endsley2022}
{Endsley}, R., {Stark}, D.~P., {Whitler}, L., {et~al.} 2022, arXiv e-prints,
  arXiv:2208.14999.
\newblock \doarXiv{2208.14999}

\bibitem[{{Fabian}(2012)}]{Fabian2012}
{Fabian}, A.~C. 2012, \araa, 50, 455,
  \dodoi{10.1146/annurev-astro-081811-125521}

\bibitem[{{Falkendal} {et~al.}(2019){Falkendal}, {De Breuck}, {Lehnert},
  {Drouart}, {Vernet}, {Emonts}, {Lee}, {Nesvadba}, {Seymour}, {B{\'e}thermin},
  {Kolwa}, {Gullberg}, \& {Wylezalek}}]{Falkendal2019}
{Falkendal}, T., {De Breuck}, C., {Lehnert}, M.~D., {et~al.} 2019, \aap, 621,
  A27, \dodoi{10.1051/0004-6361/201732485}

\bibitem[{{Ferland} {et~al.}(2013){Ferland}, {Porter}, {van Hoof}, {Williams},
  {Abel}, {Lykins}, {Shaw}, {Henney}, \& {Stancil}}]{Cloudy}
{Ferland}, G.~J., {Porter}, R.~L., {van Hoof}, P.~A.~M., {et~al.} 2013, \rmxaa,
  49, 137.
\newblock \doarXiv{1302.4485}

\bibitem[{{Gaia Collaboration} {et~al.}(2021){Gaia Collaboration}, {Brown},
  {Vallenari}, {Prusti}, {de Bruijne}, {Babusiaux}, {Biermann}, {Creevey},
  {Evans}, {Eyer}, {Hutton}, {Jansen}, {Jordi}, {Klioner}, {Lammers},
  {Lindegren}, {Luri}, {Mignard}, {Panem}, {Pourbaix}, {Randich}, {Sartoretti},
  {Soubiran}, {Walton}, {Arenou}, {Bailer-Jones}, {Bastian}, {Cropper},
  {Drimmel}, {Katz}, {Lattanzi}, {van Leeuwen}, {Bakker}, {Cacciari},
  {Casta{\~n}eda}, {De Angeli}, {Ducourant}, {Fabricius}, {Fouesneau},
  {Fr{\'e}mat}, {Guerra}, {Guerrier}, {Guiraud}, {Jean-Antoine Piccolo},
  {Masana}, {Messineo}, {Mowlavi}, {Nicolas}, {Nienartowicz}, {Pailler},
  {Panuzzo}, {Riclet}, {Roux}, {Seabroke}, {Sordo}, {Tanga}, {Th{\'e}venin},
  {Gracia-Abril}, {Portell}, {Teyssier}, {Altmann}, {Andrae}, {Bellas-Velidis},
  {Benson}, {Berthier}, {Blomme}, {Brugaletta}, {Burgess}, {Busso}, {Carry},
  {Cellino}, {Cheek}, {Clementini}, {Damerdji}, {Davidson}, {Delchambre},
  {Dell'Oro}, {Fern{\'a}ndez-Hern{\'a}ndez}, {Galluccio}, {Garc{\'\i}a-Lario},
  {Garcia-Reinaldos}, {Gonz{\'a}lez-N{\'u}{\~n}ez}, {Gosset}, {Haigron},
  {Halbwachs}, {Hambly}, {Harrison}, {Hatzidimitriou}, {Heiter},
  {Hern{\'a}ndez}, {Hestroffer}, {Hodgkin}, {Holl}, {Jan{\ss}en}, {Jevardat de
  Fombelle}, {Jordan}, {Krone-Martins}, {Lanzafame}, {L{\"o}ffler}, {Lorca},
  {Manteiga}, {Marchal}, {Marrese}, {Moitinho}, {Mora}, {Muinonen}, {Osborne},
  {Pancino}, {Pauwels}, {Petit}, {Recio-Blanco}, {Richards}, {Riello},
  {Rimoldini}, {Robin}, {Roegiers}, {Rybizki}, {Sarro}, {Siopis}, {Smith},
  {Sozzetti}, {Ulla}, {Utrilla}, {van Leeuwen}, {van Reeven}, {Abbas}, {Abreu
  Aramburu}, {Accart}, {Aerts}, {Aguado}, {Ajaj}, {Altavilla}, {{\'A}lvarez},
  {{\'A}lvarez Cid-Fuentes}, {Alves}, {Anderson}, {Anglada Varela}, {Antoja},
  {Audard}, {Baines}, {Baker}, {Balaguer-N{\'u}{\~n}ez}, {Balbinot}, {Balog},
  {Barache}, {Barbato}, {Barros}, {Barstow}, {Bartolom{\'e}}, {Bassilana},
  {Bauchet}, {Baudesson-Stella}, {Becciani}, {Bellazzini}, {Bernet}, {Bertone},
  {Bianchi}, {Blanco-Cuaresma}, {Boch}, {Bombrun}, {Bossini}, {Bouquillon},
  {Bragaglia}, {Bramante}, {Breedt}, {Bressan}, {Brouillet}, {Bucciarelli},
  {Burlacu}, {Busonero}, {Butkevich}, {Buzzi}, {Caffau}, {Cancelliere},
  {C{\'a}novas}, {Cantat-Gaudin}, {Carballo}, {Carlucci}, {Carnerero},
  {Carrasco}, {Casamiquela}, {Castellani}, {Castro-Ginard}, {Castro Sampol},
  {Chaoul}, {Charlot}, {Chemin}, {Chiavassa}, {Cioni}, {Comoretto}, {Cooper},
  {Cornez}, {Cowell}, {Crifo}, {Crosta}, {Crowley}, {Dafonte}, {Dapergolas},
  {David}, {David}, {de Laverny}, {De Luise}, {De March}, {De Ridder}, {de
  Souza}, {de Teodoro}, {de Torres}, {del Peloso}, {del Pozo}, {Delbo},
  {Delgado}, {Delgado}, {Delisle}, {Di Matteo}, {Diakite}, {Diener},
  {Distefano}, {Dolding}, {Eappachen}, {Edvardsson}, {Enke}, {Esquej}, {Fabre},
  {Fabrizio}, {Faigler}, {Fedorets}, {Fernique}, {Fienga}, {Figueras},
  {Fouron}, {Fragkoudi}, {Fraile}, {Franke}, {Gai}, {Garabato},
  {Garcia-Gutierrez}, {Garc{\'\i}a-Torres}, {Garofalo}, {Gavras}, {Gerlach},
  {Geyer}, {Giacobbe}, {Gilmore}, {Girona}, {Giuffrida}, {Gomel}, {Gomez},
  {Gonzalez-Santamaria}, {Gonz{\'a}lez-Vidal}, {Granvik},
  {Guti{\'e}rrez-S{\'a}nchez}, {Guy}, {Hauser}, {Haywood}, {Helmi}, {Hidalgo},
  {Hilger}, {H{\l}adczuk}, {Hobbs}, {Holland}, {Huckle}, {Jasniewicz},
  {Jonker}, {Juaristi Campillo}, {Julbe}, {Karbevska}, {Kervella}, {Khanna},
  {Kochoska}, {Kontizas}, {Kordopatis}, {Korn}, {Kostrzewa-Rutkowska},
  {Kruszy{\'n}ska}, {Lambert}, {Lanza}, {Lasne}, {Le Campion}, {Le Fustec},
  {Lebreton}, {Lebzelter}, {Leccia}, {Leclerc}, {Lecoeur-Taibi}, {Liao},
  {Licata}, {Lindstr{\o}m}, {Lister}, {Livanou}, {Lobel}, {Madrero Pardo},
  {Managau}, {Mann}, {Marchant}, {Marconi}, {Marcos Santos}, {Marinoni},
  {Marocco}, {Marshall}, {Martin Polo}, {Mart{\'\i}n-Fleitas}, {Masip},
  {Massari}, {Mastrobuono-Battisti}, {Mazeh}, {McMillan}, {Messina},
  {Michalik}, {Millar}, {Mints}, {Molina}, {Molinaro}, {Moln{\'a}r},
  {Montegriffo}, {Mor}, {Morbidelli}, {Morel}, {Morris}, {Mulone}, {Munoz},
  {Muraveva}, {Murphy}, {Musella}, {Noval}, {Ord{\'e}novic}, {Orr{\`u}},
  {Osinde}, {Pagani}, {Pagano}, {Palaversa}, {Palicio}, {Panahi}, {Pawlak},
  {Pe{\~n}alosa Esteller}, {Penttil{\"a}}, {Piersimoni}, {Pineau}, {Plachy},
  {Plum}, {Poggio}, {Poretti}, {Poujoulet}, {Pr{\v{s}}a}, {Pulone}, {Racero},
  {Ragaini}, {Rainer}, {Raiteri}, {Rambaux}, {Ramos}, {Ramos-Lerate}, {Re
  Fiorentin}, {Regibo}, {Reyl{\'e}}, {Ripepi}, {Riva}, {Rixon}, {Robichon},
  {Robin}, {Roelens}, {Rohrbasser}, {Romero-G{\'o}mez}, {Rowell}, {Royer},
  {Rybicki}, {Sadowski}, {Sagrist{\`a} Sell{\'e}s}, {Sahlmann}, {Salgado},
  {Salguero}, {Samaras}, {Sanchez Gimenez}, {Sanna}, {Santove{\~n}a},
  {Sarasso}, {Schultheis}, {Sciacca}, {Segol}, {Segovia}, {S{\'e}gransan},
  {Semeux}, {Shahaf}, {Siddiqui}, {Siebert}, {Siltala}, {Slezak}, {Smart},
  {Solano}, {Solitro}, {Souami}, {Souchay}, {Spagna}, {Spoto}, {Steele},
  {Steidelm{\"u}ller}, {Stephenson}, {S{\"u}veges}, {Szabados}, {Szegedi-Elek},
  {Taris}, {Tauran}, {Taylor}, {Teixeira}, {Thuillot}, {Tonello}, {Torra},
  {Torra}, {Turon}, {Unger}, {Vaillant}, {van Dillen}, {Vanel}, {Vecchiato},
  {Viala}, {Vicente}, {Voutsinas}, {Weiler}, {Wevers}, {Wyrzykowski}, {Yoldas},
  {Yvard}, {Zhao}, {Zorec}, {Zucker}, {Zurbach}, \& {Zwitter}}]{Gaia2021}
{Gaia Collaboration}, {Brown}, A.~G.~A., {Vallenari}, A., {et~al.} 2021, \aap,
  649, A1, \dodoi{10.1051/0004-6361/202039657}

\bibitem[{{Gaia Collaboration} {et~al.}(2022){Gaia Collaboration}, {Vallenari},
  {Brown}, {Prusti}, {de Bruijne}, {Arenou}, {Babusiaux}, {Biermann},
  {Creevey}, {Ducourant}, {Evans}, {Eyer}, {Guerra}, {Hutton}, {Jordi},
  {Klioner}, {Lammers}, {Lindegren}, {Luri}, {Mignard}, {Panem}, {Pourbaix},
  {Randich}, {Sartoretti}, {Soubiran}, {Tanga}, {Walton}, {Bailer-Jones},
  {Bastian}, {Drimmel}, {Jansen}, {Katz}, {Lattanzi}, {van Leeuwen}, {Bakker},
  {Cacciari}, {Casta{\~n}eda}, {De Angeli}, {Fabricius}, {Fouesneau},
  {Fr{\'e}mat}, {Galluccio}, {Guerrier}, {Heiter}, {Masana}, {Messineo},
  {Mowlavi}, {Nicolas}, {Nienartowicz}, {Pailler}, {Panuzzo}, {Riclet}, {Roux},
  {Seabroke}, {Sordo{\o}rcit}, {Th{\'e}venin}, {Gracia-Abril}, {Portell},
  {Teyssier}, {Altmann}, {Andrae}, {Audard}, {Bellas-Velidis}, {Benson},
  {Berthier}, {Blomme}, {Burgess}, {Busonero}, {Busso}, {C{\'a}novas}, {Carry},
  {Cellino}, {Cheek}, {Clementini}, {Damerdji}, {Davidson}, {de Teodoro},
  {Nu{\~n}ez Campos}, {Delchambre}, {Dell'Oro}, {Esquej},
  {Fern{\'a}ndez-Hern{\'a}ndez}, {Fraile}, {Garabato}, {Garc{\'\i}a-Lario},
  {Gosset}, {Haigron}, {Halbwachs}, {Hambly}, {Harrison}, {Hern{\'a}ndez},
  {Hestroffer}, {Hodgkin}, {Holl}, {Jan{\ss}en}, {Jevardat de Fombelle},
  {Jordan}, {Krone-Martins}, {Lanzafame}, {L{\"o}ffler}, {Marchal}, {Marrese},
  {Moitinho}, {Muinonen}, {Osborne}, {Pancino}, {Pauwels}, {Recio-Blanco},
  {Reyl{\'e}}, {Riello}, {Rimoldini}, {Roegiers}, {Rybizki}, {Sarro}, {Siopis},
  {Smith}, {Sozzetti}, {Utrilla}, {van Leeuwen}, {Abbas}, {{\'A}brah{\'a}m},
  {Abreu Aramburu}, {Aerts}, {Aguado}, {Ajaj}, {Aldea-Montero}, {Altavilla},
  {{\'A}lvarez}, {Alves}, {Anders}, {Anderson}, {Anglada Varela}, {Antoja},
  {Baines}, {Baker}, {Balaguer-N{\'u}{\~n}ez}, {Balbinot}, {Balog}, {Barache},
  {Barbato}, {Barros}, {Barstow}, {Bartolom{\'e}}, {Bassilana}, {Bauchet},
  {Becciani}, {Bellazzini}, {Berihuete}, {Bernet}, {Bertone}, {Bianchi},
  {Binnenfeld}, {Blanco-Cuaresma}, {Blazere}, {Boch}, {Bombrun}, {Bossini},
  {Bouquillon}, {Bragaglia}, {Bramante}, {Breedt}, {Bressan}, {Brouillet},
  {Brugaletta}, {Bucciarelli}, {Burlacu}, {Butkevich}, {Buzzi}, {Caffau},
  {Cancelliere}, {Cantat-Gaudin}, {Carballo}, {Carlucci}, {Carnerero},
  {Carrasco}, {Casamiquela}, {Castellani}, {Castro-Ginard}, {Chaoul},
  {Charlot}, {Chemin}, {Chiaramida}, {Chiavassa}, {Chornay}, {Comoretto},
  {Contursi}, {Cooper}, {Cornez}, {Cowell}, {Crifo}, {Cropper}, {Crosta},
  {Crowley}, {Dafonte}, {Dapergolas}, {David}, {David}, {de Laverny}, {De
  Luise}, {De March}, {De Ridder}, {de Souza}, {de Torres}, {del Peloso}, {del
  Pozo}, {Delbo}, {Delgado}, {Delisle}, {Demouchy}, {Dharmawardena}, {Di
  Matteo}, {Diakite}, {Diener}, {Distefano}, {Dolding}, {Edvardsson}, {Enke},
  {Fabre}, {Fabrizio}, {Faigler}, {Fedorets}, {Fernique}, {Fienga}, {Figueras},
  {Fournier}, {Fouron}, {Fragkoudi}, {Gai}, {Garcia-Gutierrez},
  {Garcia-Reinaldos}, {Garc{\'\i}a-Torres}, {Garofalo}, {Gavel}, {Gavras},
  {Gerlach}, {Geyer}, {Giacobbe}, {Gilmore}, {Girona}, {Giuffrida}, {Gomel},
  {Gomez}, {Gonz{\'a}lez-N{\'u}{\~n}ez}, {Gonz{\'a}lez-Santamar{\'\i}a},
  {Gonz{\'a}lez-Vidal}, {Granvik}, {Guillout}, {Guiraud},
  {Guti{\'e}rrez-S{\'a}nchez}, {Guy}, {Hatzidimitriou}, {Hauser}, {Haywood},
  {Helmer}, {Helmi}, {Sarmiento}, {Hidalgo}, {Hilger}, {H{\l}adczuk}, {Hobbs},
  {Holland}, {Huckle}, {Jardine}, {Jasniewicz}, {Jean-Antoine Piccolo},
  {Jim{\'e}nez-Arranz}, {Jorissen}, {Juaristi Campillo}, {Julbe}, {Karbevska},
  {Kervella}, {Khanna}, {Kontizas}, {Kordopatis}, {Korn}, {K{\'o}sp{\'a}l},
  {Kostrzewa-Rutkowska}, {Kruszy{\'n}ska}, {Kun}, {Laizeau}, {Lambert},
  {Lanza}, {Lasne}, {Le Campion}, {Lebreton}, {Lebzelter}, {Leccia}, {Leclerc},
  {Lecoeur-Taibi}, {Liao}, {Licata}, {Lindstr{\o}m}, {Lister}, {Livanou},
  {Lobel}, {Lorca}, {Loup}, {Madrero Pardo}, {Magdaleno Romeo}, {Managau},
  {Mann}, {Manteiga}, {Marchant}, {Marconi}, {Marcos}, {Marcos Santos},
  {Mar{\'\i}n Pina}, {Marinoni}, {Marocco}, {Marshall}, {Polo},
  {Mart{\'\i}n-Fleitas}, {Marton}, {Mary}, {Masip}, {Massari},
  {Mastrobuono-Battisti}, {Mazeh}, {McMillan}, {Messina}, {Michalik}, {Millar},
  {Mints}, {Molina}, {Molinaro}, {Moln{\'a}r}, {Monari}, {Mongui{\'o}},
  {Montegriffo}, {Montero}, {Mor}, {Mora}, {Morbidelli}, {Morel}, {Morris},
  {Muraveva}, {Murphy}, {Musella}, {Nagy}, {Noval}, {Oca{\~n}a}, {Ogden},
  {Ordenovic}, {Osinde}, {Pagani}, {Pagano}, {Palaversa}, {Palicio},
  {Pallas-Quintela}, {Panahi}, {Payne-Wardenaar}, {Pe{\~n}alosa Esteller},
  {Penttil{\"a}}, {Pichon}, {Piersimoni}, {Pineau}, {Plachy}, {Plum}, {Poggio},
  {Pr{\v{s}}a}, {Pulone}, {Racero}, {Ragaini}, {Rainer}, {Raiteri}, {Rambaux},
  {Ramos}, {Ramos-Lerate}, {Re Fiorentin}, {Regibo}, {Richards}, {Rios Diaz},
  {Ripepi}, {Riva}, {Rix}, {Rixon}, {Robichon}, {Robin}, {Robin}, {Roelens},
  {Rogues}, {Rohrbasser}, {Romero-G{\'o}mez}, {Rowell}, {Royer}, {Ruz Mieres},
  {Rybicki}, {Sadowski}, {S{\'a}ez N{\'u}{\~n}ez}, {Sagrist{\`a} Sell{\'e}s},
  {Sahlmann}, {Salguero}, {Samaras}, {Sanchez Gimenez}, {Sanna},
  {Santove{\~n}a}, {Sarasso}, {Schultheis}, {Sciacca}, {Segol}, {Segovia},
  {S{\'e}gransan}, {Semeux}, {Shahaf}, {Siddiqui}, {Siebert}, {Siltala},
  {Silvelo}, {Slezak}, {Slezak}, {Smart}, {Snaith}, {Solano}, {Solitro},
  {Souami}, {Souchay}, {Spagna}, {Spina}, {Spoto}, {Steele},
  {Steidelm{\"u}ller}, {Stephenson}, {S{\"u}veges}, {Surdej}, {Szabados},
  {Szegedi-Elek}, {Taris}, {Taylo}, {Teixeira}, {Tolomei}, {Tonello}, {Torra},
  {Torra}, {Torralba Elipe}, {Trabucchi}, {Tsounis}, {Turon}, {Ulla}, {Unger},
  {Vaillant}, {van Dillen}, {van Reeven}, {Vanel}, {Vecchiato}, {Viala},
  {Vicente}, {Voutsinas}, {Weiler}, {Wevers}, {Wyrzykowski}, {Yoldas}, {Yvard},
  {Zhao}, {Zorec}, {Zucker}, \& {Zwitter}}]{Gaia2022}
{Gaia Collaboration}, {Vallenari}, A., {Brown}, A.~G.~A., {et~al.} 2022, arXiv
  e-prints, \url{https://arxiv.org/abs/2208.00211}.
\newblock \doarXiv{2208.00211}

\bibitem[{{Gardner} {et~al.}(2017){Gardner}, {Jones}, {Scannapieco}, \&
  {Windhorst}}]{Gardner2017}
{Gardner}, C.~L., {Jones}, J.~R., {Scannapieco}, E., \& {Windhorst}, R.~A.
  2017, \apj, 835, 232, \dodoi{10.3847/1538-4357/835/2/232}

\bibitem[{{Gordon} {et~al.}(2003){Gordon}, {Clayton}, {Misselt}, {Landolt}, \&
  {Wolff}}]{Gordon2003}
{Gordon}, K.~D., {Clayton}, G.~C., {Misselt}, K.~A., {Landolt}, A.~U., \&
  {Wolff}, M.~J. 2003, \apj, 594, 279, \dodoi{10.1086/376774}

\bibitem[{{Hardcastle}(2018)}]{Hardcastle2018}
{Hardcastle}, M.~J. 2018, \mnras, 475, 2768, \dodoi{10.1093/mnras/stx3358}

\bibitem[{{Hodge} \& {da Cunha}(2020)}]{Hodge2020}
{Hodge}, J.~A., \& {da Cunha}, E. 2020, Royal Society Open Science, 7, 200556,
  \dodoi{10.1098/rsos.200556}

\bibitem[{{Inskip} {et~al.}(2002){Inskip}, {Best}, {Rawlings}, {Longair},
  {Cotter}, {R{\"o}ttgering}, \& {Eales}}]{Inskip2002}
{Inskip}, K.~J., {Best}, P.~N., {Rawlings}, S., {et~al.} 2002, \mnras, 337,
  1381, \dodoi{10.1046/j.1365-8711.2002.06012.x}

\bibitem[{{Intema} {et~al.}(2006){Intema}, {Venemans}, {Kurk}, {Ouchi},
  {Kodama}, {R{\"o}ttgering}, {Miley}, \& {Overzier}}]{Intema2006}
{Intema}, H.~T., {Venemans}, B.~P., {Kurk}, J.~D., {et~al.} 2006, \aap, 456,
  433, \dodoi{10.1051/0004-6361:20064812}

\bibitem[{Johnson {et~al.}(2021)Johnson, Foreman-Mackey, Sick, Leja, Byler,
  Walmsley, Tollerud, Leung, \& Scott}]{pythonFSPS}
Johnson, B., Foreman-Mackey, D., Sick, J., {et~al.} 2021, dfm/python-fsps:
  python-fsps v0.4.1rc1, v0.4.1rc1,  Zenodo, \dodoi{10.5281/zenodo.4737461}

\bibitem[{{Johnson} {et~al.}(2021){Johnson}, {Leja}, {Conroy}, \&
  {Speagle}}]{Prospector}
{Johnson}, B.~D., {Leja}, J., {Conroy}, C., \& {Speagle}, J.~S. 2021, \apjs,
  254, 22, \dodoi{10.3847/1538-4365/abef67}

\bibitem[{{Keel} {et~al.}(2015){Keel}, {Maksym}, {Bennert}, {Lintott},
  {Chojnowski}, {Moiseev}, {Smirnova}, {Schawinski}, {Urry}, {Evans},
  {Pancoast}, {Scott}, {Showley}, \& {Flatland}}]{Keel2015}
{Keel}, W.~C., {Maksym}, W.~P., {Bennert}, V.~N., {et~al.} 2015, \aj, 149, 155,
  \dodoi{10.1088/0004-6256/149/5/155}

\bibitem[{{Kere{\v{s}}} {et~al.}(2005){Kere{\v{s}}}, {Katz}, {Weinberg}, \&
  {Dav{\'e}}}]{Keres2005}
{Kere{\v{s}}}, D., {Katz}, N., {Weinberg}, D.~H., \& {Dav{\'e}}, R. 2005,
  \mnras, 363, 2, \dodoi{10.1111/j.1365-2966.2005.09451.x}

\bibitem[{King \& Pounds(2015)}]{King2015}
King, A., \& Pounds, K. 2015, Annual Review of Astronomy and Astrophysics, 53,
  115, \dodoi{10.1146/annurev-astro-082214-122316}

\bibitem[{{Koekemoer} {et~al.}(2013){Koekemoer}, {Ellis}, {McLure}, {Dunlop},
  {Robertson}, {Ono}, {Schenker}, {Ouchi}, {Bowler}, {Rogers}, {Curtis-Lake},
  {Schneider}, {Charlot}, {Stark}, {Furlanetto}, {Cirasuolo}, {Wild}, \&
  {Targett}}]{Koekemoer2013}
{Koekemoer}, A.~M., {Ellis}, R.~S., {McLure}, R.~J., {et~al.} 2013, \apjs, 209,
  3, \dodoi{10.1088/0067-0049/209/1/3}

\bibitem[{{Kroupa}(2001)}]{Kroupa2001}
{Kroupa}, P. 2001, \mnras, 322, 231, \dodoi{10.1046/j.1365-8711.2001.04022.x}

\bibitem[{{Laigle} {et~al.}(2016){Laigle}, {McCracken}, {Ilbert}, {Hsieh},
  {Davidzon}, {Capak}, {Hasinger}, {Silverman}, {Pichon}, {Coupon}, {Aussel},
  {Le Borgne}, {Caputi}, {Cassata}, {Chang}, {Civano}, {Dunlop}, {Fynbo},
  {Kartaltepe}, {Koekemoer}, {Le F{\`e}vre}, {Le Floc'h}, {Leauthaud}, {Lilly},
  {Lin}, {Marchesi}, {Milvang-Jensen}, {Salvato}, {Sanders}, {Scoville},
  {Smolcic}, {Stockmann}, {Taniguchi}, {Tasca}, {Toft}, {Vaccari}, \&
  {Zabl}}]{Laigle2016}
{Laigle}, C., {McCracken}, H.~J., {Ilbert}, O., {et~al.} 2016, \apjs, 224, 24,
  \dodoi{10.3847/0067-0049/224/2/24}

\bibitem[{{Leauthaud} {et~al.}(2007){Leauthaud}, {Massey}, {Kneib}, {Rhodes},
  {Johnston}, {Capak}, {Heymans}, {Ellis}, {Koekemoer}, {Le F{\`e}vre},
  {Mellier}, {R{\'e}fr{\'e}gier}, {Robin}, {Scoville}, {Tasca}, {Taylor}, \&
  {Van Waerbeke}}]{Leauthaud2007}
{Leauthaud}, A., {Massey}, R., {Kneib}, J.-P., {et~al.} 2007, \apjs, 172, 219,
  \dodoi{10.1086/516598}

\bibitem[{{Legrand} {et~al.}(2019){Legrand}, {McCracken}, {Davidzon}, {Ilbert},
  {Coupon}, {Aghanim}, {Douspis}, {Capak}, {Le F{\`e}vre}, \&
  {Milvang-Jensen}}]{Legrand2019}
{Legrand}, L., {McCracken}, H.~J., {Davidzon}, I., {et~al.} 2019, \mnras, 486,
  5468, \dodoi{10.1093/mnras/stz1198}

\bibitem[{{Leja} {et~al.}(2019){Leja}, {Carnall}, {Johnson}, {Conroy}, \&
  {Speagle}}]{Leja2019}
{Leja}, J., {Carnall}, A.~C., {Johnson}, B.~D., {Conroy}, C., \& {Speagle},
  J.~S. 2019, \apj, 876, 3, \dodoi{10.3847/1538-4357/ab133c}

\bibitem[{{M{\'a}rmol-Queralt{\'o}} {et~al.}(2016){M{\'a}rmol-Queralt{\'o}},
  {McLure}, {Cullen}, {Dunlop}, {Fontana}, \& {McLeod}}]{MarmolQueralto2016}
{M{\'a}rmol-Queralt{\'o}}, E., {McLure}, R.~J., {Cullen}, F., {et~al.} 2016,
  \mnras, 460, 3587, \dodoi{10.1093/mnras/stw1212}

\bibitem[{{Marshall} {et~al.}(2021){Marshall}, {Wyithe}, {Windhorst}, {Di
  Matteo}, {Ni}, {Wilkins}, {Croft}, \& {Mechtley}}]{Marshall2021}
{Marshall}, M.~A., {Wyithe}, J. S.~B., {Windhorst}, R.~A., {et~al.} 2021,
  \mnras, 506, 1209, \dodoi{10.1093/mnras/stab1763}

\bibitem[{{Mathews} \& {Ferland}(1987)}]{Mathews1987}
{Mathews}, W.~G., \& {Ferland}, G.~J. 1987, \apj, 323, 456,
  \dodoi{10.1086/165843}

\bibitem[{{McDonald} {et~al.}(2012){McDonald}, {Veilleux}, \&
  {Rupke}}]{McDonald2012}
{McDonald}, M., {Veilleux}, S., \& {Rupke}, D. S.~N. 2012, \apj, 746, 153,
  \dodoi{10.1088/0004-637X/746/2/153}

\bibitem[{{McDonald} {et~al.}(2010){McDonald}, {Veilleux}, {Rupke}, \&
  {Mushotzky}}]{McDonald2010}
{McDonald}, M., {Veilleux}, S., {Rupke}, D. S.~N., \& {Mushotzky}, R. 2010,
  \apj, 721, 1262, \dodoi{10.1088/0004-637X/721/2/1262}

\bibitem[{{McDonald} {et~al.}(2013){McDonald}, {Benson}, {Vikhlinin},
  {Stalder}, {Bleem}, {de Haan}, {Lin}, {Aird}, {Ashby}, {Bautz}, {Bayliss},
  {Bocquet}, {Brodwin}, {Carlstrom}, {Chang}, {Cho}, {Clocchiatti}, {Crawford},
  {Crites}, {Desai}, {Dobbs}, {Dudley}, {Foley}, {Forman}, {George},
  {Gettings}, {Gladders}, {Gonzalez}, {Halverson}, {High}, {Holder},
  {Holzapfel}, {Hoover}, {Hrubes}, {Jones}, {Joy}, {Keisler}, {Knox}, {Lee},
  {Leitch}, {Liu}, {Lueker}, {Luong-Van}, {Mantz}, {Marrone}, {McMahon},
  {Mehl}, {Meyer}, {Miller}, {Mocanu}, {Mohr}, {Montroy}, {Murray},
  {Nurgaliev}, {Padin}, {Plagge}, {Pryke}, {Reichardt}, {Rest}, {Ruel}, {Ruhl},
  {Saliwanchik}, {Saro}, {Sayre}, {Schaffer}, {Shirokoff}, {Song},
  {{\v{S}}uhada}, {Spieler}, {Stanford}, {Staniszewski}, {Stark}, {Story}, {van
  Engelen}, {Vanderlinde}, {Vieira}, {Williamson}, {Zahn}, \&
  {Zenteno}}]{McDonald2013}
{McDonald}, M., {Benson}, B.~A., {Vikhlinin}, A., {et~al.} 2013, \apj, 774, 23,
  \dodoi{10.1088/0004-637X/774/1/23}

\bibitem[{{Mehta} {et~al.}(2018){Mehta}, {Scarlata}, {Capak}, {Davidzon},
  {Faisst}, {Hsieh}, {Ilbert}, {Jarvis}, {Laigle}, {Phillips}, {Silverman},
  {Strauss}, {Tanaka}, {Bowler}, {Coupon}, {Foucaud}, {Hemmati}, {Masters},
  {McCracken}, {Mobasher}, {Ouchi}, {Shibuya}, \& {Wang}}]{Mehta2018}
{Mehta}, V., {Scarlata}, C., {Capak}, P., {et~al.} 2018, \apjs, 235, 36,
  \dodoi{10.3847/1538-4365/aab60c}

\bibitem[{{Miley} \& {De Breuck}(2008)}]{Miley2008}
{Miley}, G., \& {De Breuck}, C. 2008, \aapr, 15, 67,
  \dodoi{10.1007/s00159-007-0008-z}

\bibitem[{{Miley} {et~al.}(2004){Miley}, {Overzier}, {Tsvetanov}, {Bouwens},
  {Ben{\'\i}tez}, {Blakeslee}, {Ford}, {Illingworth}, {Postman}, {Rosati},
  {Clampin}, {Hartig}, {Zirm}, {R{\"o}ttgering}, {Venemans}, {Ardila},
  {Bartko}, {Broadhurst}, {Brown}, {Burrows}, {Cheng}, {Cross}, {De Breuck},
  {Feldman}, {Franx}, {Golimowski}, {Gronwall}, {Infante}, {Martel},
  {Menanteau}, {Meurer}, {Sirianni}, {Kimble}, {Krist}, {Sparks}, {Tran},
  {White}, \& {Zheng}}]{Miley2004}
{Miley}, G.~K., {Overzier}, R.~A., {Tsvetanov}, Z.~I., {et~al.} 2004, \nat,
  427, 47, \dodoi{10.1038/nature02125}

\bibitem[{{Moy} \& {Rocca-Volmerange}(2002)}]{Moy2002}
{Moy}, E., \& {Rocca-Volmerange}, B. 2002, \aap, 383, 46,
  \dodoi{10.1051/0004-6361:20011727}

\bibitem[{{Nesvadba} {et~al.}(2020){Nesvadba}, {Bicknell}, {Mukherjee}, \&
  {Wagner}}]{Nesvadba2020}
{Nesvadba}, N.~P.~H., {Bicknell}, G.~V., {Mukherjee}, D., \& {Wagner}, A.~Y.
  2020, \aap, 639, L13, \dodoi{10.1051/0004-6361/202038269}

\bibitem[{{Oke} \& {Gunn}(1983)}]{OkeGunn1983}
{Oke}, J.~B., \& {Gunn}, J.~E. 1983, \apj, 266, 713, \dodoi{10.1086/160817}

\bibitem[{{Overzier} {et~al.}(2008){Overzier}, {Bouwens}, {Cross}, {Venemans},
  {Miley}, {Zirm}, {Ben{\'\i}tez}, {Blakeslee}, {Coe}, {Demarco}, {Ford},
  {Homeier}, {Illingworth}, {Kurk}, {Martel}, {Mei}, {Oliveira},
  {R{\"o}ttgering}, {Tsvetanov}, \& {Zheng}}]{Overzier2008}
{Overzier}, R.~A., {Bouwens}, R.~J., {Cross}, N.~J.~G., {et~al.} 2008, \apj,
  673, 143, \dodoi{10.1086/524342}

\bibitem[{{Overzier} {et~al.}(2009){Overzier}, {Shu}, {Zheng}, {Rettura},
  {Zirm}, {Bouwens}, {Ford}, {Illingworth}, {Miley}, {Venemans}, \&
  {White}}]{Overzier2009}
{Overzier}, R.~A., {Shu}, X., {Zheng}, W., {et~al.} 2009, \apj, 704, 548,
  \dodoi{10.1088/0004-637X/704/1/548}

\bibitem[{{Pentericci} {et~al.}(2000){Pentericci}, {Van Reeven}, {Carilli},
  {R{\"o}ttgering}, \& {Miley}}]{Pentericci2000}
{Pentericci}, L., {Van Reeven}, W., {Carilli}, C.~L., {R{\"o}ttgering},
  H.~J.~A., \& {Miley}, G.~K. 2000, \aaps, 145, 121,
  \dodoi{10.1051/aas:2000104}

\bibitem[{{Robotham} {et~al.}(2018){Robotham}, {Davies}, {Driver}, {Koushan},
  {Taranu}, {Casura}, \& {Liske}}]{Robotham2018}
{Robotham}, A.~S.~G., {Davies}, L.~J.~M., {Driver}, S.~P., {et~al.} 2018,
  \mnras, 476, 3137, \dodoi{10.1093/mnras/sty440}

\bibitem[{{Robotham} {et~al.}(2017){Robotham}, {Taranu}, {Tobar}, {Moffett}, \&
  {Driver}}]{Robotham2017}
{Robotham}, A.~S.~G., {Taranu}, D.~S., {Tobar}, R., {Moffett}, A., \& {Driver},
  S.~P. 2017, \mnras, 466, 1513, \dodoi{10.1093/mnras/stw3039}

\bibitem[{{Saito} {et~al.}(2015){Saito}, {Matsuda}, {Lacey}, {Smail}, {Orsi},
  {Baugh}, {Inoue}, {Tanaka}, {Yamada}, {Ohta}, {De Breuck}, {Kodama}, \&
  {Taniguchi}}]{Saito2015}
{Saito}, T., {Matsuda}, Y., {Lacey}, C.~G., {et~al.} 2015, \mnras, 447, 3069,
  \dodoi{10.1093/mnras/stu2538}

\bibitem[{{S{\'a}nchez-Bl{\'a}zquez} {et~al.}(2006){S{\'a}nchez-Bl{\'a}zquez},
  {Peletier}, {Jim{\'e}nez-Vicente}, {Cardiel}, {Cenarro},
  {Falc{\'o}n-Barroso}, {Gorgas}, {Selam}, \& {Vazdekis}}]{MILES}
{S{\'a}nchez-Bl{\'a}zquez}, P., {Peletier}, R.~F., {Jim{\'e}nez-Vicente}, J.,
  {et~al.} 2006, \mnras, 371, 703, \dodoi{10.1111/j.1365-2966.2006.10699.x}

\bibitem[{{Santos} {et~al.}(2008){Santos}, {Rosati}, {Tozzi}, {B{\"o}hringer},
  {Ettori}, \& {Bignamini}}]{Santos2008}
{Santos}, J.~S., {Rosati}, P., {Tozzi}, P., {et~al.} 2008, \aap, 483, 35,
  \dodoi{10.1051/0004-6361:20078815}

\bibitem[{{Smith} {et~al.}(2016){Smith}, {Best}, {Duncan}, {Hatch}, {Jarvis},
  {R{\"o}ttgering}, {Simpson}, {Stott}, {Cochrane}, {Coppin}, {Dannerbauer},
  {Davis}, {Geach}, {Hale}, {Hardcastle}, {Hatfield}, {Houghton}, {Maddox},
  {McGee}, {Morabito}, {Nisbet}, {Pandey-Pommier}, {Prandoni}, {Saxena},
  {Shimwell}, {Tarr}, {van Bemmel}, {Verma}, {White}, \&
  {Williams}}]{WEAVELOFAR}
{Smith}, D.~J.~B., {Best}, P.~N., {Duncan}, K.~J., {et~al.} 2016, in SF2A-2016:
  Proceedings of the Annual meeting of the French Society of Astronomy and
  Astrophysics, ed. C.~{Reyl{\'e}}, J.~{Richard}, L.~{Cambr{\'e}sy},
  M.~{Deleuil}, E.~{P{\'e}contal}, L.~{Tresse}, \& I.~{Vauglin}, 271--280.
\newblock \doarXiv{1611.02706}

\bibitem[{{Song} {et~al.}(2016){Song}, {Finkelstein}, {Ashby}, {Grazian}, {Lu},
  {Papovich}, {Salmon}, {Somerville}, {Dickinson}, {Duncan}, {Faber}, {Fazio},
  {Ferguson}, {Fontana}, {Guo}, {Hathi}, {Lee}, {Merlin}, \&
  {Willner}}]{Song2016}
{Song}, M., {Finkelstein}, S.~L., {Ashby}, M. L.~N., {et~al.} 2016, \apj, 825,
  5, \dodoi{10.3847/0004-637X/825/1/5}

\bibitem[{{Stanway} {et~al.}(2016){Stanway}, {Eldridge}, \&
  {Becker}}]{Stanway2016}
{Stanway}, E.~R., {Eldridge}, J.~J., \& {Becker}, G.~D. 2016, \mnras, 456, 485,
  \dodoi{10.1093/mnras/stv2661}

\bibitem[{{Steinbring}(2014)}]{Steinbring2014}
{Steinbring}, E. 2014, \aj, 148, 10, \dodoi{10.1088/0004-6256/148/1/10}

\bibitem[{{Swinbank} {et~al.}(2015){Swinbank}, {Vernet}, {Smail}, {De Breuck},
  {Bacon}, {Contini}, {Richard}, {R{\"o}ttgering}, {Urrutia}, \&
  {Venemans}}]{Swinbank2015}
{Swinbank}, A.~M., {Vernet}, J.~D.~R., {Smail}, I., {et~al.} 2015, \mnras, 449,
  1298, \dodoi{10.1093/mnras/stv366}

\bibitem[{{Tacchella} {et~al.}(2022){Tacchella}, {Johnson}, {Robertson},
  {Carniani}, {D'Eugenio}, {Kumar}, {Maiolino}, {Nelson}, {Suess}, {{\"U}bler},
  {Williams}, {Adebusola}, {Alberts}, {Arribas}, {Bhatawdekar}, {Bonaventura},
  {Bowler}, {Bunker}, {Cameron}, {Curti}, {Egami}, {Eisenstein}, {Frye},
  {Hainline}, {Helton}, {Ji}, {Looser}, {Lyu}, {Perna}, {Rawle}, {Rieke},
  {Rieke}, {Saxena}, {Sandles}, {Shivaei}, {Simmonds}, {Sun}, {Willmer},
  {Willott}, \& {Witstok}}]{Tacchella2022b}
{Tacchella}, S., {Johnson}, B.~D., {Robertson}, B.~E., {et~al.} 2022, arXiv
  e-prints, arXiv:2208.03281.
\newblock \doarXiv{2208.03281}

\bibitem[{{Topping} {et~al.}(2022){Topping}, {Stark}, {Endsley}, {Plat},
  {Whitler}, {Chen}, \& {Charlot}}]{Topping2022}
{Topping}, M.~W., {Stark}, D.~P., {Endsley}, R., {et~al.} 2022, arXiv e-prints,
  arXiv:2208.01610.
\newblock \doarXiv{2208.01610}

\bibitem[{{Tremblay} {et~al.}(2015){Tremblay}, {O'Dea}, {Baum}, {Mittal},
  {McDonald}, {Combes}, {Li}, {McNamara}, {Bremer}, {Clarke}, {Donahue},
  {Edge}, {Fabian}, {Hamer}, {Hogan}, {Oonk}, {Quillen}, {Sanders},
  {Salom{\'e}}, \& {Voit}}]{Tremblay2015}
{Tremblay}, G.~R., {O'Dea}, C.~P., {Baum}, S.~A., {et~al.} 2015, \mnras, 451,
  3768, \dodoi{10.1093/mnras/stv1151}

\bibitem[{{Trenti} {et~al.}(2011){Trenti}, {Bradley}, {Stiavelli}, {Oesch},
  {Treu}, {Bouwens}, {Shull}, {MacKenty}, {Carollo}, \&
  {Illingworth}}]{Trenti2011}
{Trenti}, M., {Bradley}, L.~D., {Stiavelli}, M., {et~al.} 2011, \apjl, 727,
  L39, \dodoi{10.1088/2041-8205/727/2/L39}

\bibitem[{{Tsvetanov} {et~al.}(1996){Tsvetanov}, {Morse}, {Wilson}, \&
  {Cecil}}]{Tsvetanov1996}
{Tsvetanov}, Z.~I., {Morse}, J.~A., {Wilson}, A.~S., \& {Cecil}, G. 1996, \apj,
  458, 172, \dodoi{10.1086/176801}

\bibitem[{{Vayner} {et~al.}(2023){Vayner}, {Zakamska}, {Ishikawa}, {Sankar},
  {Wylezalek}, {Rupke}, {Veilleux}, {Bertemes}, {Barrera-Ballesteros}, {Chen},
  {Diachenko}, {Goulding}, {Greene}, {Hainline}, {Hamann}, {Heckman},
  {Johnson}, {Murphree}, {Lim}, {Liu}, {Lutz}, {L{\"u}tzgendorf}, {McCrory},
  {Mainieri}, {Nesvadba}, {Ogle}, {Sturm}, \& {Whitesell}}]{Vayner2023}
{Vayner}, A., {Zakamska}, N.~L., {Ishikawa}, Y., {et~al.} 2023, arXiv e-prints,
  arXiv:2303.06970, \dodoi{10.48550/arXiv.2303.06970}

\bibitem[{{Venemans} {et~al.}(2002){Venemans}, {Kurk}, {Miley},
  {R{\"o}ttgering}, {van Breugel}, {Carilli}, {De Breuck}, {Ford}, {Heckman},
  {McCarthy}, \& {Pentericci}}]{Venemans2002}
{Venemans}, B.~P., {Kurk}, J.~D., {Miley}, G.~K., {et~al.} 2002, \apjl, 569,
  L11, \dodoi{10.1086/340563}

\bibitem[{{Whitler} {et~al.}(2022{\natexlab{a}}){Whitler}, {Endsley}, {Stark},
  {Topping}, {Chen}, \& {Charlot}}]{Whitler2022b}
{Whitler}, L., {Endsley}, R., {Stark}, D.~P., {et~al.} 2022{\natexlab{a}},
  arXiv e-prints, arXiv:2208.01599.
\newblock \doarXiv{2208.01599}

\bibitem[{{Whitler} {et~al.}(2022{\natexlab{b}}){Whitler}, {Stark}, {Endsley},
  {Leja}, {Charlot}, \& {Chevallard}}]{Whitler2022a}
{Whitler}, L., {Stark}, D.~P., {Endsley}, R., {et~al.} 2022{\natexlab{b}},
  arXiv e-prints, arXiv:2206.05315.
\newblock \doarXiv{2206.05315}

\bibitem[{{Windhorst} {et~al.}(2022){Windhorst}, {Cohen}, {Jansen}, {Summers},
  {Tompkins}, {Conselice}, {Driver}, {Yan}, {Coe}, {Frye}, {Grogin},
  {Koekemoer}, {Marshall}, {Pirzkal}, {Robotham}, {Ryan}, {Willmer},
  {Carleton}, {Diego}, {Keel}, {O'Brien}, {Porto}, {Redshaw}, {Scheller},
  {Swirbul}, {Wilkins}, {Willner}, {Zitrin}, {Adams}, {Austin}, {Arendt},
  {Beacom}, {Bhatawdekar}, {Bradley}, {Broadhurst}, {Cheng}, {Civano}, {Dai},
  {Dole}, {D'Silva}, {Duncan}, {Fazio}, {Ferrami}, {Ferreira}, {Finkelstein},
  {Furtak}, {Griffiths}, {Hammel}, {Harrington}, {Hathi}, {Holwerda}, {Huang},
  {Hyun}, {Im}, {Joshi}, {Kamieneski}, {Kelly}, {Larson}, {Li}, {Lim}, {Ma},
  {Maksym}, {Manzoni}, {Meena}, {Milam}, {Nonino}, {Pascale}, {Pierel},
  {Petric}, {Polletta}, {Rottgering}, {Rutkowski}, {Smail}, {Straughn},
  {Strolger}, {Trussler}, {Wang}, {Welch}, {Wyithe}, {Yun}, {Zackrisson},
  {Zhang}, \& {Zhao}}]{PearlsOverview}
{Windhorst}, R.~A., {Cohen}, S.~H., {Jansen}, R.~A., {et~al.} 2022, arXiv
  e-prints, arXiv:2209.04119.
\newblock \doarXiv{2209.04119}

\bibitem[{{Zirm} {et~al.}(2005){Zirm}, {Overzier}, {Miley}, {Blakeslee},
  {Clampin}, {De Breuck}, {Demarco}, {Ford}, {Hartig}, {Homeier},
  {Illingworth}, {Martel}, {R{\"o}ttgering}, {Venemans}, {Ardila}, {Bartko},
  {Ben{\'\i}tez}, {Bouwens}, {Bradley}, {Broadhurst}, {Brown}, {Burrows},
  {Cheng}, {Cross}, {Feldman}, {Franx}, {Golimowski}, {Goto}, {Gronwall},
  {Holden}, {Infante}, {Kimble}, {Krist}, {Lesser}, {Mei}, {Menanteau},
  {Meurer}, {Motta}, {Postman}, {Rosati}, {Sirianni}, {Sparks}, {Tran},
  {Tsvetanov}, {White}, \& {Zheng}}]{Zirm2005}
{Zirm}, A.~W., {Overzier}, R.~A., {Miley}, G.~K., {et~al.} 2005, \apj, 630, 68,
  \dodoi{10.1086/431921}

\end{thebibliography}
\bibliographystyle{aasjournal}

\appendix
\section{Resolved SEDs for TNJ1338}
Fig.~\ref{fig:sed_galaxy} presents the observed SEDs and best-fit stellar population models for the resolved segments not presented in Fig.~\ref{fig:sed_shocks} or \ref{fig:sed_core}.

\begin{figure*}
\centering
    \includegraphics[width=0.47\textwidth]{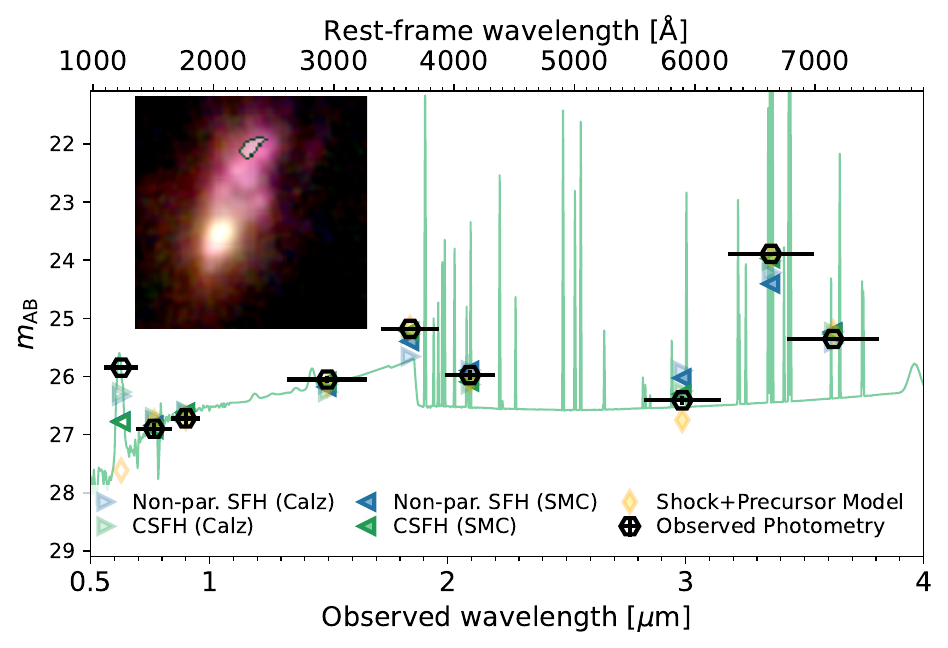}
    \includegraphics[width=0.47\textwidth]{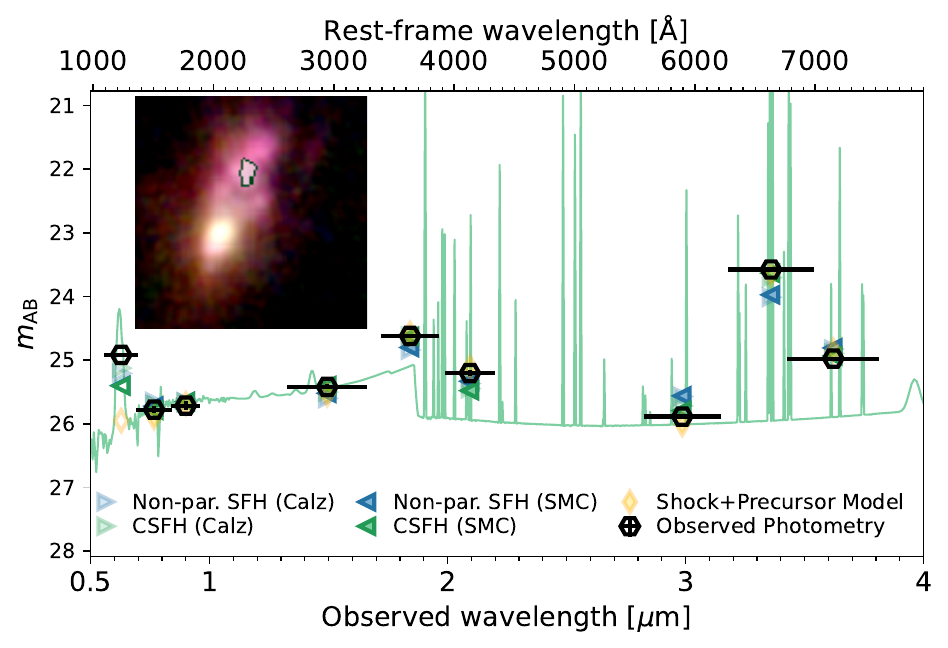}
    \includegraphics[width=0.47\textwidth]{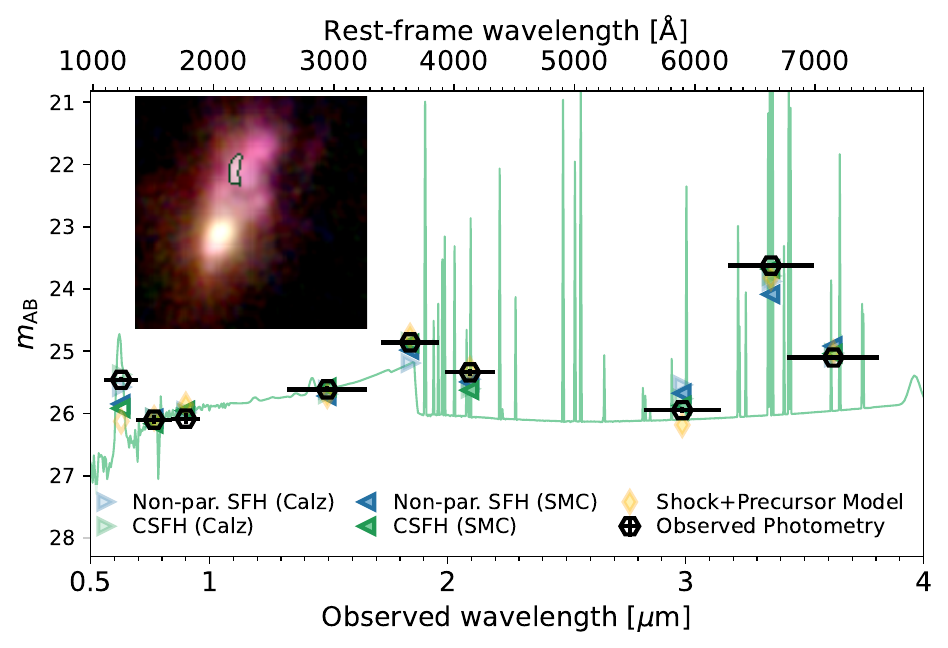}
    \includegraphics[width=0.47\textwidth]{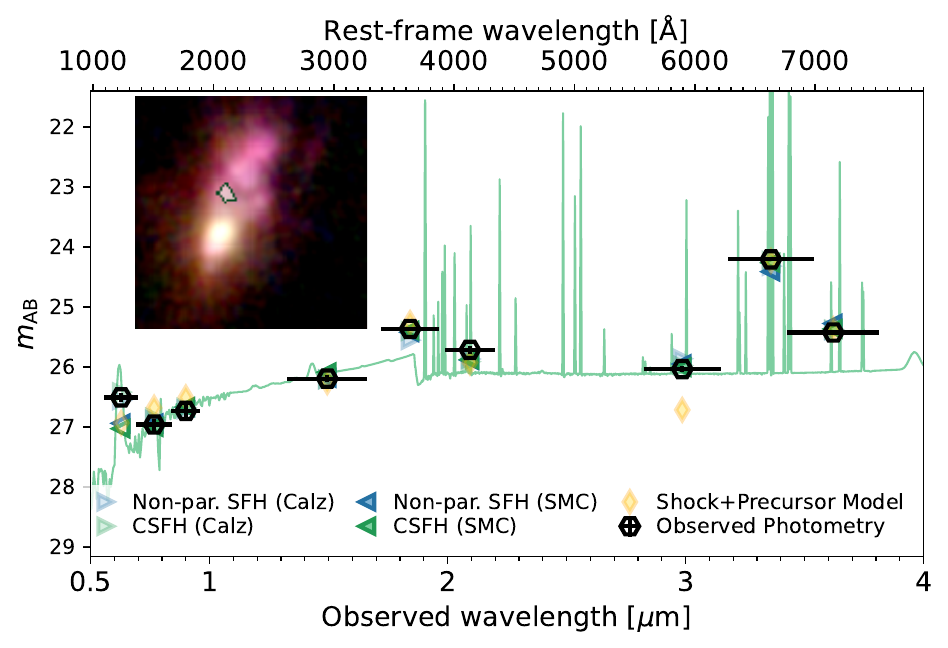}
    \includegraphics[width=0.47\textwidth]{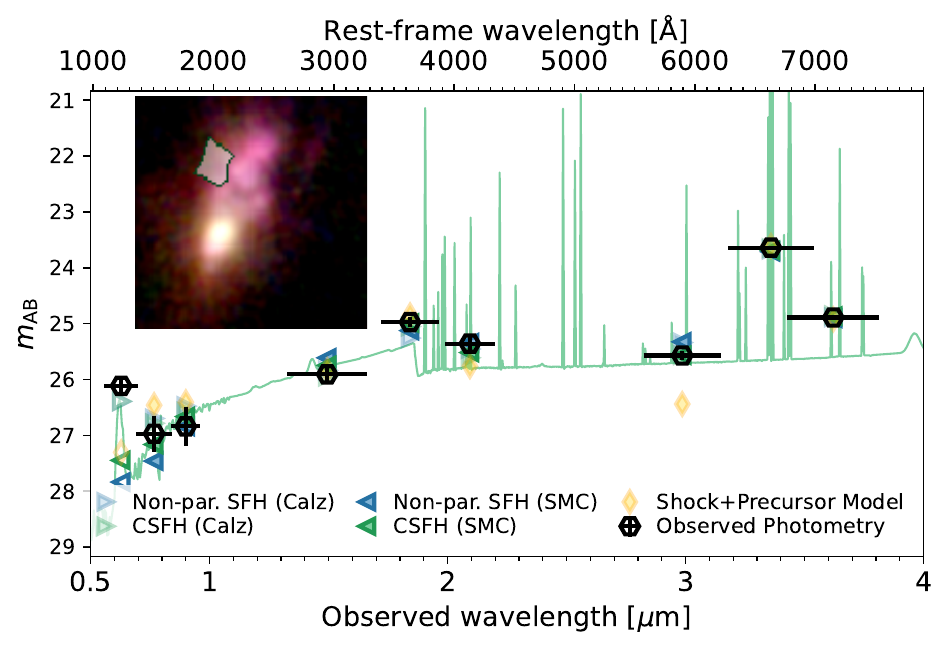}
    \includegraphics[width=0.47\textwidth]{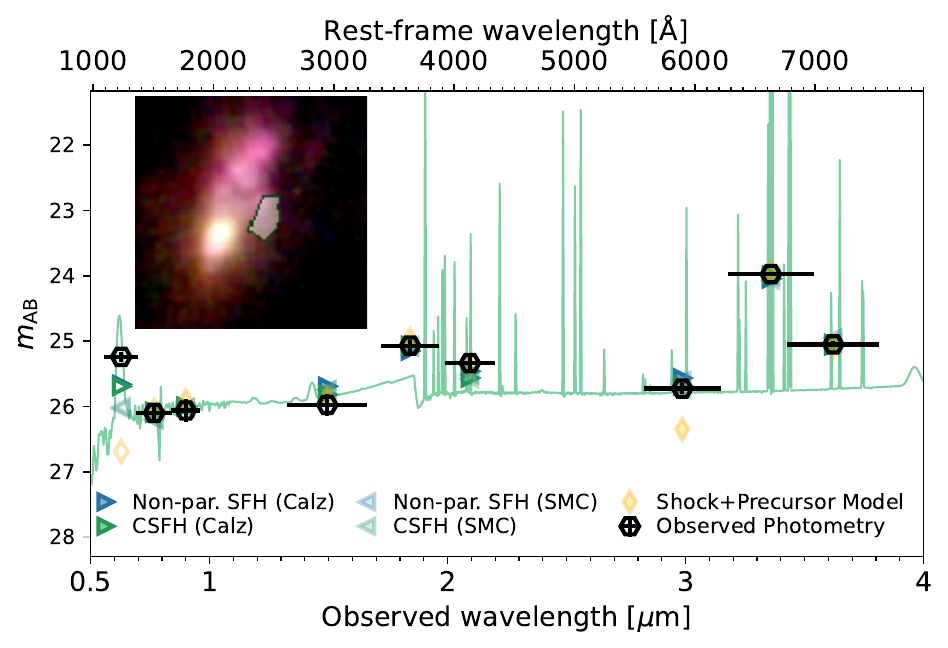}
	\caption{Observed SED and best-fitting model photometry for the regions of TNJ1338 not shown in Fig.~\ref{fig:sed_shocks} or ~\ref{fig:sed_core}. Plot symbols and colours are defined as in Fig.~\ref{fig:sed_shocks}.}
    \label{fig:sed_galaxy}
\end{figure*}

\begin{figure*}\ContinuedFloat
\centering
    \includegraphics[width=0.47\textwidth]{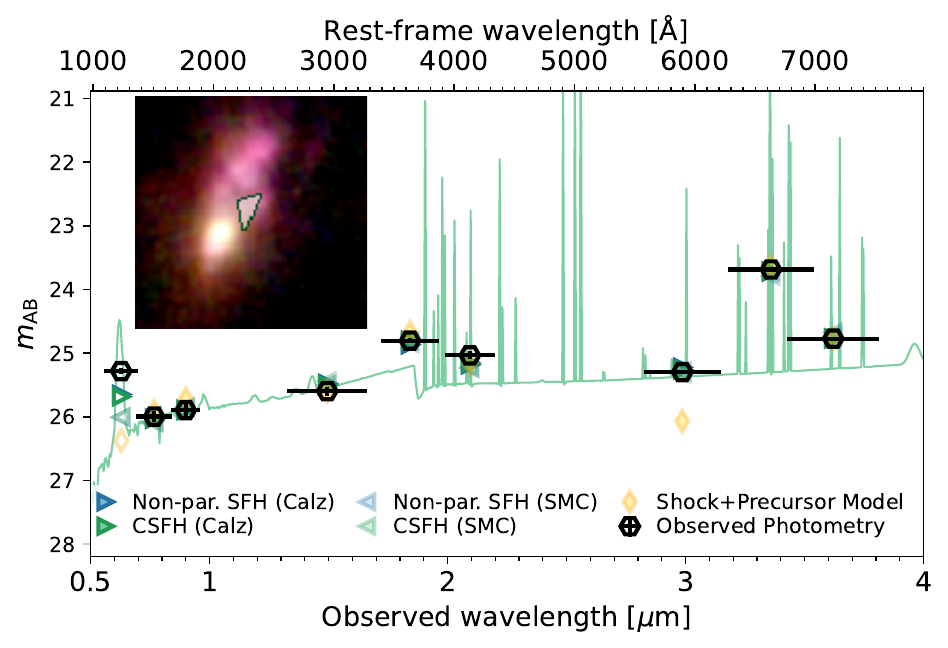}
    \includegraphics[width=0.47\textwidth]{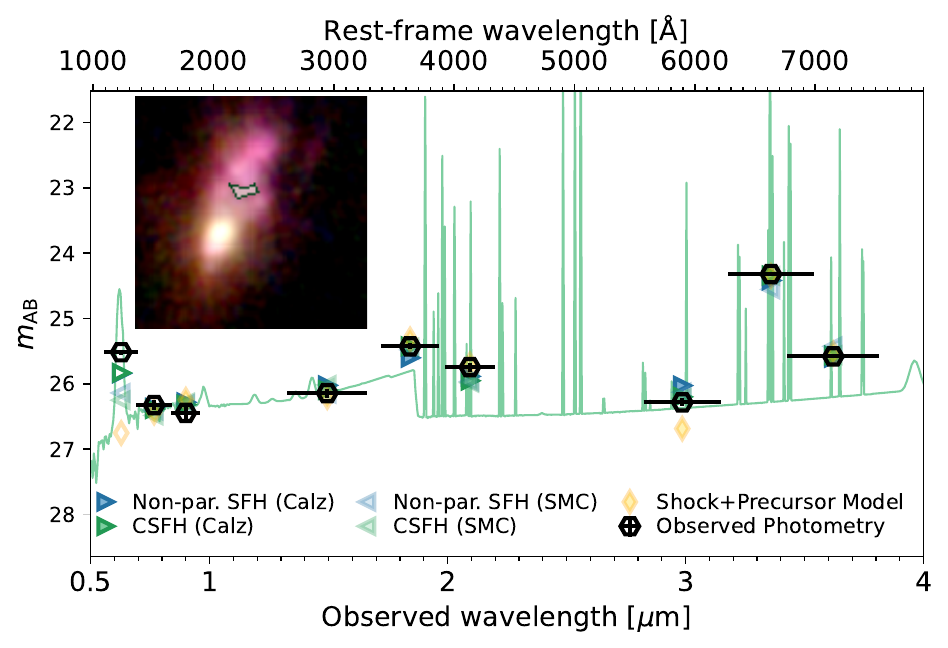}
    \includegraphics[width=0.47\textwidth]{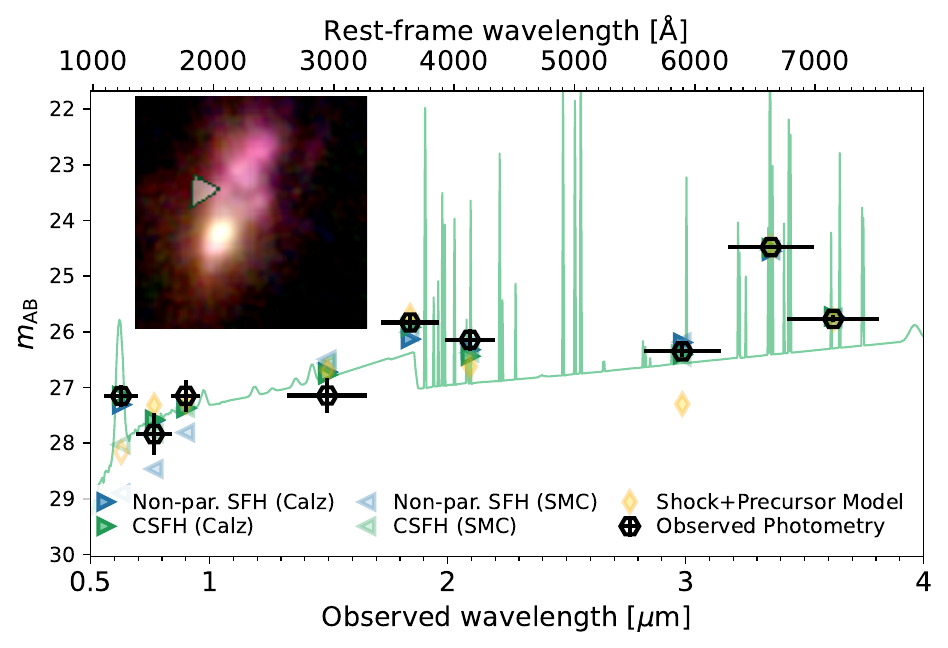}
    \includegraphics[width=0.47\textwidth]{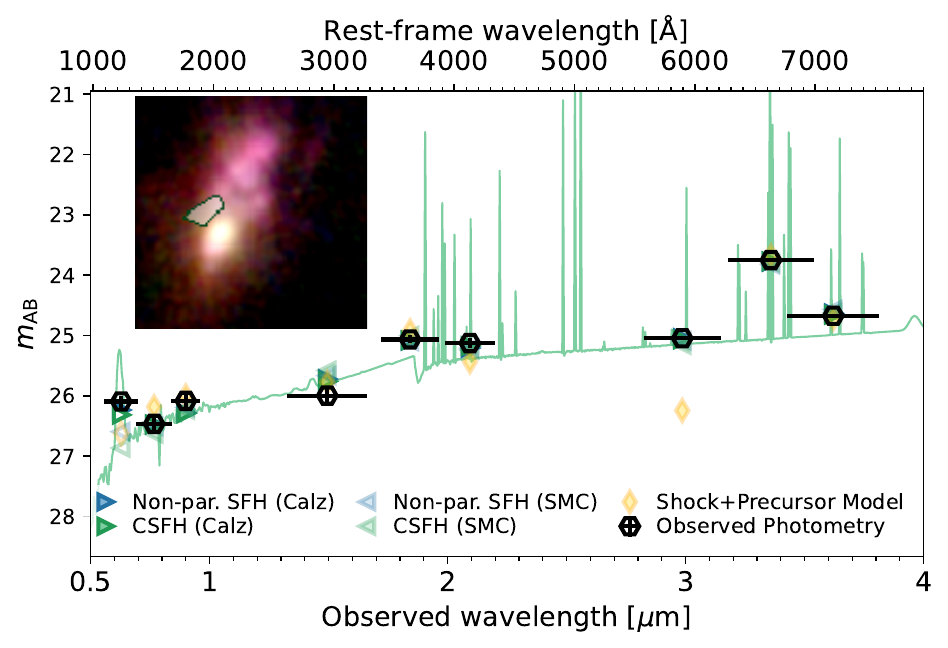}
    \includegraphics[width=0.47\textwidth]{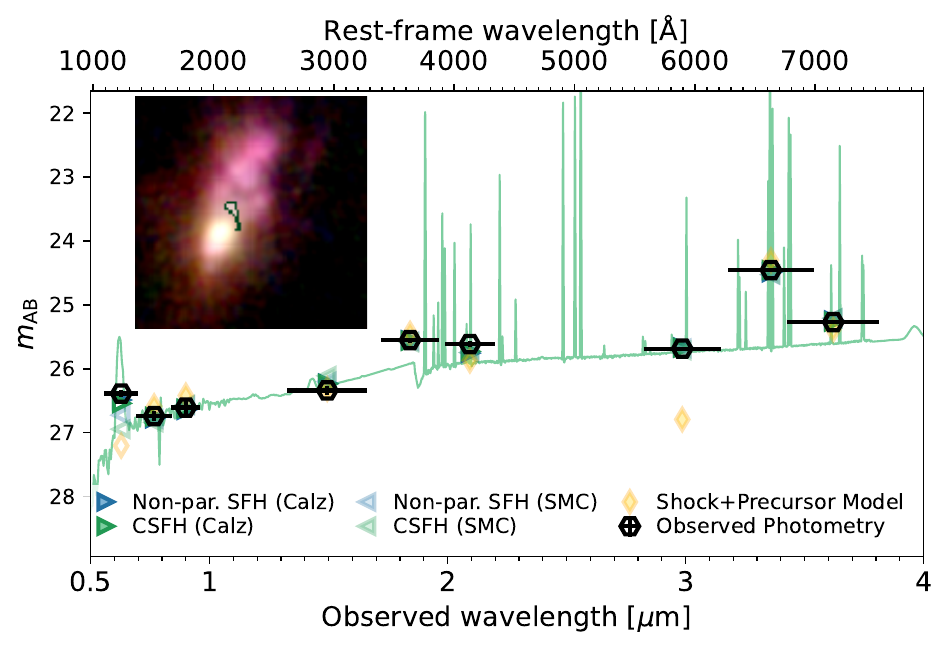}
    \includegraphics[width=0.47\textwidth]{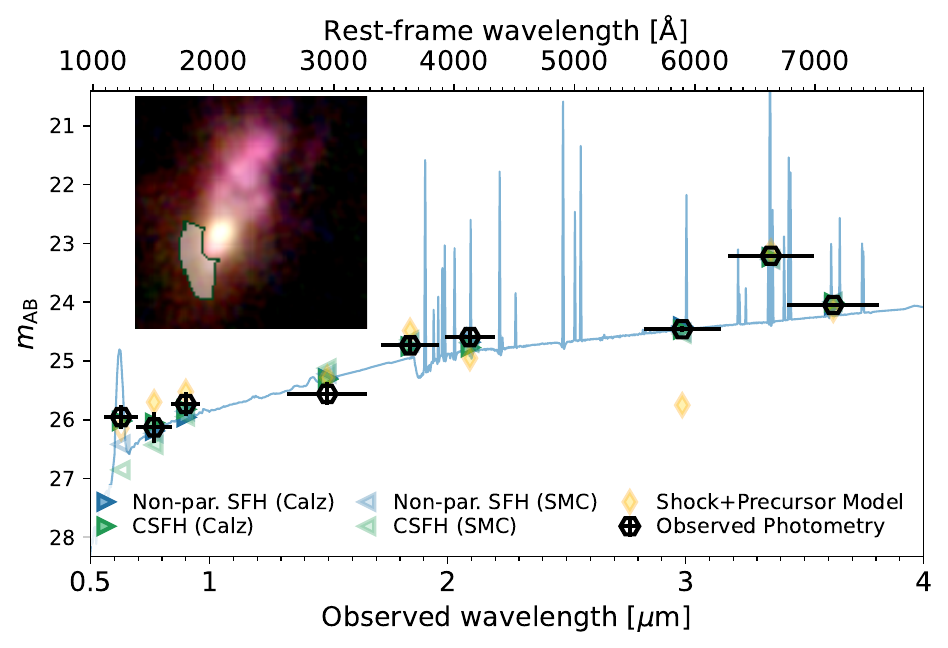}
    \includegraphics[width=0.47\textwidth]{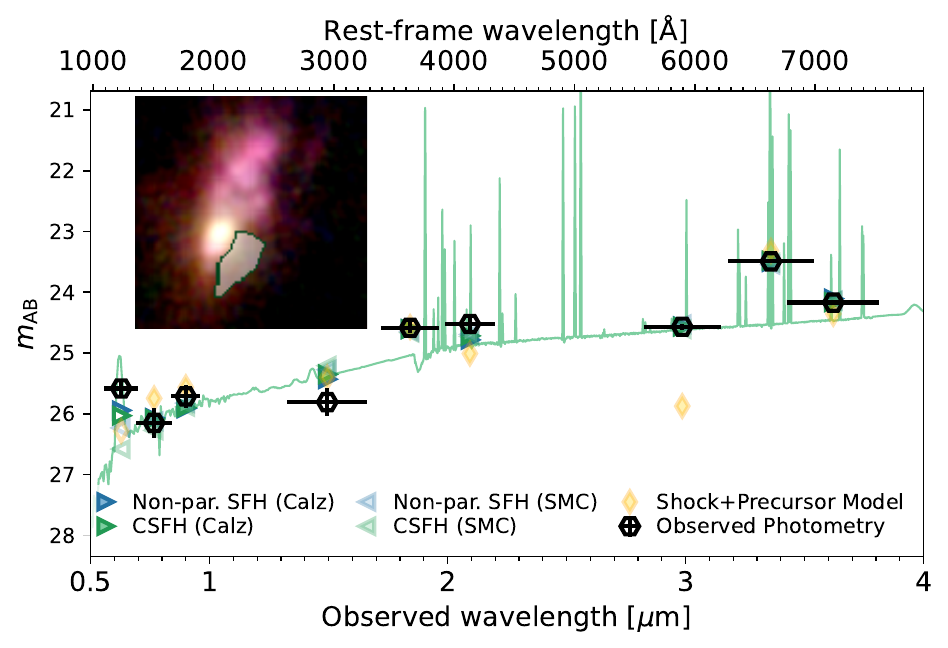}
	\caption{Continued.}
\end{figure*}



% Don't change these lines
\bsp	% typesetting comment
\label{lastpage}
\end{document}